\pdfoutput=1
\newcommand*{\ATLASLATEXPATH}{latex/}
\documentclass[cernpreprint,texlive=2014,txfonts,UKenglish]{\ATLASLATEXPATH atlasdoc} 

\usepackage{\ATLASLATEXPATH atlaspackage}
\usepackage{\ATLASLATEXPATH atlasbiblatex}

\usepackage{\ATLASLATEXPATH atlascontribute}

\usepackage{\ATLASLATEXPATH atlasphysics}

\graphicspath{{logos/}{figures/}}

\usepackage{ktSplittingsPaper-defs}
\usepackage{tikz}

\AtlasTitle{Measurement of the \kt\ splitting scales in \Zll\ events in $pp$ collisions at $\rts = 8\,\TeV$ with the ATLAS detector}

\author{The ATLAS Collaboration}

\AtlasRefCode{STDM-2015-14}

\PreprintIdNumber{CERN-EP-2017-033}

\AtlasJournalRef{JHEP 08 (2017) 026}
\AtlasDOI{10.1007/JHEP08(2017)026}

\AtlasAbstract{
  A measurement of the splitting scales occuring in the \kt\ jet-clustering algorithm is presented for final states containing 
  a \Z\ boson. The measurement is done using 20.2\,\ifb\ of proton--proton collision data collected at a centre-of-mass energy 
  of $\rts = 8\,\TeV$ 
  by the ATLAS experiment at the LHC in 2012. The measurement is based on charged-particle track information, which is measured with 
  excellent precision in the \pT\ region relevant for the transition between the perturbative and the non-perturbative regimes. The 
  data distributions are corrected for detector effects, and are found to deviate from state-of-the-art predictions in various regions 
  of the observables.
}

\hypersetup{pdftitle={ATLAS draft},pdfauthor={The ATLAS Collaboration}}

\begin{document}

\maketitle

\tableofcontents

\section{Introduction}
\label{sec:intro}

A collimated spray of particles arising from a cascade of strong interactions
is commonly referred to as a hadronic jet. 
Jet production in association with other final-state particles can constitute signal as well as an important background process in
many of the precision measurements and new-physics searches conducted at CERN’s Large Hadron Collider (LHC). A good understanding
of processes initiated by strong interactions is therefore crucial.

Jet production in association with a leptonically decaying heavy gauge boson ($V = W,Z$) allows effects of the strong interaction to be studied in a 
relatively clean environment. Good progress has been made in recent years towards higher-order calculations of these processes in quantum chromodynamics (QCD),
e.g.\ fixed-order calculations of high jet multiplicities at next-to-leading order (NLO)~\cite{Ita:2011wn,Bern:2013gka} and the inclusive 
as well as $V + \mbox{1-jet}$ processes at next-to-next-to-leading order (NNLO)~\cite{Boughezal:2015dva}. Furthermore, new methods have 
been published for matching NLO $V + \mbox{multijet}$ predictions with the parton shower in a merged 
sample~\cite{Hoeche:2012yf,Frederix:2012ps,Lonnblad:2012ix}, or to match NNLO calculations to a parton shower for the inclusive 
processes~\cite{Hoeche:2014aia,Karlberg:2014qua}. Comparisons of precision measurements to these accurate predictions are a powerful 
means by which to study aspects of QCD.

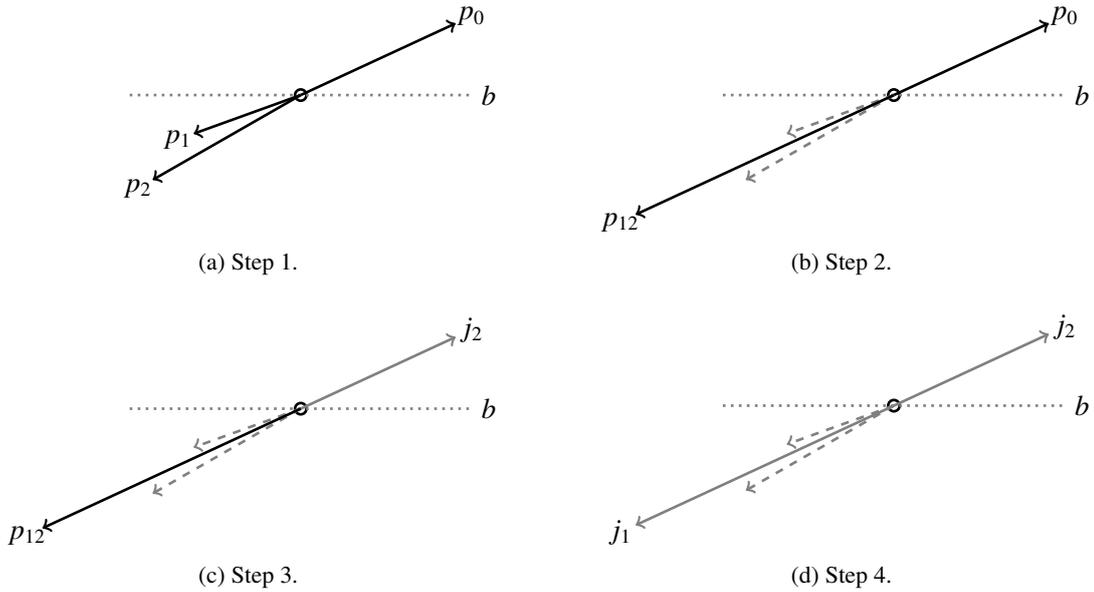
\begin{figure}[htbp]
\centering
\subfloat[Step 1.]{
\begin{tikzpicture}[scale=0.75,every node/.style={transform shape}]
  \draw[line width=1pt] (0,0) circle (0.1cm);
  \draw[line width=1pt,color=gray,dotted] (-3cm,0) -- (3cm,0);
  \node at (3.3cm,0) {\Large{$b$}};
  \draw[line width=1pt,->] (0,0) -- (25:3cm);
  \node at (25:3.3cm) {\Large{$p_0$}};
  \draw[line width=1pt,color=white,->] (0,0) -- (205:5cm);
  \node at (205:5.3cm) {\color{white}{\Large{$p_{12}$}}};
  \draw[line width=1pt,->] (0,0) -- (200:2cm);
  \node at (200:2.3cm) {\Large{$p_1$}};
  \draw[line width=1pt,->] (0,0) -- (210:3cm);
  \node at (210:3.3cm) {\Large{$p_2$}};
\end{tikzpicture}}\hspace{1cm}
\subfloat[Step 2.]{
\begin{tikzpicture}[scale=0.75,every node/.style={transform shape}]
  \draw[line width=1pt] (0,0) circle (0.1cm);
  \draw[line width=1pt,color=gray,dotted] (-3cm,0) -- (3cm,0);
  \node at (3.3cm,0) {\Large{$b$}};
  \draw[line width=1pt,->] (0,0) -- (25:3cm);
  \node at (25:3.3cm) {\Large{$p_0$}};
  \draw[line width=1pt,color=gray,dashed,->] (0,0) -- (200:2cm);
  \draw[line width=1pt,color=gray,dashed,->] (0,0) -- (210:3cm);
  \draw[line width=1pt,->] (0,0) -- (205:5cm);
  \node at (205:5.3cm) {\Large{$p_{12}$}};
\end{tikzpicture}}\\
\subfloat[Step 3.]{
\begin{tikzpicture}[scale=0.75,every node/.style={transform shape}]
  \draw[line width=1pt] (0,0) circle (0.1cm);
  \draw[line width=1pt,color=gray,dotted] (-3cm,0) -- (3cm,0);
  \node at (3.3cm,0) {\Large{$b$}};
  \draw[line width=1pt,color=gray,->] (0,0) -- (25:3cm);
  \node at (25:3.3cm) {\Large{$j_2$}};
  \draw[line width=1pt,color=gray,dashed,->] (0,0) -- (200:2cm);
  \draw[line width=1pt,color=gray,dashed,->] (0,0) -- (210:3cm);
  \draw[line width=1pt,->] (0,0) -- (205:5cm);
  \node at (205:5.3cm) {\Large{$p_{12}$}};
\end{tikzpicture}}\hspace{1cm}
\subfloat[Step 4.]{
\begin{tikzpicture}[scale=0.75,every node/.style={transform shape}]
  \draw[line width=1pt] (0,0) circle (0.1cm);
  \draw[line width=1pt,color=gray,dotted] (-3cm,0) -- (3cm,0);
  \node at (3.3cm,0) {\Large{$b$}};
  \draw[line width=1pt,color=gray,->] (0,0) -- (25:3cm);
  \node at (25:3.3cm) {\Large{$j_2$}};
  \draw[line width=1pt,color=gray,dashed,->] (0,0) -- (200:2cm);
  \draw[line width=1pt,color=gray,dashed,->] (0,0) -- (210:3cm);
  \draw[line width=1pt,color=gray,->] (0,0) -- (205:5cm);
  \node at (205:5.3cm) {\color{white}{\Large{$p_{12}$}}};
  \node at (205:5.3cm) {\Large{$j_1$}};
\end{tikzpicture}}
\caption{Simplified illustration of the \kt\ clustering algorithm, starting with three input momenta
$p_0$, $p_1$ and $p_2$ (step 1). The dotted line labelled $b$ represents the beam line. 
In step~2, the minimum distance measure is the one between two input momenta $p_1$ and $p_2$, so that the two input momenta
are replaced by their vector combination. In step~3, the minimum distance measure is between the $p_0$ and the beam line, so
that $p_0$ is declared a jet ($j_2$) and removed from the input list. Finally in step~4, there is only the combined input momentum
$p_{12}$ left and so it will be declared a jet ($j_1$) and the algorithm terminates.}
\label{fig:kt-sequence}
\end{figure}

While properties of the jets can be studied directly using the jet momenta, a complementary approach is taken
in this paper by studying the jet production rates at different resolution scales. To this end, splitting scales of jets are constructed using 
an infrared-safe clustering algorithm based on sequential combination of the input momenta.
In this analysis the \kt\ algorithm~\cite{Catani:1993hr,Salam:2009jx} is used, with distance measures defined for every iteration as follows:
\begin{eqnarray}
d_{ij} &=& \mbox{min}\left( p_{\text{T},i}^2, p_{\text{T},j}^2 \right) \times \dfrac{ \Delta R_{ij}^2 }{ R^2 }, \\
d_{ib} &=& p_{\text{T},i}^2,
\end{eqnarray}
where the transverse momentum \pT\ carries an index corresponding to the $i^\mathrm{th}$ and $j^\mathrm{th}$ constituent momentum 
in the input list, for all possible permutations of $i$ and $j$ in the given clustering step.
The input momenta separation $\Delta R_{ij}$ 
is defined in terms of the rapidity $y$ and the azimuthal angle $\phi$ via the relation 
$\left(\Delta R_{ij}\right)^2 = \left(y_i - y_j\right)^2 + \left(\phi_i - \phi_j\right)^2$. The index $b$ denotes the beam line and the parameter
$R$ governs the average cone size in $y$--$\phi$ space around the jet axis. For a given iteration of the algorithm in which the
number of input momenta drops from $k + 1$ to $k$, the associated squared splitting scale $d_k$ is given by the minimum of all the $d_{ij}$ and
$d_{ib}$ scales defined for that iteration step:
\begin{equation}
  d_k = \min_{i,j} (d_{ij}, d_{ib}).
\end{equation}
If this minimum is a $d_{ij}$, the $i^\mathrm{th}$ and $j^\mathrm{th}$ momenta in the input list are replaced by their combination. 
If the minimum is a $d_{ib}$, the $i^\mathrm{th}$ momentum is removed from the input collection and is declared a jet.
The index $k$ defines the order of the splitting scale, with $k=0$ being the last iteration step before the algorithm terminates.
Hence the zeroth-order splitting scale, \dsub{0}, corresponds to the \pT\ of the leading \kt-jet,
and one can regard the $N^\mathrm{th}$ splitting scale, \dsub{N}, as the distance measure at which an $N$-jet event is resolved as an $(N+1)$-jet event.
The steps of a \kt\ clustering sequence 
using three input momenta are illustrated in Figure~\ref{fig:kt-sequence}.

In this paper, measurements of differential distributions of the splitting scales occurring in the \kt\ clustering
algorithm using charged-particle tracks in events with \Zjets\ are presented. The aim is to constrain the theoretical 
modelling of strong-interaction effects, and charged-particle tracks are used instead of calorimeter cells to reduce the 
systematic uncertainties of the measurements significantly.
In addition to these primary results using only charged-particle information, less precise extrapolated results including neutral
particles are also provided to allow comparisons to fixed-order calculations.

The measurements are performed independently in the \Zee\ and \Zmm\ decay channels as well as for
jet-radius parameters of $R=0.4$ and $R=1.0$ in each decay channel. The presented analysis is complementary to 
the ATLAS measurement of the \kt\ splitting scales in \Wjets\ events at $\rts = 7\,\TeV$~\cite{STDM-2012-12}.

\section{ATLAS detector}
\label{sec:detector}

The ATLAS detector is described in detail in Ref.~\cite{PERF-2007-01}. Tracks and interaction vertices are reconstructed 
with the inner detector (ID) tracking system, consisting of a silicon pixel detector, a silicon microstrip detector (SCT) and 
a transition radiation tracker. The ID is immersed in a 2\,T axial magnetic field, providing charged-particle tracking in 
the pseudorapidity range $\abseta < 2.5$.\footnote{ATLAS uses a right-handed coordinate system with its origin at the nominal
interaction point (IP) in the centre of the detector and the $z$-axis along the beam pipe. The $x$-axis points from the IP to the
centre of the LHC ring, and the $y$-axis points upward. Cylindrical coordinates $(r,\phi)$ are used in the transverse plane, $\phi$
being the azimuthal angle around the $z$-axis. The pseudorapidity is defined in terms of the polar angle $\theta$ as
$\eta=-\ln\tan(\theta/2)$.} The ATLAS calorimeter system provides fine-grained measurements of shower energy depositions over a 
wide range of $\eta$. An electromagnetic liquid-argon sampling calorimeter covers the region $\abseta < 3.2$ and is divided into 
a barrel part ($\abseta < 1.475$) and an endcap part ($1.375 < \abseta < 3.2$). The hadronic barrel calorimeter ($\abseta < 1.7$)
consists of steel absorbers and active scintillator tiles. The hadronic endcap calorimeter ($1.5 < \abseta < 3.2$) and forward 
electromagnetic and hadronic calorimeters ($3.1 < \abseta < 4.9$) use liquid argon as the active medium. The muon spectrometer 
comprises separate trigger and high-precision tracking chambers measuring the deflection of muons in a magnetic field generated by
superconducting air-core toroids. The precision chamber system covers the region $\abseta < 2.7$ with three layers of monitored drift 
tube chambers, complemented by cathode strip chambers in the forward region. The muon trigger system
covers the range $\abseta < 2.4$ with resistive plate chambers in the barrel, and thin gap chambers in the endcap regions. A three-level 
trigger system is used to select events of interest~\cite{PERF-2011-02}. The Level-1 trigger reduces the event rate to less than 75\,kHz 
using hardware-based trigger algorithms acting on a subset of the available detector information. Two software-based trigger levels 
then reduce the event rate further to about 400\,Hz using the complete detector information.

\section{Event selection}

The measurement is performed using proton--proton collision data recorded at a centre-of-mass energy
of $\rts = 8\,\TeV$. The data were collected between April and December 2012 in data-taking periods
where the detector was fully operational, resulting in an integrated luminosity of $20.2\,\ifb$.

\subsection{Object reconstruction and event selection at detector level}
\label{sec:reconstruction}

Events containing a dilepton candidate were retained for 
further analysis using dedicated triggers
requiring the presence of two oppositely charged electrons with transverse momentum above 12\,\GeV{} or two oppositely charged muons with transverse momenta above 18\,\GeV{} and 8\,\GeV{} respectively.

Inner detector tracks are selected in the phase-space region $\pT > 400\,\MeV$ and $\abseta < 2.5$. The track candidates 
are required to have at least one hit in the pixel detector and at least five SCT hits. A hit in the innermost pixel layer is 
required in cases where the track candidates have passed through an active region of that layer. The track reconstructed from 
the hits is then extrapolated and combined with information from the transition radiation tracker~\cite{ATLAS-CONF-2012-042}. The 
reduced $\chi^2$ of the track fit is required to be less than 3 in order to remove mismeasured tracks or 
combinatorial background. In order to reject backgrounds stemming from other proton--proton collisions in the same or different
bunch crossings (pileup), the transverse and longitudinal impact parameters are required to be $|d_0| < 1.0$\,mm and
$|z_0\times\sin\theta|<0.6$\,mm with respect to the primary vertex, respectively.
The primary vertex in the event is defined as the collision vertex with the highest sum of squared
transverse momenta of the associated ID tracks. 

Electron candidates are identified as clusters of energy in the electromagnetic calorimeter which 
are associated with a corresponding ID track. They are required to have $\pT > 25\,\GeV$ and 
$\abseta < 2.47$, excluding the transition regions between the barrel and endcap electromagnetic 
calorimeters ($1.37 < \abseta < 1.52$). The electron candidates must satisfy a set of \emph{medium}
selection criteria~\cite{PERF-2016-01} that have been optimised for the high rate of proton--proton 
collisions per beam crossing observed in the 2012 data. Electron candidates are required to be isolated, 
meaning that the scalar sum of the \pT\ of those tracks within $\Delta R = 0.2$ around the electron track is required 
to be less than 13\,\% of the \pT\ of the electron. (The definition of $\Delta R$ is the same as for the \kt\ algorithm, 
except that it makes use of pseudorapidity rather than rapidity.) Impact parameter requirements
are imposed to ensure that the electron candidates originate from the primary vertex. 

Muon candidates are reconstructed using the \emph{combined muon} algorithm~\cite{PERF-2014-05}, which involves matching 
and combining ID tracks with tracks in the muon spectrometer. The muon candidates are required to have 
$\pT > 25\,\GeV$ and $\abseta < 2.4$. Muon track quality requirements are imposed to suppress backgrounds, along with 
impact parameter requirements to ensure that the muon candidates originate from the primary vertex. Muon candidates are 
also required to be isolated, meaning that the scalar sum of the \pT\ of those tracks within $\Delta R = 0.2$ around the 
muon track (using pseudorapidity again) is required to be less than 10\,\% of the \pT\ of the muon.

A $Z$-boson candidate is selected by requiring exactly two opposite-charge same-flavour leptons (electrons or muons) 
and requiring the invariant mass of the dilepton system to satisfy $71\,\GeV < m_{\ell\ell} < 111\,\GeV$. 
The momenta of all selected ID tracks -- apart from the two lepton tracks -- are then passed into the \kt\ clustering 
algorithm introduced in Section~\ref{sec:intro} to construct the splitting scales.

\subsection{Particle-level selection and phase-space definition}
\label{sec:particlelevel}

Particle-level predictions are obtained using final-state objects with a mean decay length ($c\tau$)
longer than 10\,mm. Leptons are defined at the dressed level, i.e.\ they are given by the
four-momentum combination of the respective lepton (an electron or a muon) and all nearby photons
within a cone of size $\Delta R = 0.1$ centred on the lepton.
Electrons are required to pass $|\eta| < 2.47$, excluding the transition region between the barrel and end cap
electromagnetic calorimeters $1.37 < |\eta| < 1.52$, whilst muons are selected with $|\eta| < 2.4$.
Furthermore, a transverse momentum requirement of $\pT > 25\,\GeV$ is imposed for either lepton flavour.

Events are required to contain a $Z$-boson candidate, defined as exactly two oppositely charged, same-flavour leptons (electrons or muons) 
with a dilepton invariant mass of $71\,\GeV < m_{\ell\ell} < 111\,\GeV$.

All \textit{charged} final-state particles with $\pT > 400\,\MeV$ and $|\eta| < 2.5$ -- excluding the selected leptons
-- serve as input to the \kt\ clustering algorithm to construct the splitting-scale observables.
The measurement is performed twice, using jet-radius parameters of 0.4 and 1.0 respectively. This allows studies of the resolution
scales of both narrow and broad jets, which have different sensitivity to the hadronisation and underlying-event modelling.

\section{Monte Carlo simulation}

\subsection{Samples of simulated events}
\label{sec:samples}

Signal event samples for \Zee\ and \Zmm\ production in association with jets (QCD \Zjets) were generated 
in order to correct the data for detector effects in an unfolding procedure and to estimate systematic uncertainties.
The samples were produced using the \sherpa~v1.4.3 and
\powhegbox~\cite{Nason:2004rx,Frixione:2007vw,Alioli:2010xd} (SVN revision r1556) event generators.
The \sherpa\ samples are based on matrix elements with up to four additional hard emissions at leading order (LO), 
using parton distribution functions (PDFs) from the CT10 set~\cite{Lai:2010vv}, which were matched and merged to the
\sherpa\ parton shower using the set of tuned parameters developed by the \sherpa\ authors.
The \powhegbox\ samples (hereafter referred to as the \powheg\ samples) use the CT10 PDF set and are passed through
\textsc{Pythia}~v8.175 and subsequently through \textsc{PHOTOS}$++$~\cite{Golonka:2005pn} for parton showering and
radiative quantum-electrodynamical corrections, respectively.

\Ztt\ as well as \Wjets\ production were generated with \sherpa\ using the \menlops\ prescription~\cite{Hoche:2010kg}
to merge the results of inclusive \Ztt\ and $W\rightarrow\ell\nu$ calculations performed at NLO accuracy with the LO 
multi-leg prediction. All QCD $V$ $+$ jets samples are normalised using an inclusive NNLO cross section~\cite{Anastasiou:2003ds}.

Processes involing matrix elements with up to one additional parton emission for electroweak \Zjets\ production 
(including all diagrams with three electroweak couplings at tree level) were simulated using \sherpa.
The background contribution from $WW$ production with both bosons decaying leptonically is estimated using \powheg$+$\pythia.

Background contributions stemming from top-quark interactions (\ttbar, $t$-channel single top and $Wt$) 
were generated using \powheg\ with the CT10 PDF set in the hard scattering in conjunction with \textsc{Pythia}~v6.427~\cite{Sjostrand:2006za} 
for parton showering and hadronisation using the CTEQ6L1~\cite{Pumplin:2002vw} PDF set and the corresponding Perugia~2011C~\cite{Skands:2010ak} 
set of tuned parameters. The \texttt{hdamp} parameter, which controls the \pT\ of the first additional emission 
beyond the Born configuration, is set to the mass of the top quark. The \ttbar\ sample is normalised using a NNLO calculation 
in QCD including resummation of next-to-next-to-leading logarithmic (NNLL) soft gluon terms~\cite{Czakon:2013goa}, while the 
single-top samples are normalised using an approximate NNLO calculation including NNLL-accurate resummation of soft gluon 
terms~\cite{Kidonakis:2010ux,Kidonakis:2010tc,Kidonakis:2011wy}.

The Monte Carlo event samples mentioned above were passed through \textsc{GEANT4}~\cite{Agostinelli:2002hh,Allison:2006ve}
for a full simulation~\cite{SOFT-2010-01} of the ATLAS detector and are reconstructed with the same analysis chain as used for 
the measured data. Pileup is simulated with \textsc{Pythia}~v8.175~\cite{Sjostrand:2007gs} using the A2~\cite{ATL-PHYS-PUB-2011-014} 
set of tuned parton shower parameters and the MSTW2008lo~\cite{Martin:2009iq} set of parton distribution functions.

\subsection{Theoretical predictions}
\label{sec:predictions}

In addition to the samples of fully simulated events, additional particle-level predictions were
generated to provide a state-of-the-art comparison to the unfolded measurements.

Predictions for $Z$ boson production in association with jets are obtained using the \sherpa~v2.2.1
generator~\cite{Gleisberg:2008ta}. Matrix elements are calculated for up to two additional parton emissions 
at NLO accuracy and up to four additional parton emissions at LO accuracy using the Comix~\cite{Gleisberg:2008fv} 
and \textsc{OpenLoops}~\cite{Cascioli:2011va} matrix element generators, and merged 
with the \sherpa\ parton shower~\cite{Schumann:2007mg} which is based on Catani--Seymour subtraction terms.
The merging of multi-parton matrix elements with the parton shower is achieved using an improved CKKW matching 
procedure~\cite{Catani:2001cc,Hoeche:2009rj}, which is extended to NLO accuracy using the MEPS@NLO prescription~\cite{Hoeche:2012yf}.
The PDFs are provided by the NNPDF3.0nnlo set~\cite{Ball:2014uwa} and the dedicated 
set of tuned parton shower parameters developed by the \sherpa\ authors is used.

Predictions for \Zjets\ production at NNLO were provided by the 
DY@NNLOPS authors~\cite{Karlberg:2014qua}. The calculations are obtained using
DYNNLO~\cite{Catani:2009sm,Catani:2007vq} along with the multi-scale improved NLO (MiNLO) prescription~\cite{Hamilton:2012rf}
for the scale choices as implemented in the \powhegbox\ package~\cite{Alioli:2010qp,Alioli:2010xd}. For the hard scattering the
PDF4LHC15nnlo PDF set~\cite{Badger:2016bpw} is used, and parton showering is
provided by \textsc{Pythia}~v8.185 using the Monash~\cite{Skands:2014pea} set of tuned parameters.

\section{Analysis method}
\label{sec:method}

Distributions of the \kt\ splitting scales up to the seventh order were constructed from track momenta
selected according to Section~\ref{sec:reconstruction}.

\subsection{Background estimation}

Events from the signal process and the background contributions from $\Ztt$, $\Wjets$, top-quark pair and single 
top quark production as well as diboson and electroweak $\Zjets$ production are obtained by applying the 
analysis chain to the dedicated simulated samples introduced in Section~\ref{sec:samples}.
The multijet background can also contribute if two jets are misidentified as leptons or if they contain 
leptons from $b$- or $c$-hadron decays. A multijet-enhanced sample is obtained from the data by reversing 
some of the lepton identification criteria. A two-component template fit to the dilepton 
invariant mass spectrum is then employed to determine the normalisation in the fiducial measurement region, 
using the multijet template and a template formed from all other processes. Table~\ref{tab:process-composition} 
shows the estimated composition in each lepton channel. The relative sizes of the various background contributions
are also illustrated in Figure~\ref{fig:reco-level-d1}, which shows the $\sqrt{d_1}$ distribution
at the detector level in both the electron and the muon channel. The purity of the signal is close to 99\,\%.

\begin{table}[htbp]
\centering
\caption{Observed and expected numbers of events in the electron and the muon channels. 
The signal as well as the electroweak \Zjets, $WW$, \Wjets, \ttbar\ and single-top background rates are estimated 
using dedicated Monte Carlo samples. The multijet background is estimated using a data-driven technique.}
\label{tab:process-composition}
\begin{tabular*}{0.84\textwidth} { l r r r r }
\hline
 & \multicolumn{2}{c}{$\Z \rightarrow e^+e^-$} & \multicolumn{2}{c}{$\Z \rightarrow \mu^+\mu^-$} \\
 Process & Events & Contribution [\%] & Events & Contribution [\%] \\
\hline
QCD \Zjets         & 5\,090\,000 &  98.93\,\% & 7\,220\,000 &  99.40\,\% \\
Multijet           &     42\,000 &   0.81\,\% &     25\,000 &   0.34\,\% \\ 
Electroweak \Zjets &      5\,350 &   0.10\,\% &      7\,340 &   0.10\,\% \\ 
Top quarks         &      6\,190 &   0.12\,\% &      8\,440 &   0.12\,\% \\ 
$W(W)$             &      1\,100 &   0.02\,\% &      1\,460 &   0.02\,\% \\ 
\Ztt               &      1\,100 &   0.02\,\% &      1\,700 &   0.02\,\% \\
\hline
Total expected     & 5\,150\,000 & 100.00\,\% & 7\,260\,000 & 100.00\,\% \\
\hline
Total observed     & 5\,196\,858 &            & 7\,349\,195 &            \\
\hline
\end{tabular*}
\end{table}

\begin{figure}[htbp]
  \centering
  \includegraphics[width=0.5\textwidth]{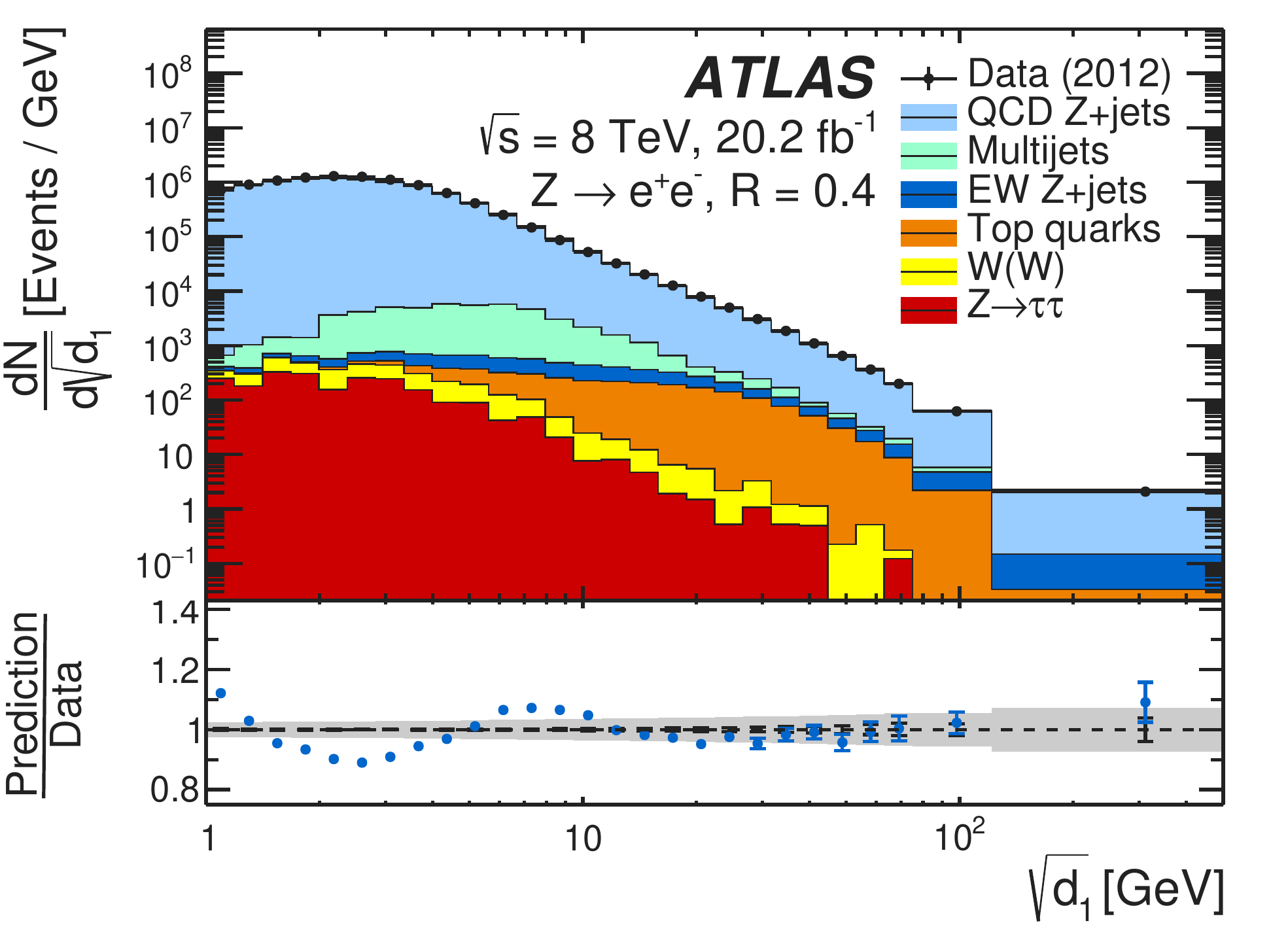}%
  \includegraphics[width=0.5\textwidth]{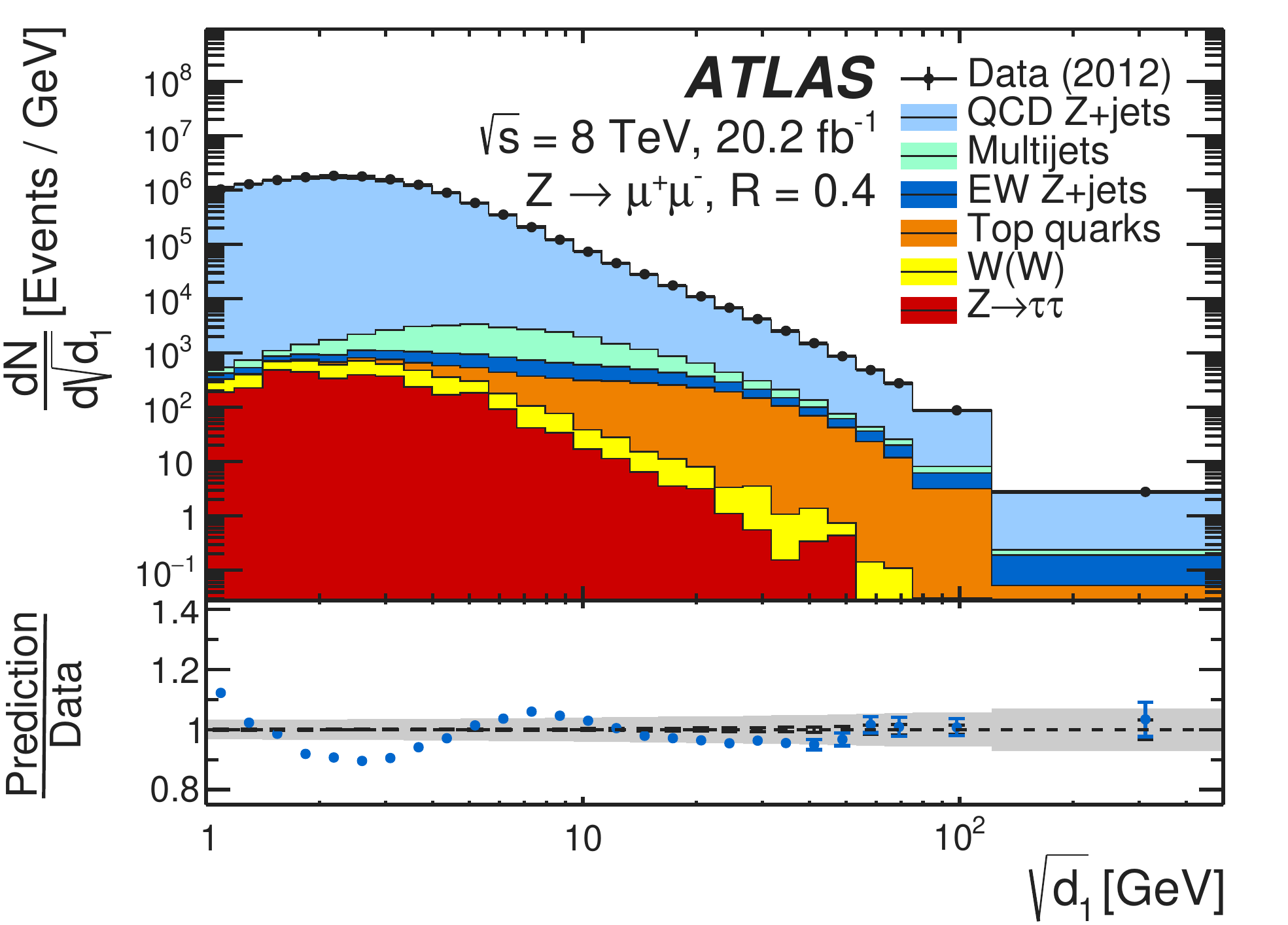}
  \caption{Detector-level splitting scale distributions in the electron and muon channels using the jet-radius parameter $R=0.4$.
    The lower panel shows the ratio of the combined Monte Carlo predictions to the data. The size of the error bars reflects
    the statistical uncertainty for predictions and data, while the combined experimental and systematic data uncertainty is 
    indicated by a grey band around unity.}
  \label{fig:reco-level-d1}
\end{figure}

\subsection{Unfolding}
\label{sec:method-unfolding}

The estimated number of background events is subtracted from the data in each bin of a given distribution. 
The background-subtracted data are then unfolded back to the particle level using an iterative procedure based 
on Bayes' theorem~\cite{D'Agostini:1994zf,Adye:2011gm}, which makes use of an unfolding prior that is updated at
each iteration. In the first iteration, the nominal particle-level predictions from \sherpa\ are used as the 
unfolding prior. The detector resolution causes bin-to-bin migrations, which are corrected for using a detector 
response matrix in the unfolding procedure. Acceptance and efficiency losses in the fiducial measurement region 
as well as fake contributions (e.g.\ due to pileup) are also accounted for. The unfolded bin values are found to 
converge after five iterations in most cases. The statistical uncertainty in data and simulation is estimated 
using pseudoexperiments.

In addition to the nominal charged-particle-level results (``\emph{charged-only}''),
the unfolding procedure is repeated using particle-level predictions that include the neutral 
particles as well, thereby extrapolating the data to a particle-level
including all particles (``\emph{charged+neutral}'').
The requirements on the $Z$-boson candidate
and the particles entering the clustering sequence remain identical to those of the charged-only
analysis. 
These extrapolated results are provided
for the benefit of theoretical calculations which cannot account for hadronisation processes,
and depend strongly on the modelling of hadronisation processes in the generator used during
the unfolding.
Since the extrapolation and the corresponding uncertainty estimate could change with other
hadronisation models in the future, the ``charged-only'' measurements should be considered the
primary results of this publication.

\subsection{Systematic uncertainties}
\label{sec:method-systs}

The systematic variations outlined below are propagated through the unfolding procedure by creating a 
new response matrix constructed from the simulation after reweighting, smearing or shifting the relevant 
event weights or objects. The shift in the unfolded spectrum is symmetrised and taken as a systematic 
uncertainty in the final result. The binning is chosen to be uniform in logarithmic space, with some bins 
merged towards the tails of the distributions to compensate for statistical fluctuations in the data.

Experimental uncertainties arise from the lepton-based and luminosity systematic uncertainties, the pileup modelling as well as
the track reconstruction efficiency. The systematic uncertainties in the lepton reconstruction, identification, isolation and
trigger efficiencies, as well as the lepton momentum scale and resolution, are defined in
Refs.~\cite{PERF-2016-01,PERF-2014-05}. The total impact of the lepton-based systematic uncertainties on the final results
is typically 1\,\% or less. The uncertainty in the integrated luminosity is 1.9\,\%, as determined from beam separation
scans~\cite{Aaboud:2016hhf} performed in November 2012. The difference in the pileup profile between the detector-level simulation and the 
data is corrected for by reweighting the simulation to match the average number of proton--proton collisions observed in the data.
The impact of the remaining mismodelling on the measurement is found to be at most 1\,\%, which is assigned as a systematic uncertainty. 
The efficiency to reconstruct an ID track depends on the ID material distribution.
A corresponding uncertainty is derived by comparing the nominal simulation to a dedicated simulation using a distorted ID geometry
and a 15\,\% increase in the material budget. The resulting systematic uncertainty associated with the track reconstruction efficiency is 
typically at the level of 5\,\% but can rise to 10\,\% in the tails of the distributions, in particular for higher-order splitting scales.
Uncertainties associated with the track momentum are found to be negligible for the range of track momenta probed in this measurement.
As a cross-check, splitting scales are also constructed in data and simulation using tracks that do not pass the nominal longitudinal 
impact parameter requirements in order to assess the modelling of pileup and fake tracks in the simulation. Good agreement is observed
between data and simulation, with the data being generally described to within 10\,\%. The differences are covered by the assigned 
uncertainties, which are larger in that region due to the larger fraction of fake and pileup tracks, and so no additional 
systematic uncertainties in the pileup modelling or fake-track rates are assigned. 

A data-driven uncertainty associated with the imperfect modelling of the unfolded observables at the detector level is obtained by
reweighting the simulation such that the level of agreement between the detector-level distributions and the data is improved before
performing the unfolding procedure. The reweighting is applied at the particle level using the ratio of the distributions in data 
and detector-level simulation and results in a systematic uncertainty of at most 5\,\%. An algorithmic uncertainty associated with 
the number of iterations chosen in the unfolding procedure is taken to be the difference between the results unfolded using ten 
iterations and the nominal results obtained using five unfolding iterations. The algorithmic unfolding uncertainty is at the 
subpercent level.

An uncertainty due to the choice of generator used to unfold the data, which is the dominant systematic uncertainty, is obtained 
by replacing the nominal \sherpa\ prediction in the full analysis chain with a prediction from \powheg$+$\pythia, which uses a very 
different parton shower and hadronisation model compared to the nominal generator. The difference between the results unfolded with 
the nominal predictions and the alternative predictions is then symmetrised and assigned as an uncertainty in the choice of 
unfolding prior. This uncertainty estimate might be reduced by reweighting both \sherpa\ and \powheg$+$\pythia\ to
the experimental data at reconstruction level before constructing the response matrices.
Since it is not clear that this would still cover the full impact of non-perturbative effects on the unfolding, 
no such reweighting is applied when assessing the unfolding-prior uncertainty.

The extrapolated ``charged+neutral'' results are also obtained using the generator combinations described above
as unfolding priors, providing an estimate of the extrapolation uncertainty.

To cover further systematic effects of the unfolding algorithm on the extrapolated results,
an MC-closure uncertainty was derived by unfolding \powheg$+$\pythia\ from detector level
to the charged+neutral level using response matrices generated from the \sherpa\ samples. The
difference with respect to the direct charged+neutral predictions from \powheg$+$\pythia\ is then added as an
MC-based closure uncertainty in analogy to the data-driven reweighting uncertainty described above.

\begin{figure}[htbp]
  \centering
  \includegraphics[width=0.50\textwidth]{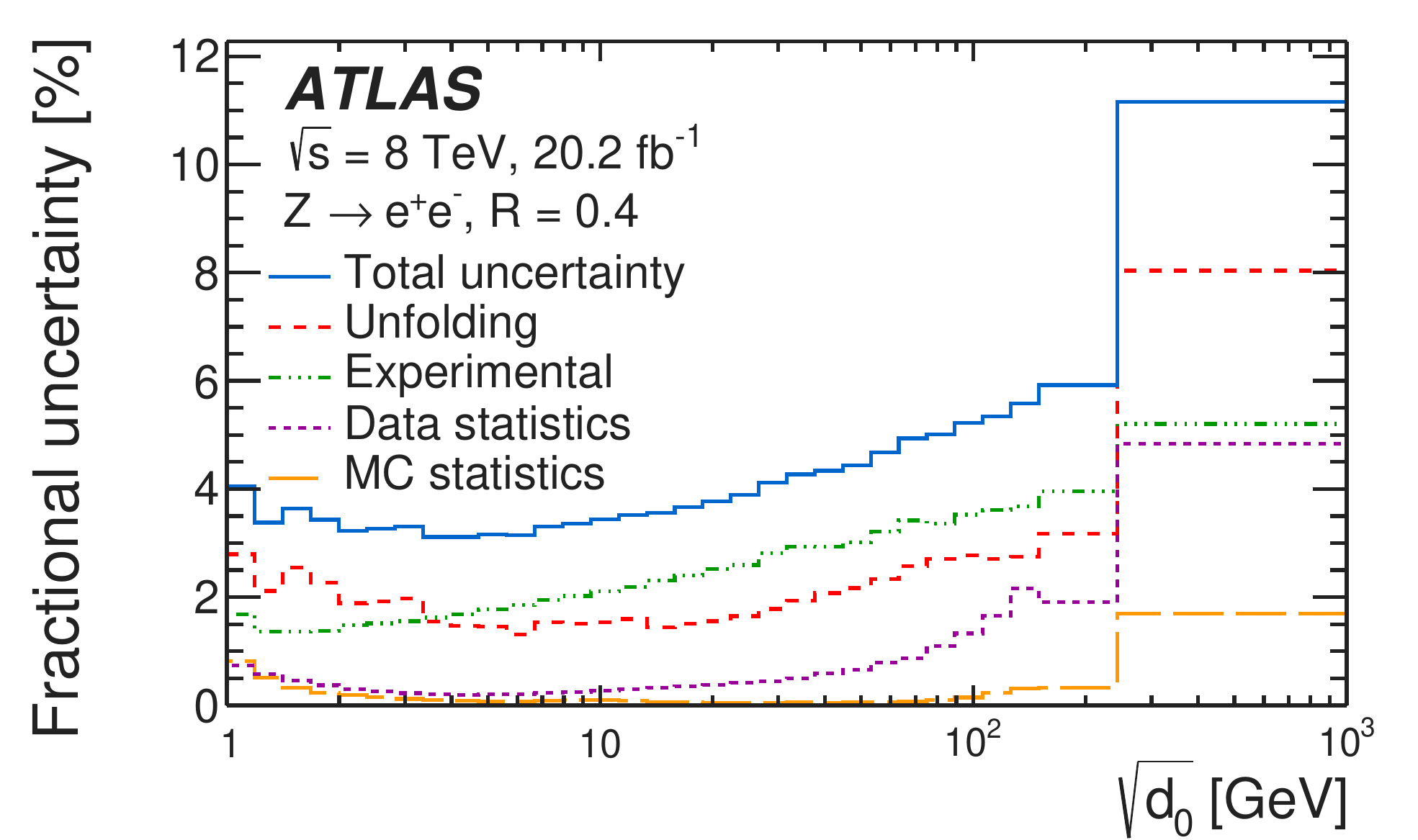}%
  \includegraphics[width=0.50\textwidth]{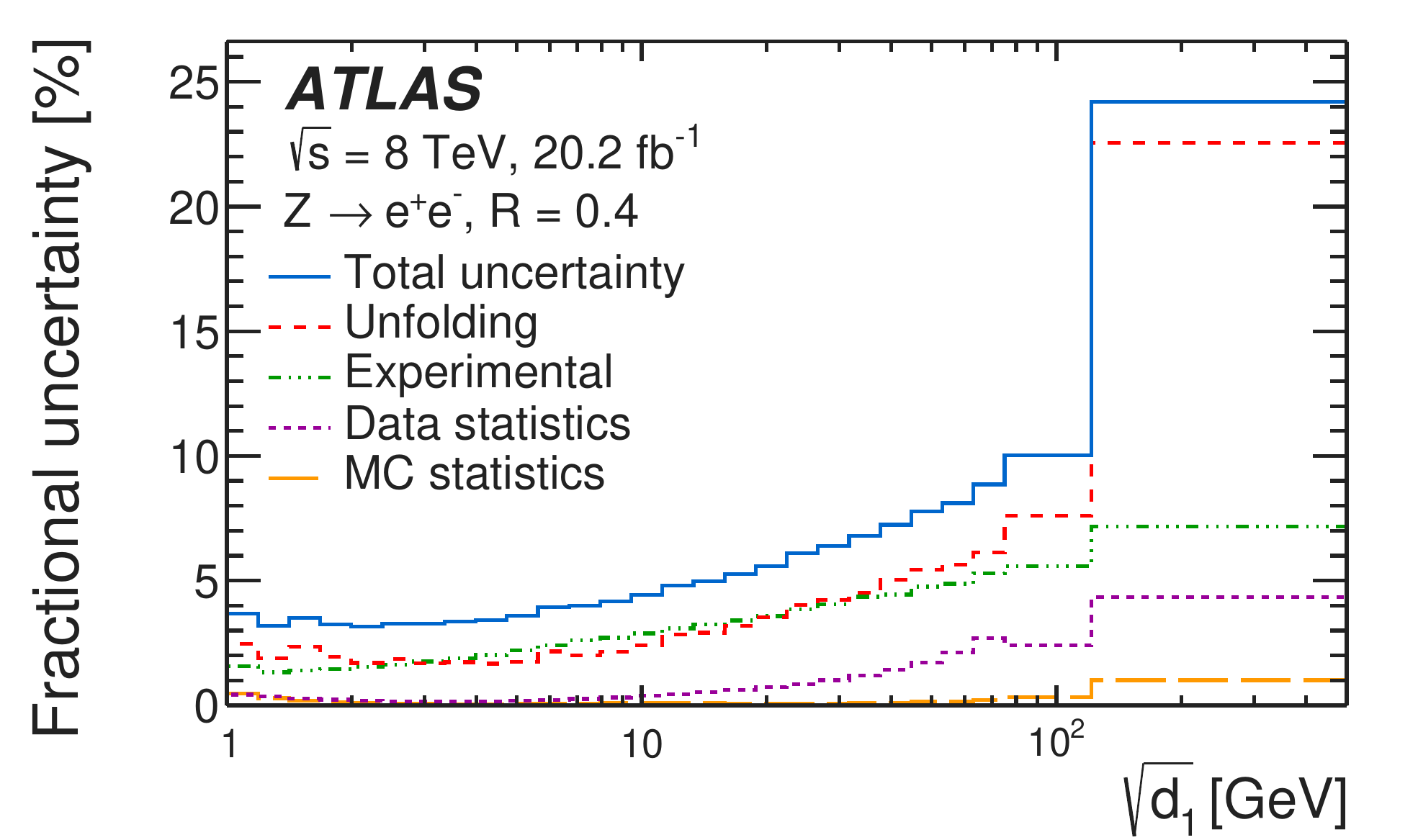}\\
  \includegraphics[width=0.50\textwidth]{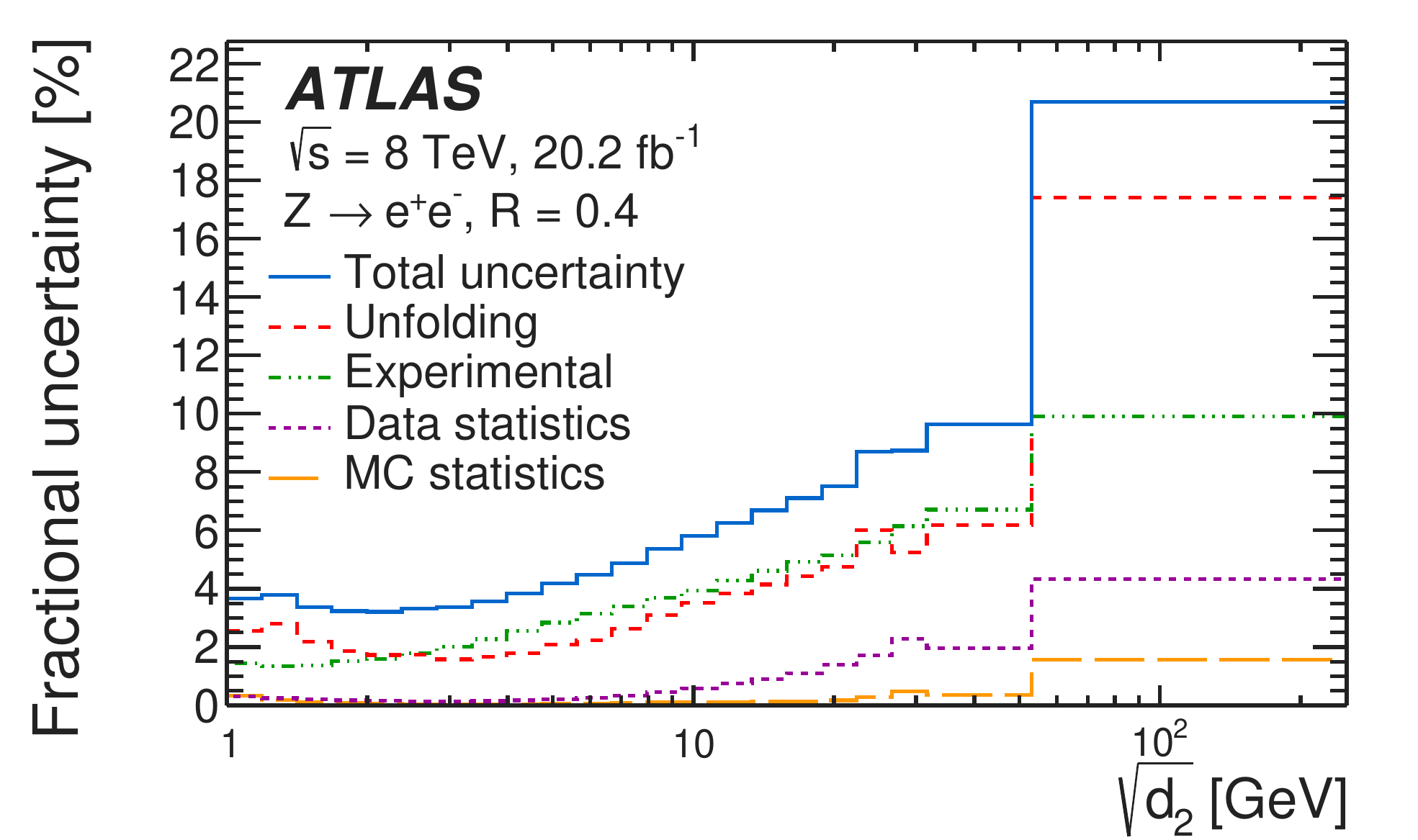}%
  \includegraphics[width=0.50\textwidth]{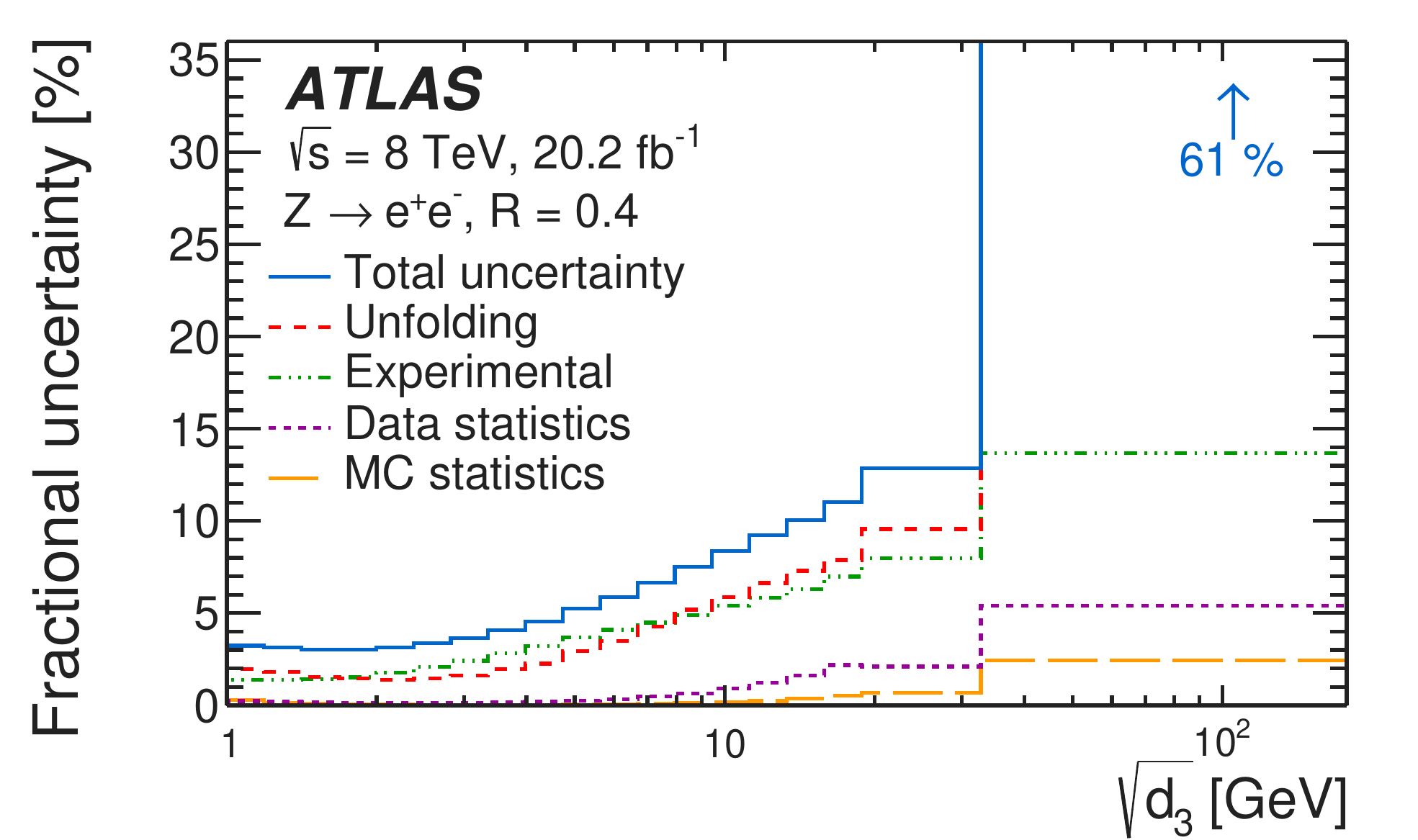}\\
  \includegraphics[width=0.50\textwidth]{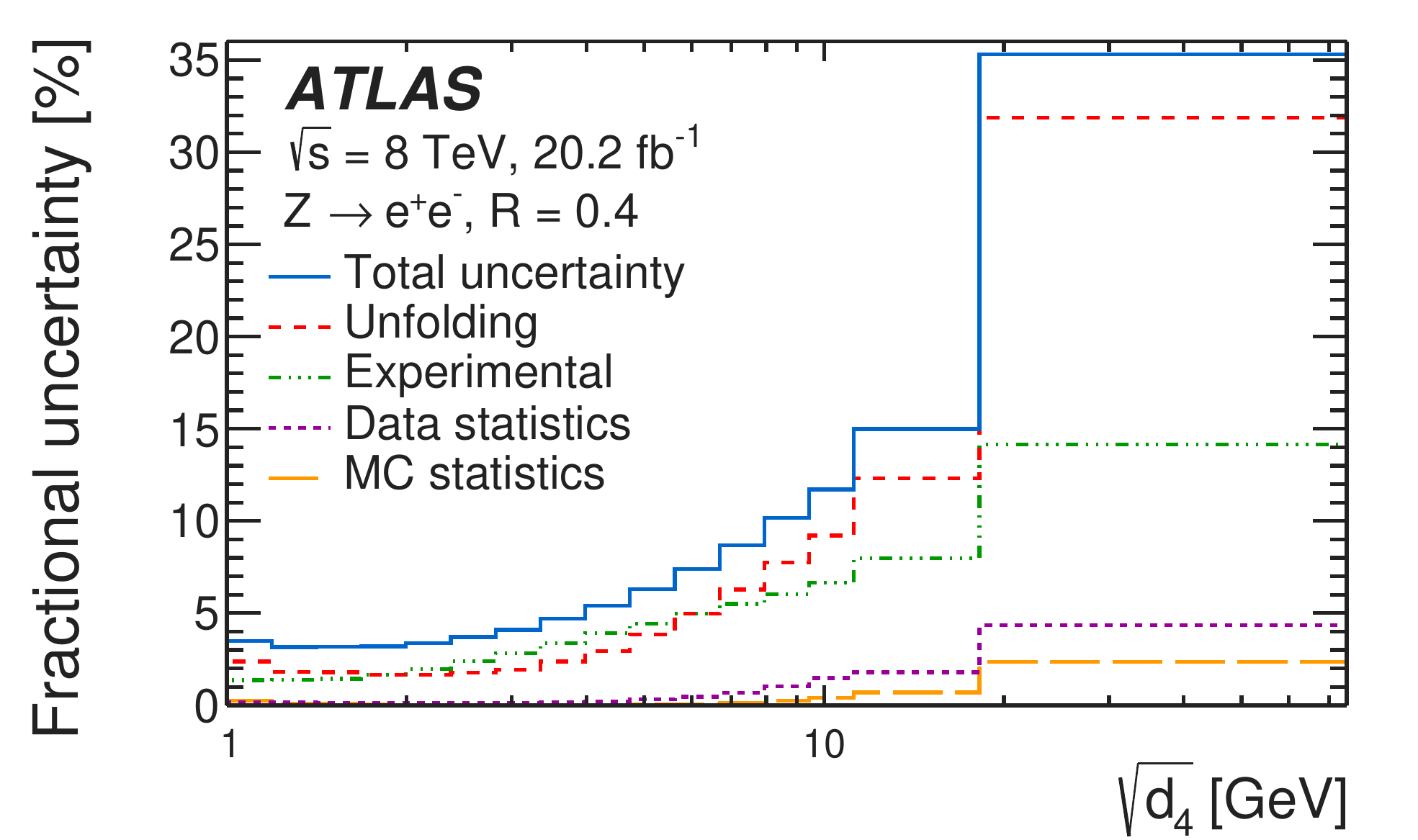}%
  \includegraphics[width=0.50\textwidth]{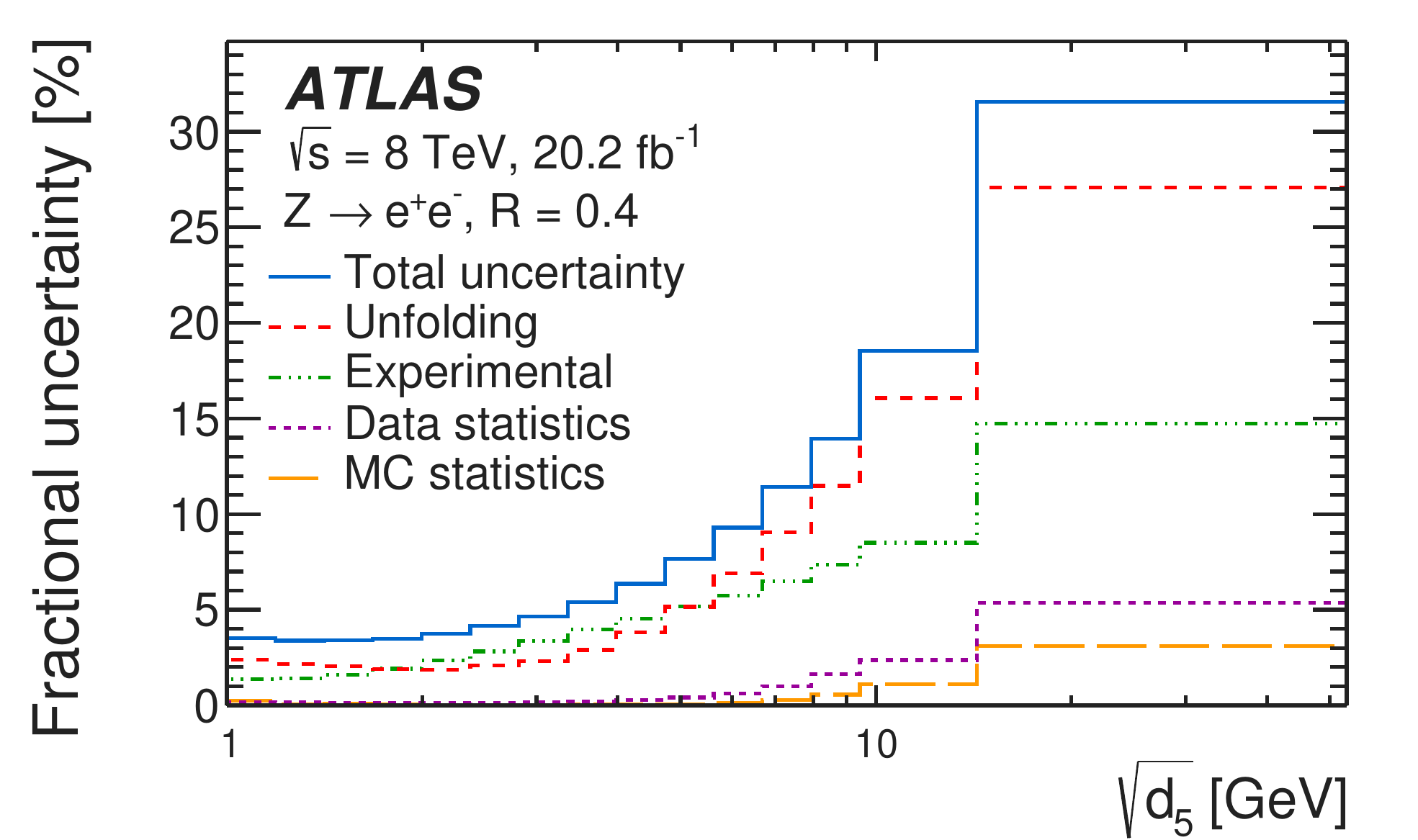}\\
  \includegraphics[width=0.50\textwidth]{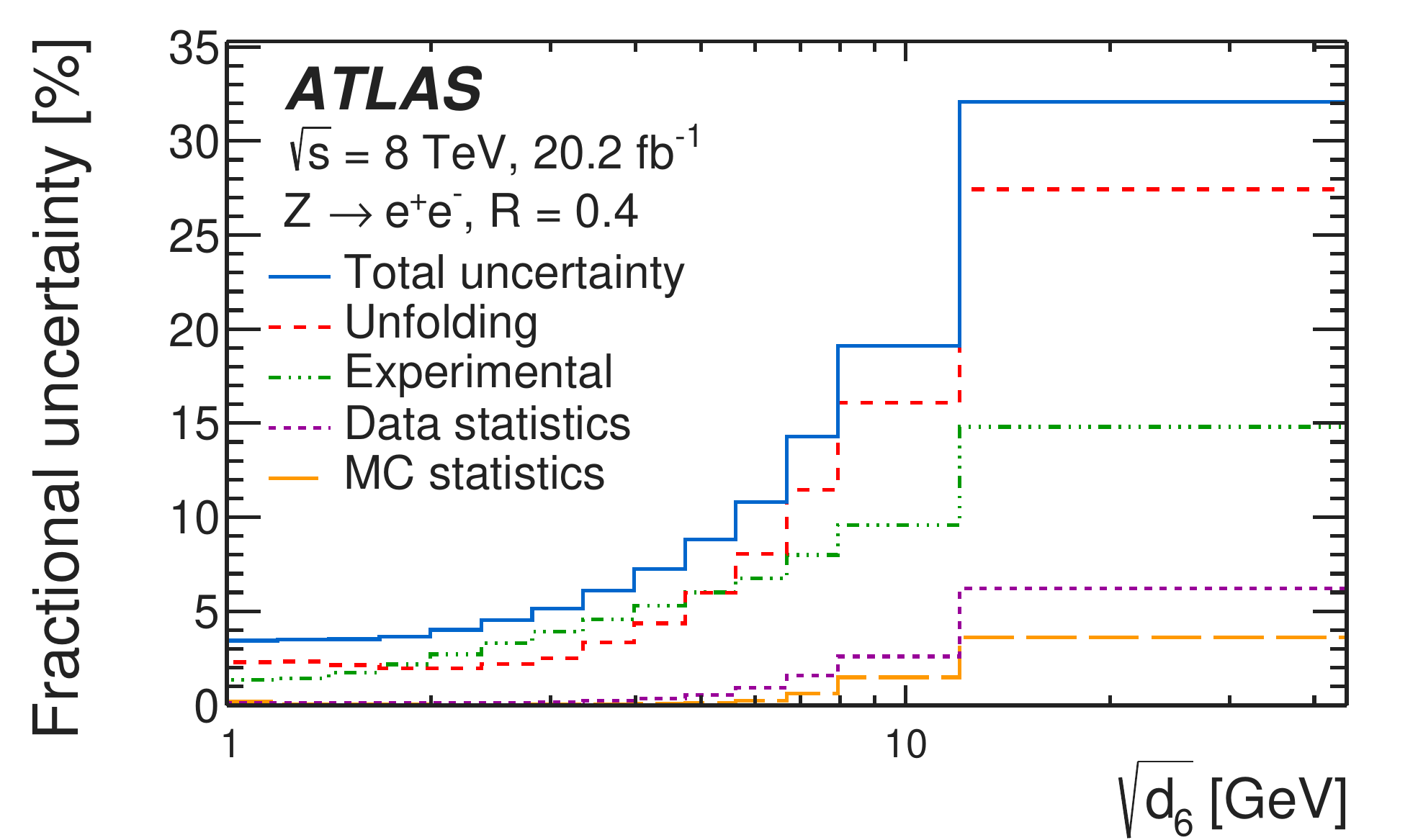}%
  \includegraphics[width=0.50\textwidth]{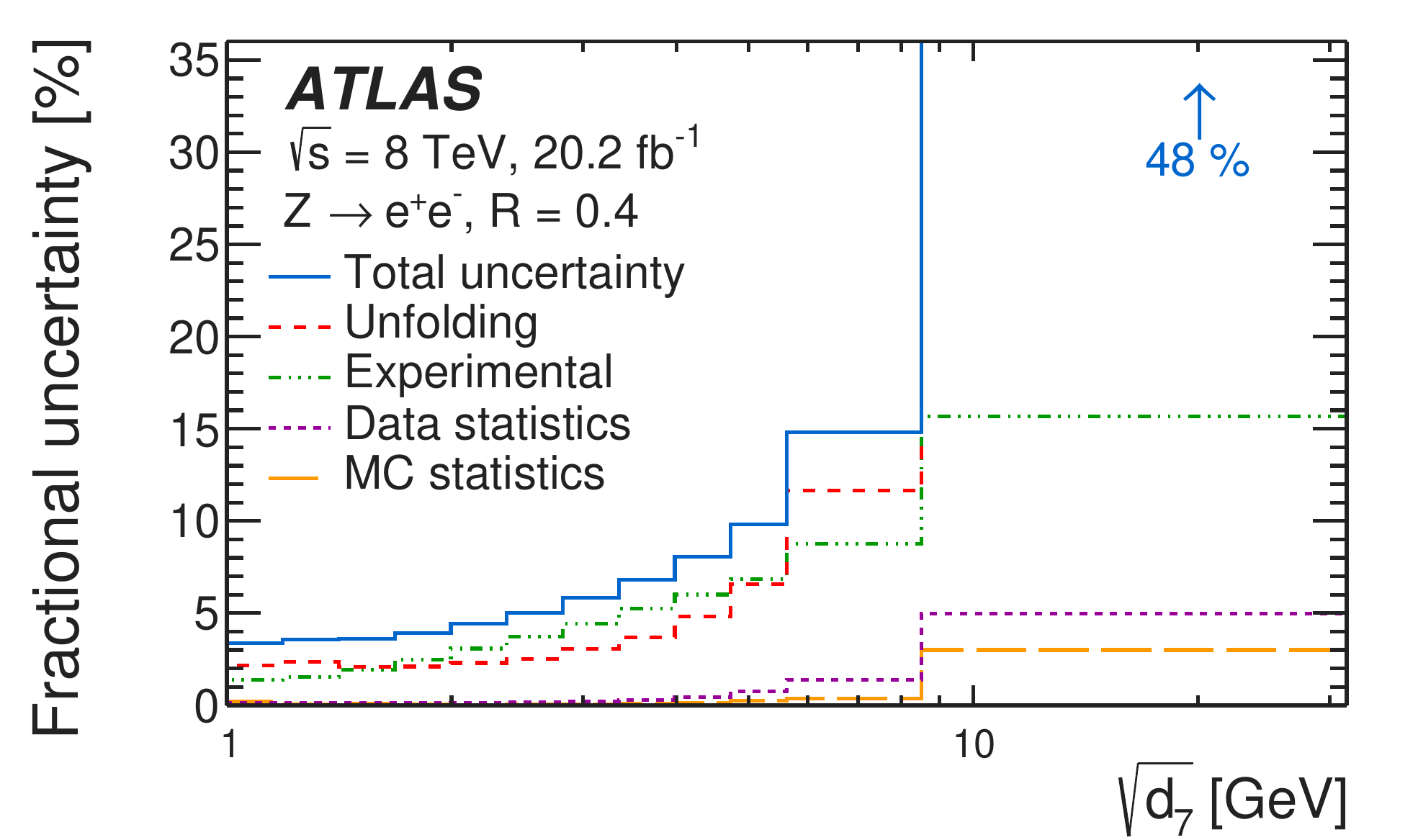}
  \caption{Breakdowns of the total systematic uncertainty into its contributions
    in the electron channel using the jet-radius parameter $R=0.4$.
    The total uncertainty shown is the full quadrature sum of all the individual
    statistical and systematic uncertainty components, including a constant
    $\pm 1.9\,\%$ uncertainty in the luminosity estimate.
}
  \label{fig:el04nominal_systs}
\end{figure}

The breakdown of the corresponding systematic uncertainties can be found in
Figure~\ref{fig:el04nominal_systs}.
In the breakdowns, the total 
uncertainty shown is the full quadrature sum of all the individual statistical and systematic uncertainty
components. This includes a constant $\pm 1.9$\,\% uncertainty in the luminosity estimate,
which is omitted from the breakdown for clarity. The curve labelled `experimental' contains 
the tracking and pileup uncertainties as well as the systematic uncertainties associated with the 
\Z-boson candidate selection. The curve labelled `unfolding' contains the uncertainties associated with the 
unfolding procedure, namely the unfolding closure uncertainty, the uncertainty associated with the number of iterations 
as well as the unfolding-prior uncertainty.

\section{Results}
\label{sec:result}

In this analysis, distributions for eight splitting scales, $\sqrt{d_0}\dots\sqrt{d_7}$, are obtained for all combinations of
\begin{itemize}
\item two lepton flavours, $Z\to e^+e^-$ and $Z\to \mu^+\mu^-$,
\item two radius parameters in the jet algorithm, $R=0.4$ and $R=1.0$, and
\item two particle-level definitions: ``charged-only'' and ``charged+neutral'',
  as introduced in Section~\ref{sec:method-unfolding}.
\end{itemize}

A full set considering all splitting scales
is shown only for the electron-channel analysis with $R=0.4$ in Figure~\ref{fig:el04nominal},
while other combinations can be found in
Appendices~\ref{sec:app_el10nominal} and \ref{sec:muonresults}.

The measured differential cross sections are compared to state-of-the-art predictions obtained from 
\sherpa\ (`MEPS@NLO') and DY@NNLO$+$\powheg$+$\pythia\ (`NNLOPS'). Details of the 
respective generator setups are given in Section~\ref{sec:predictions}. PDF uncertainties for the MEPS@NLO 
prediction are estimated with the NNPDF3.0nnlo~\cite{Ball:2014uwa} replicas using LHAPDF~\cite{Buckley:2014ana}.
They are combined with an uncertainty 
estimate based on 7-point scale variations, i.e.\ all combinations of factors of $0.5$, $1$ and $2$ in the 
factorisation and renormalisation scales are taken into account except the opposite combinations 
of $(0.5, 2)$ or $(2, 0.5)$. The scale uncertainty envelope for the NNLOPS prediction is obtained 
from 21-point scale variations using the prescription described in Ref.~\cite{Karlberg:2014qua}.

Neither of the generators provides a fully satisfactory description of the experimental data.
In the ratio plots of the lower-order splitting scales it is seen that both predictions
underestimate the cross section in the peak region at values of around 3\,\GeV\ by typically 10--20\,\%,
but are consistent with the data at 10\,\GeV.
At higher values the MEPS@NLO prediction agrees well with the data while the NNLOPS prediction
systematically overestimates the cross section.
Both overshoot the data significantly in the region close to 1\,\GeV.
The level of agreement of the NNLOPS predictions in the soft region is improved significantly
for the higher-order splitting scales.

The generator uncertainties in these comparisons
are only estimated by varying parameters related to the perturbative
aspects of the MC generators. In the
soft region of the splitting-scale distributions other aspects become relevant,
such as hadronisation and multiple parton interactions, and even the parton shower
modelling. The discrepancies unveiled by these measurements
indicate that the data can provide new input for the tuning of parameters
in the non-perturbative stages of event generators.

Figure~\ref{fig:variations} displays again the first-order splitting scale,
$\sqrt{d_1}$, to demonstrate the different variations in the analysis due to
the choice of lepton flavour, jet-radius parameter and particle level.
A detailed breakdown of the systematic uncertainties is included here in the same style
as described earlier.

The uncertainties for the extrapolated charged$+$neutral distributions are increased due to the unfolding
procedure, which includes the correction for going from a fiducial phase space using only
charged particles to one using all particles. This
effect is most significant for low values of the lower-order splitting scales
such as $\sqrt{d_1}$, where uncertainties grow beyond 10\,\% compared to less than
5\,\% in the charged-only measurement. For higher values of the splitting scales the
differences between the charged-only and charged$+$neutral uncertainties are not as large, as this region is not affected much
by hadronisation effects. The peak of the extrapolated distributions, when
compared to the nominal (charged) distributions, is shifted
by roughly $1\,\GeV$ in the $R=0.4$ case and by about $2.5\,\GeV$ in the
$R=1.0$ case. It moves to higher values of the splitting scales as expected
given that the jet energies increase with the addition of neutral
particles.

The two lepton flavours agree to better than $1\sigma$, once differences in the fiducial
lepton selection are accounted for. The channels are not combined, because data
statistics is a subdominant uncertainty.
The $R=1.0$ comparisons reveal features similar to the $R=0.4$ case.
A full set of distributions and systematic uncertainties in the $R=1.0$ case are shown in
Appendix~\ref{sec:app_el10nominal}. Distributions of all possible combinations are furthermore provided in
HEPDATA~\cite{Buckley:2010jn} along with an associated Rivet~\cite{Buckley:2010ar} routine.

\begin{figure}[htbp]
  \centering
  \includegraphics[width=0.45\textwidth]{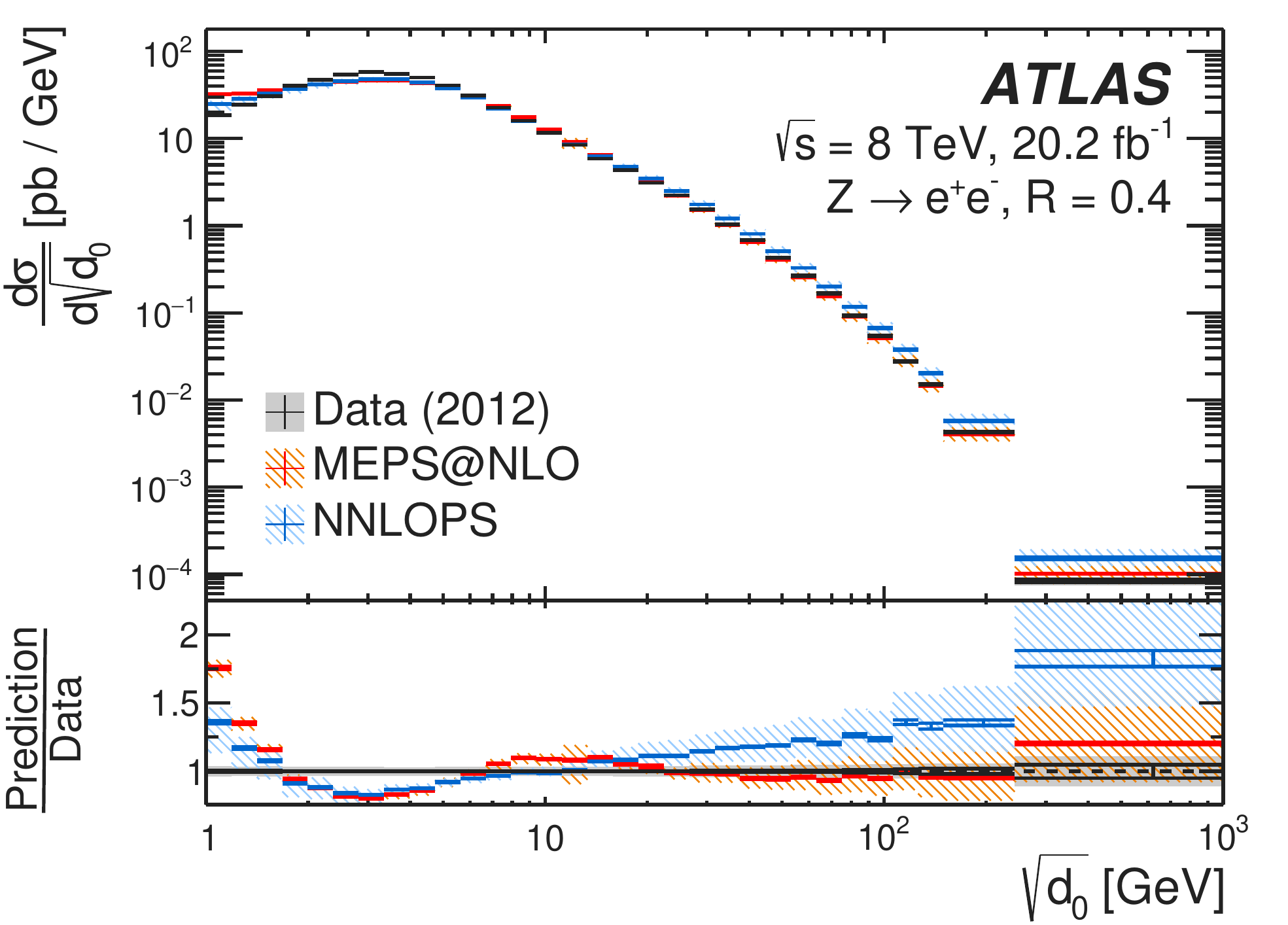}%
  \includegraphics[width=0.45\textwidth]{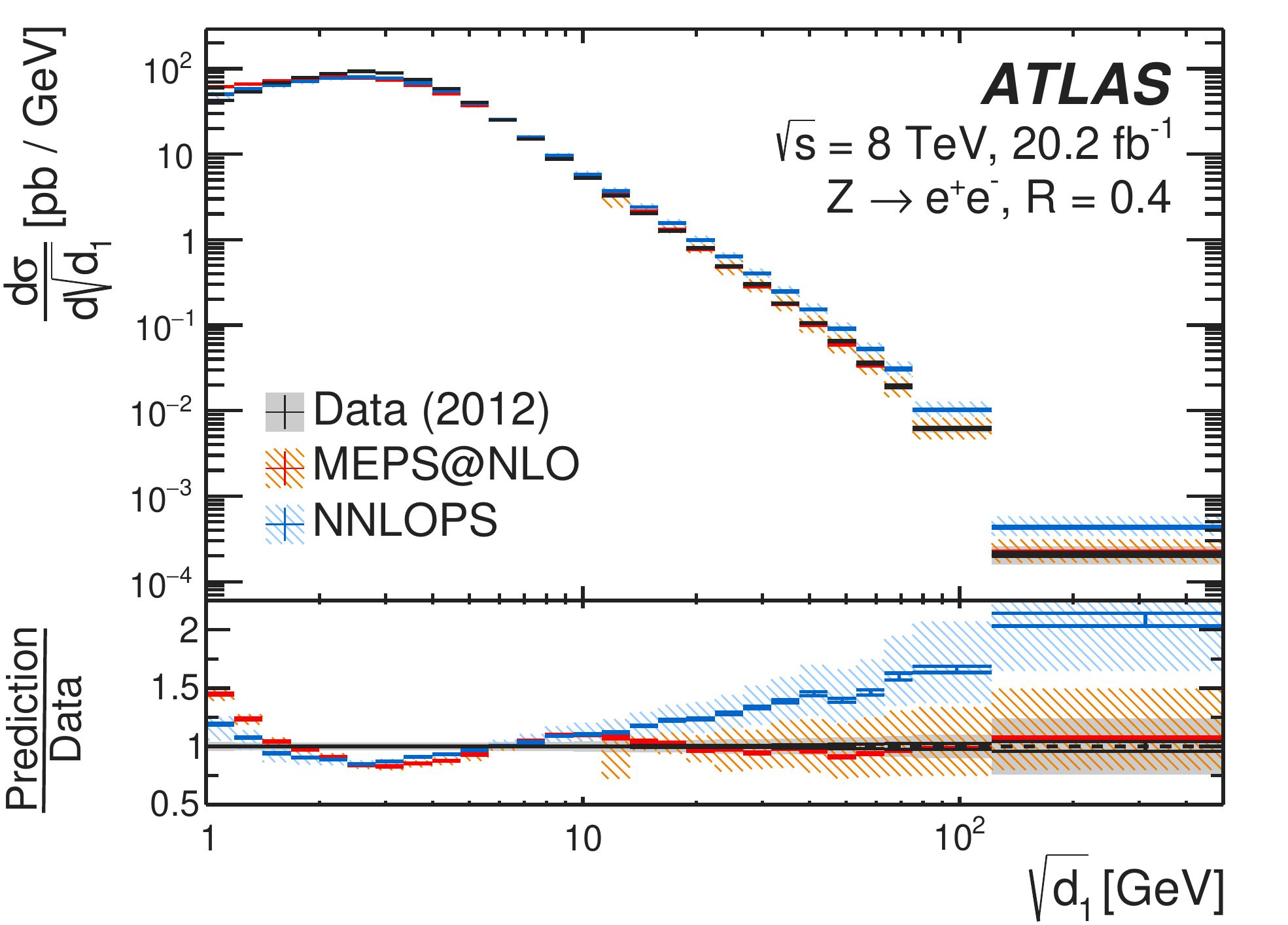}\\
  \includegraphics[width=0.45\textwidth]{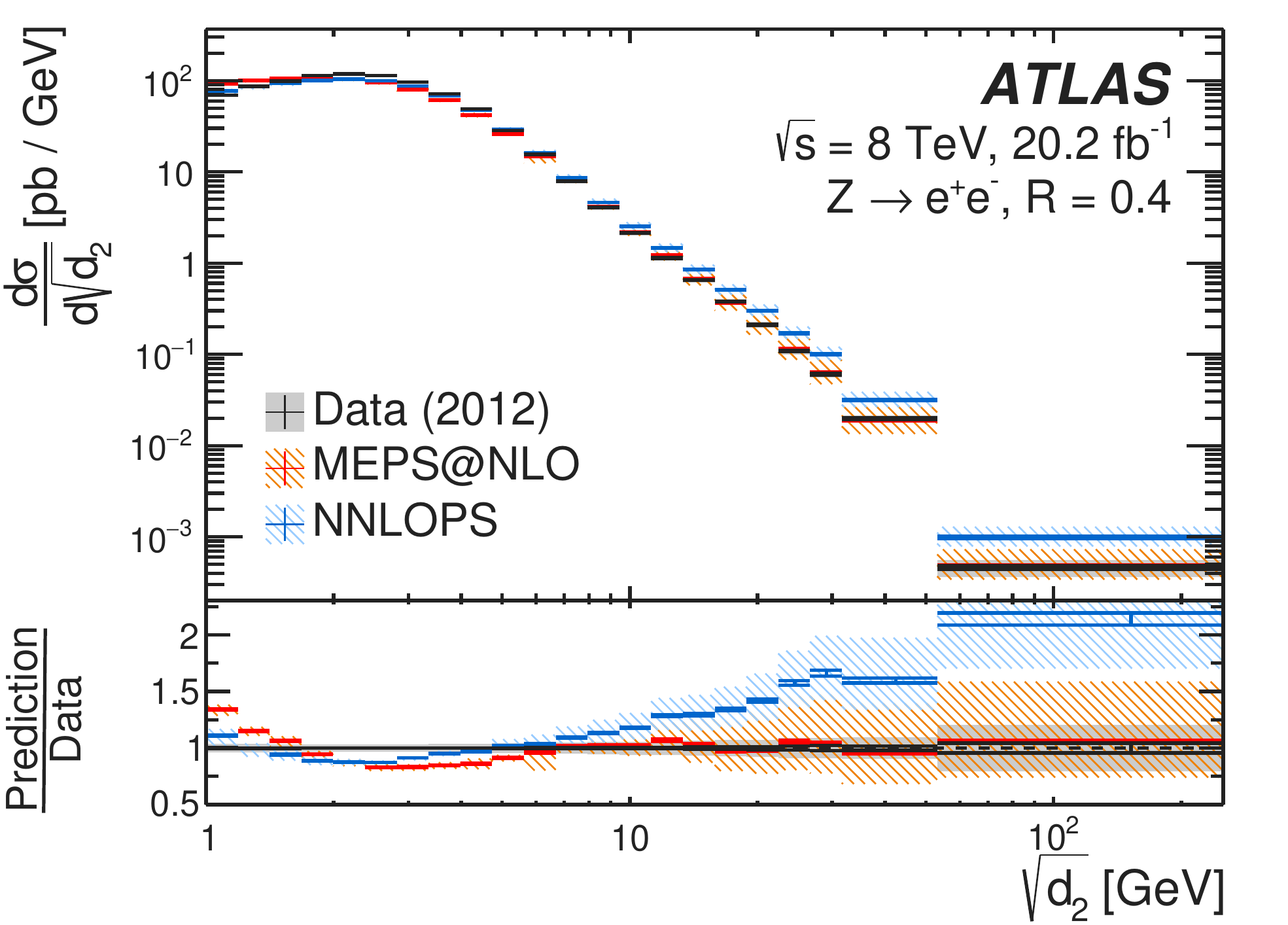}%
  \includegraphics[width=0.45\textwidth]{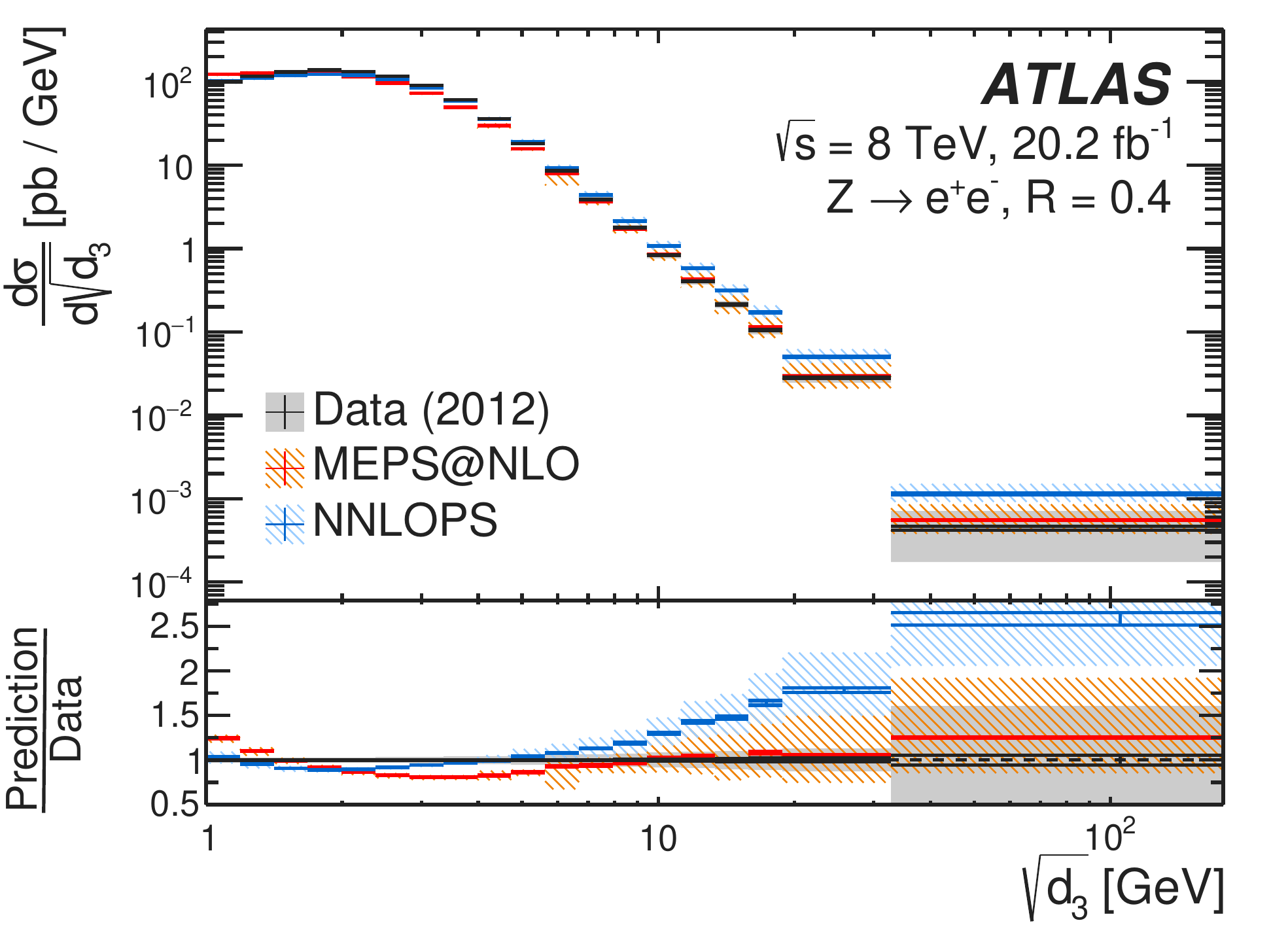}\\
  \includegraphics[width=0.45\textwidth]{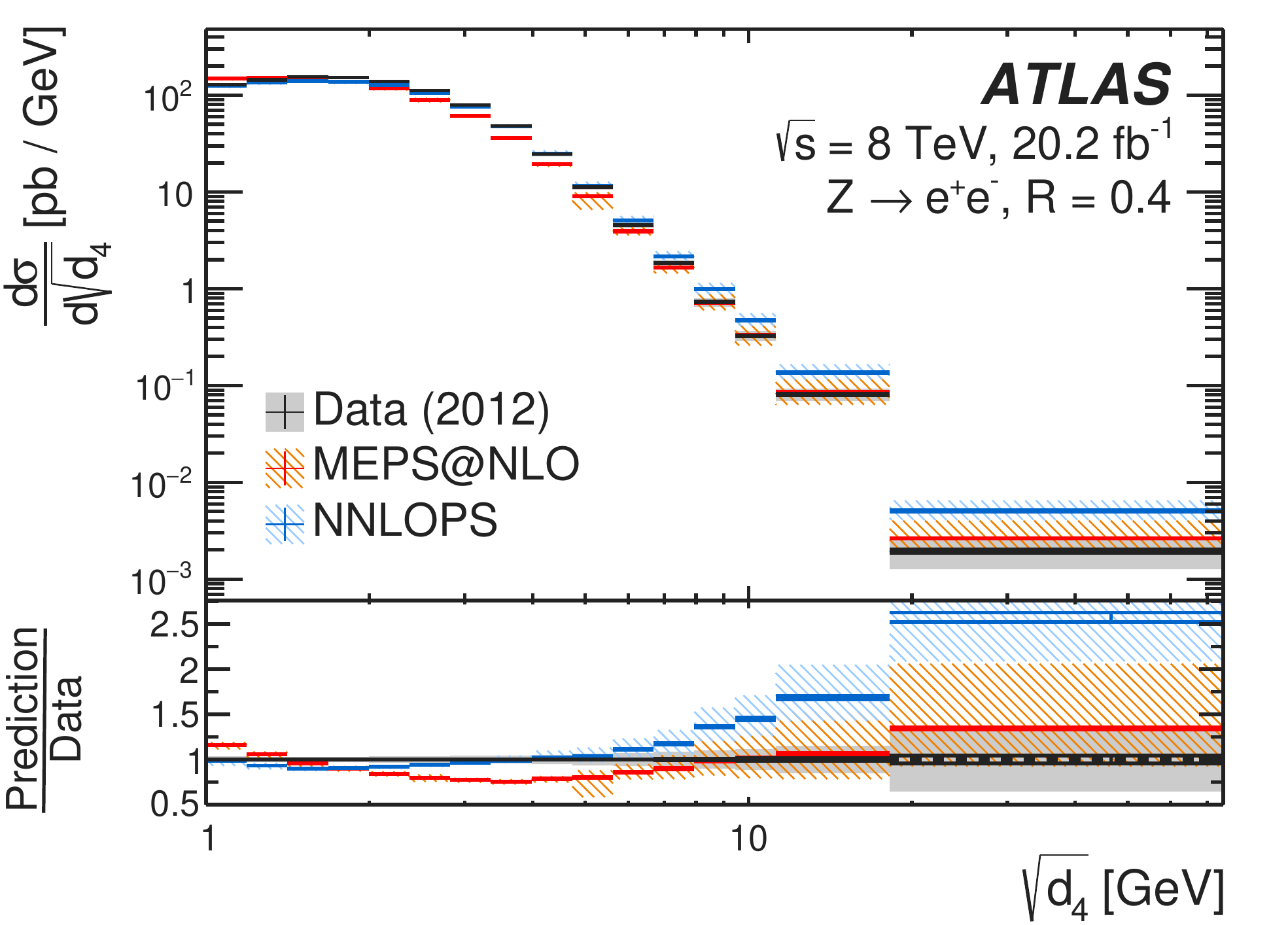}%
  \includegraphics[width=0.45\textwidth]{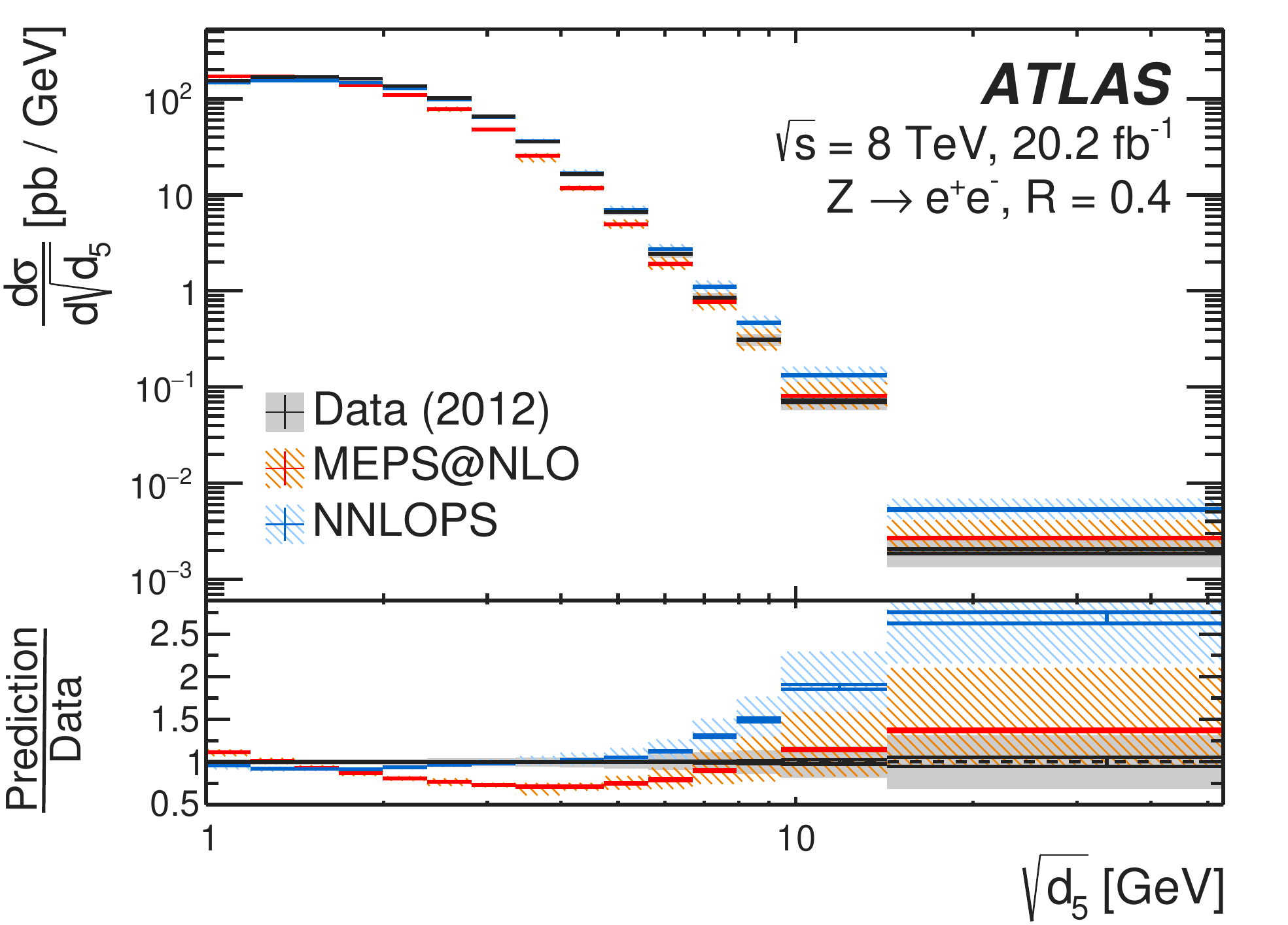}\\
  \includegraphics[width=0.45\textwidth]{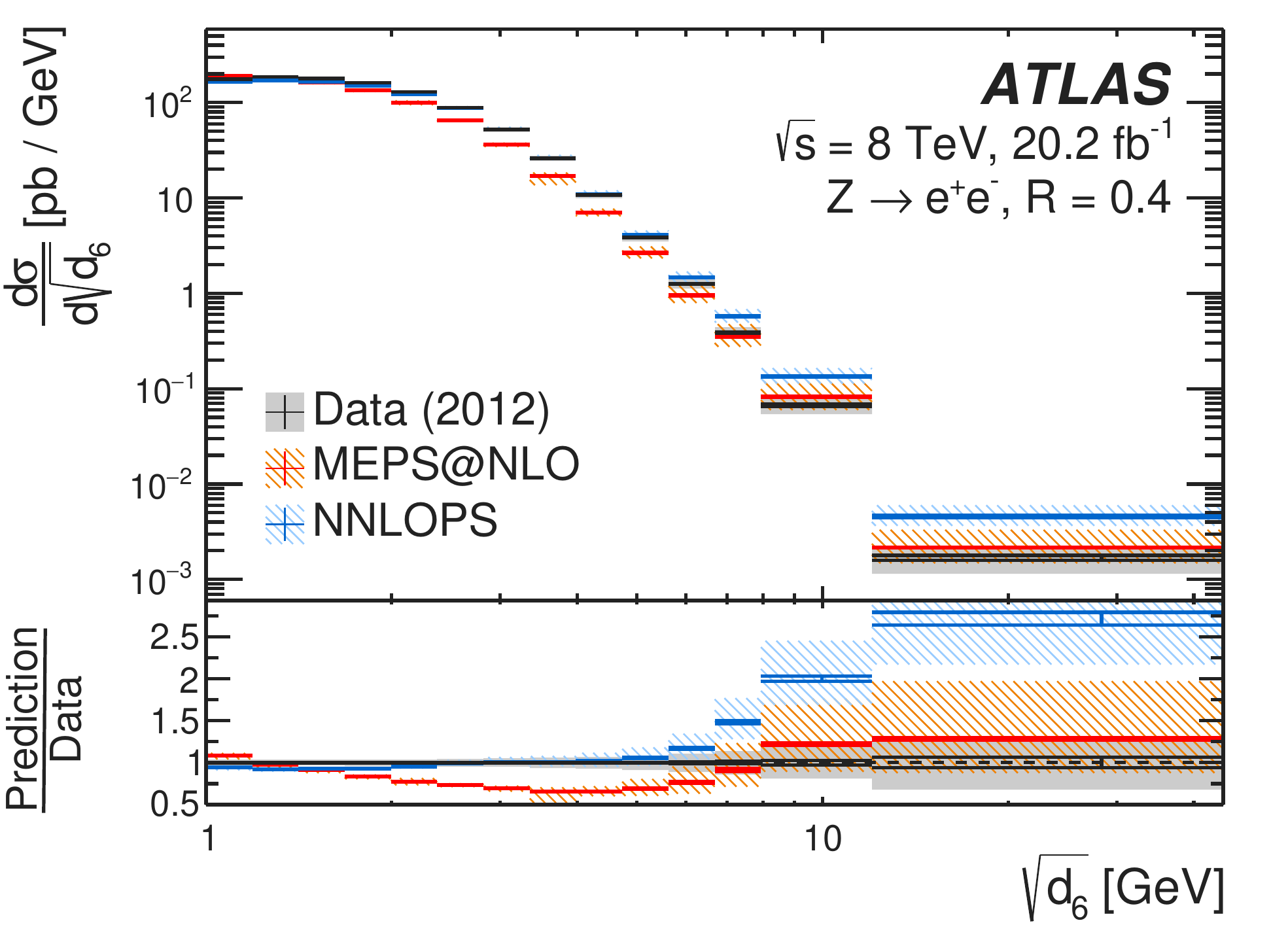}%
  \includegraphics[width=0.45\textwidth]{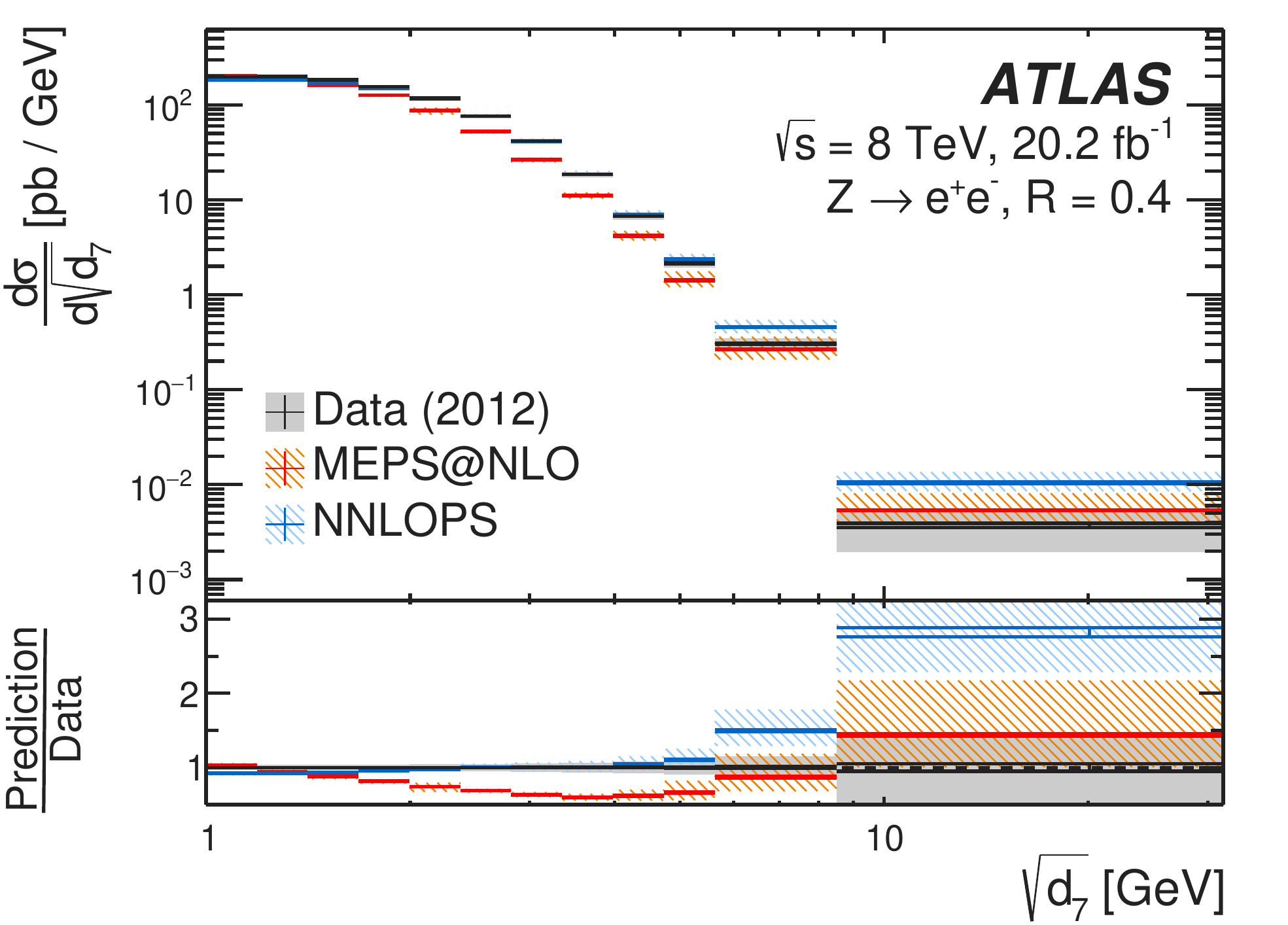}
  \caption{Charged-only distributions for the eight leading splitting scales
    in the electron channel using the jet-radius parameter $R=0.4$.
    The size of the error bars reflects the statistical uncertainty, while the combined statistical
    and systematic uncertainty is indicated by the grey band.
    Theoretical predictions from \sherpa\ with NLO multijet merging (``MEPS@NLO'') and from 
    \powheg$+$\pythia\ with NNLO matching (``NNLOPS'') are displayed including error bands for the 
    generator uncertainties.
  }
  \label{fig:el04nominal}
\end{figure}

\begin{figure}[htbp]
  \centering
  \subfloat[muon, $R=0.4$, charged-only]{\centering
    \shortstack{
      \includegraphics[width=0.45\textwidth]{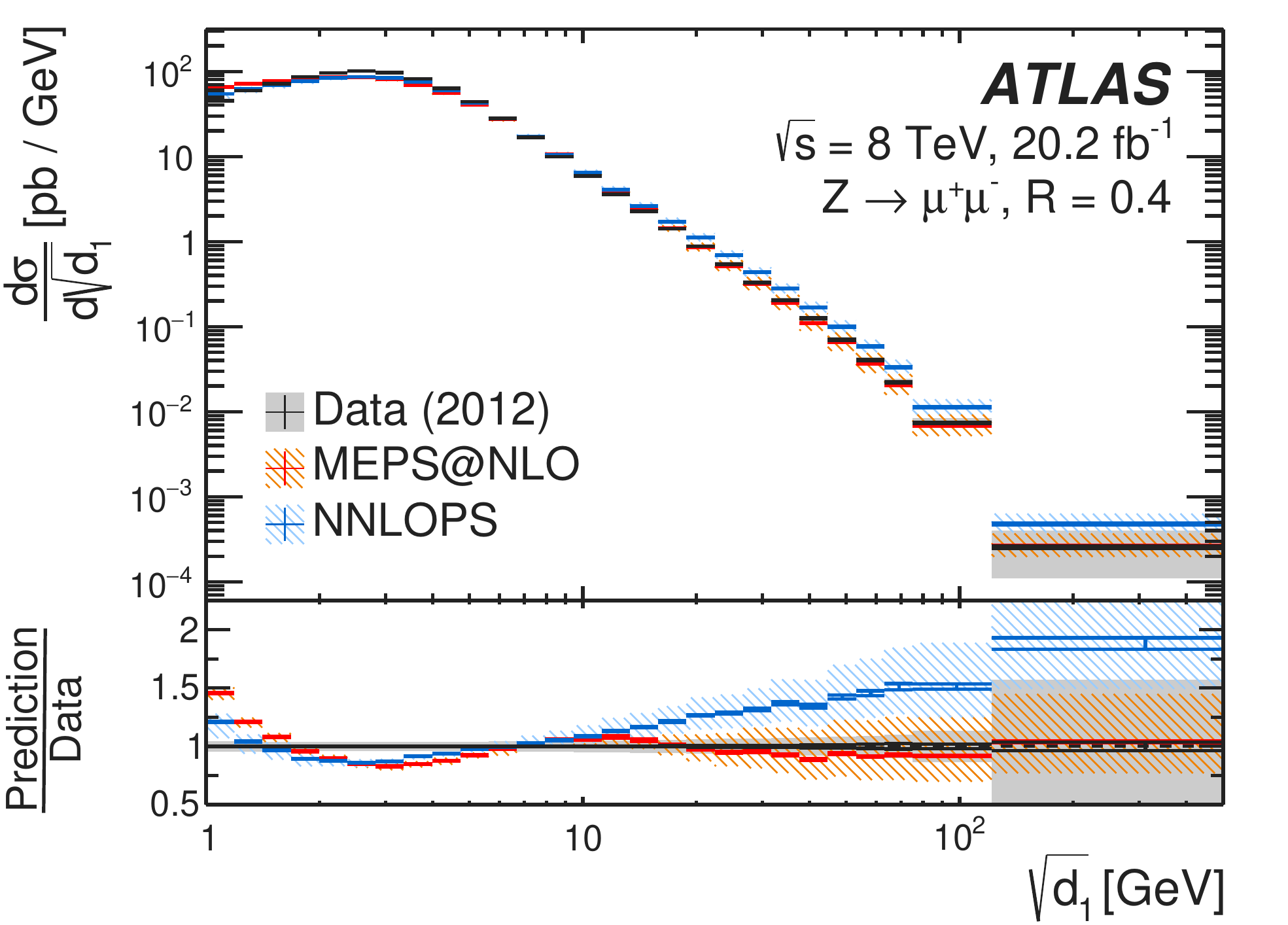}\\
      \includegraphics[width=0.45\textwidth]{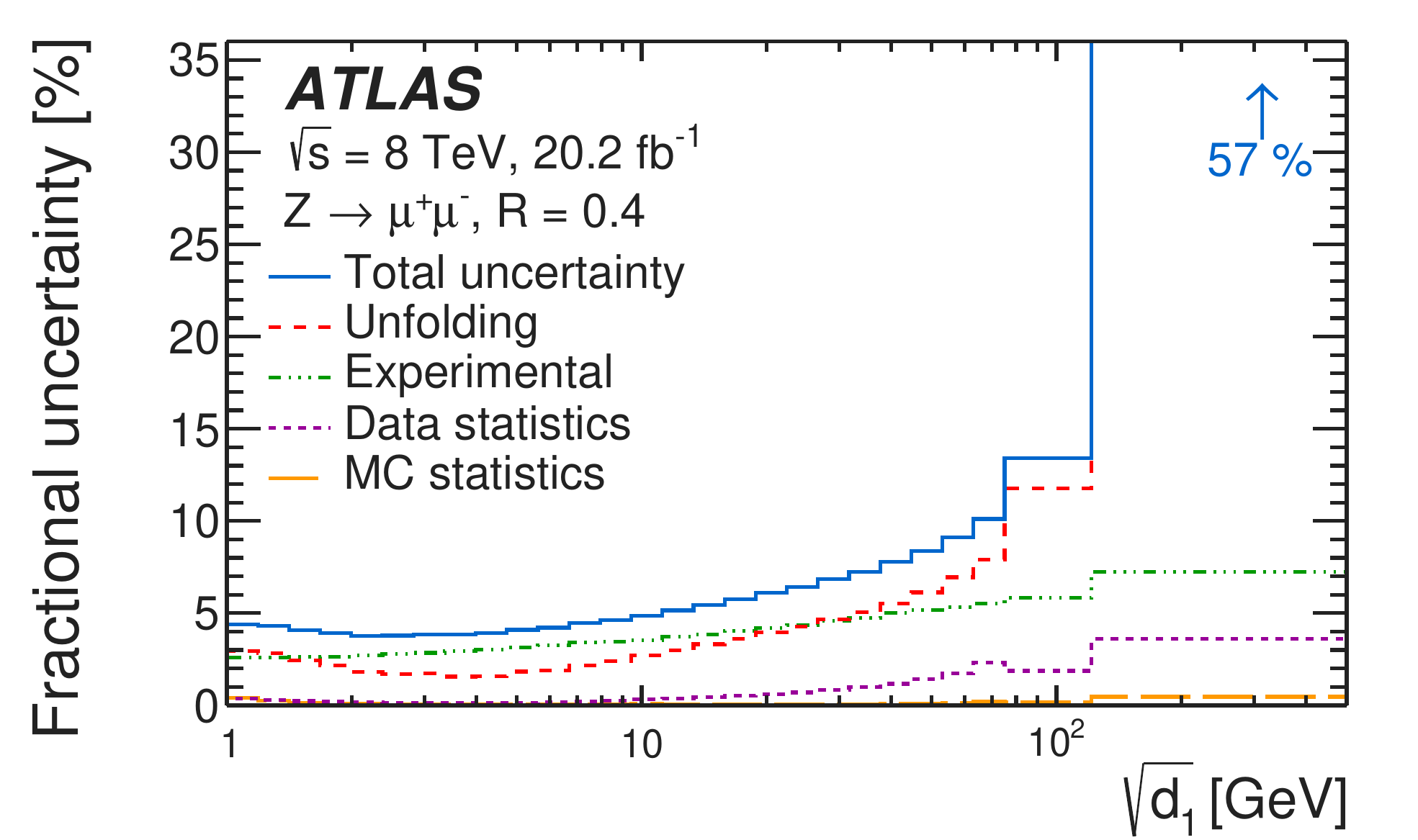}
    }
  }
  \subfloat[electron, $R=1.0$, charged-only]{\centering
    \shortstack{
      \includegraphics[width=0.45\textwidth]{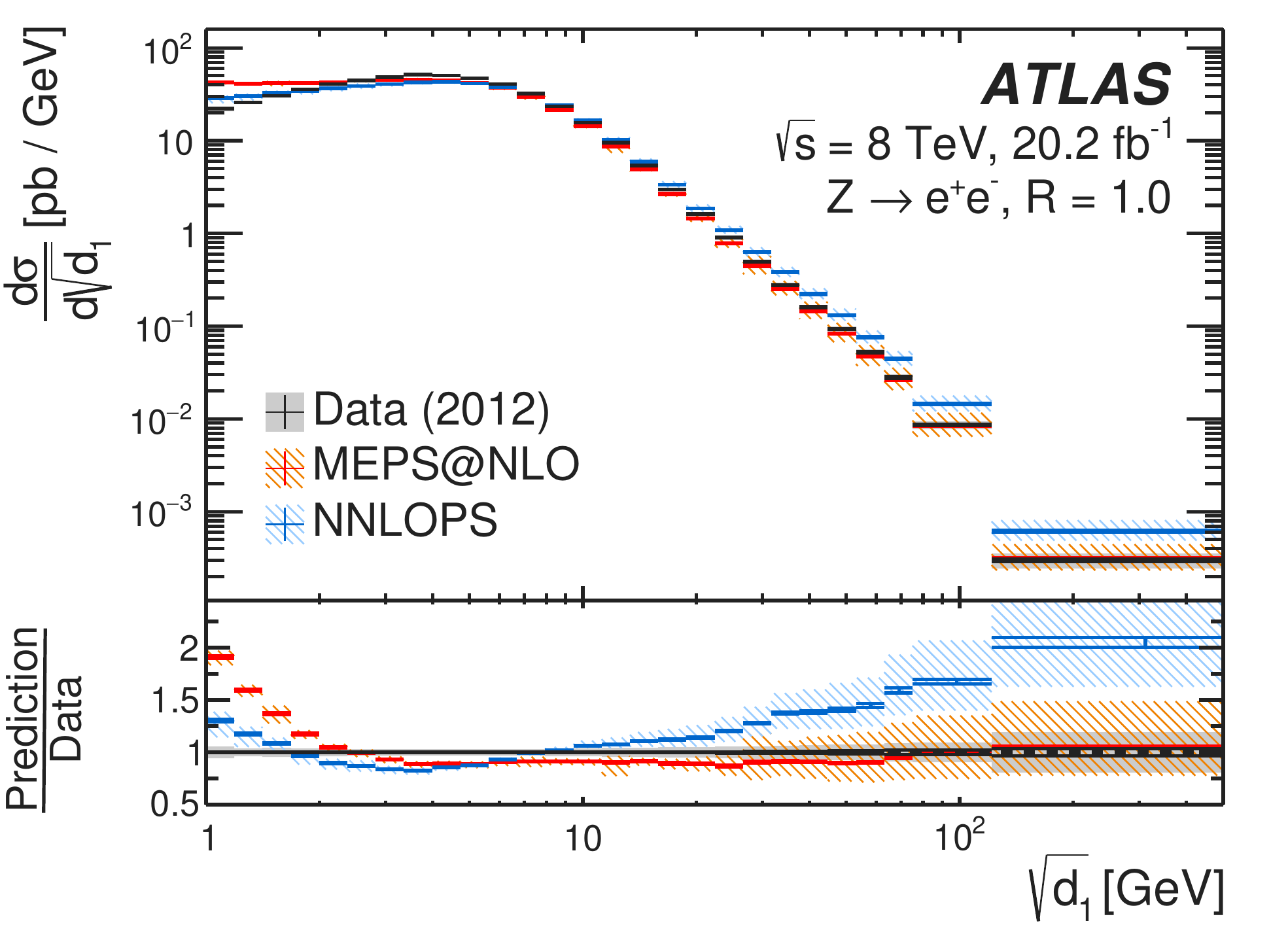}\\
      \includegraphics[width=0.45\textwidth]{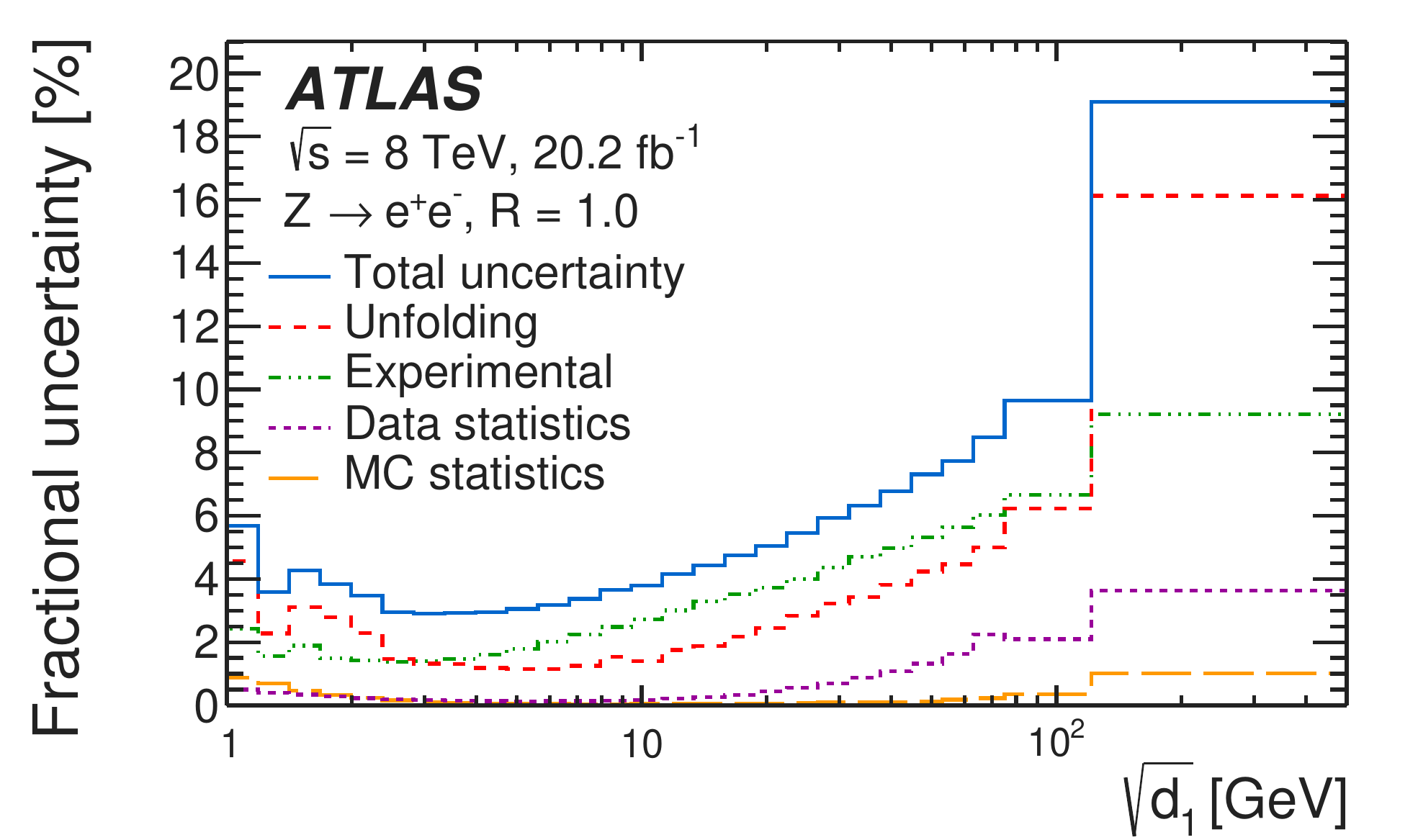}
    }
  }\\
  \subfloat[electron, $R=0.4$, charged+neutral]{
    \shortstack{
      \includegraphics[width=0.45\textwidth]{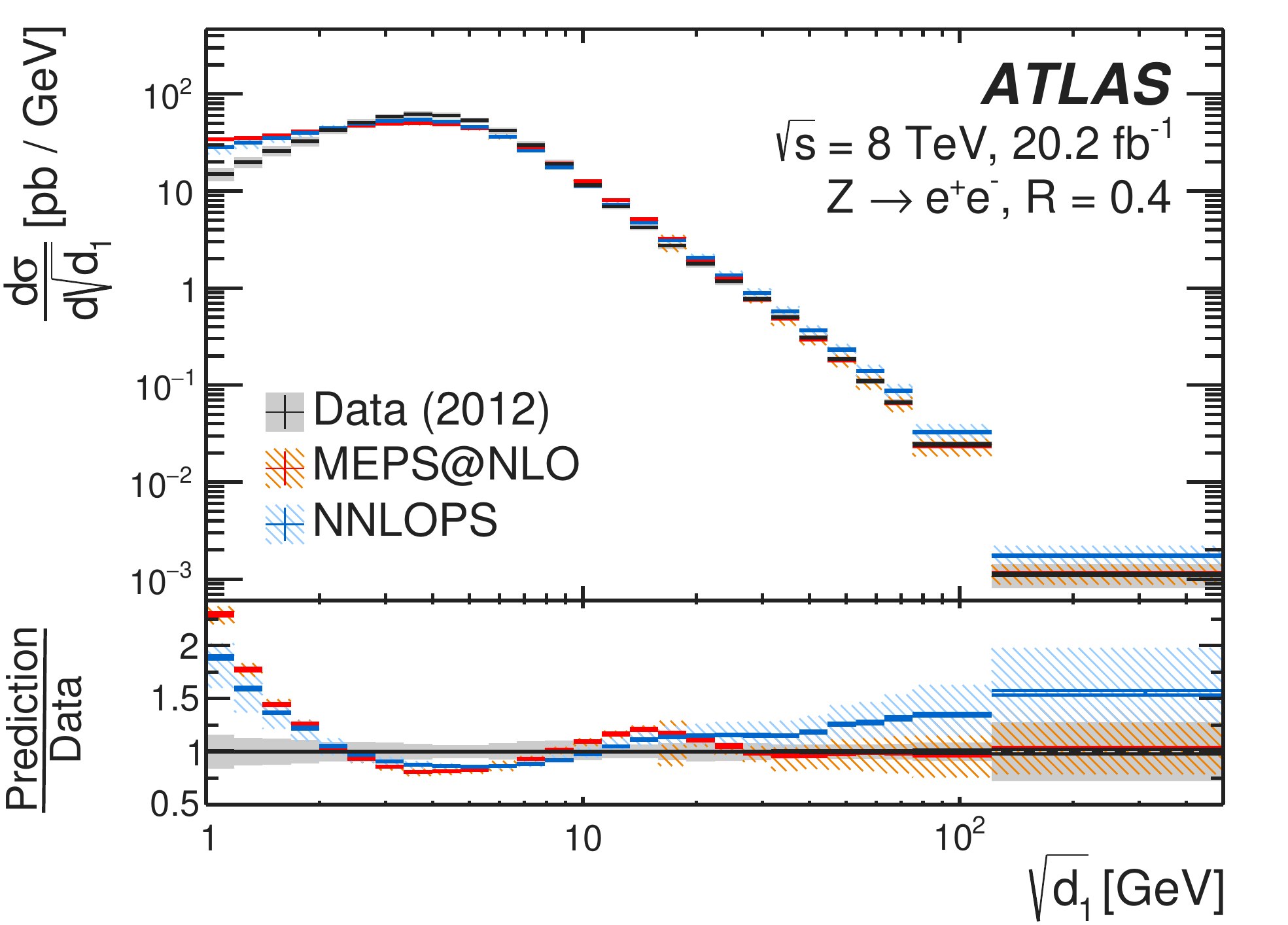}\\
      \includegraphics[width=0.45\textwidth]{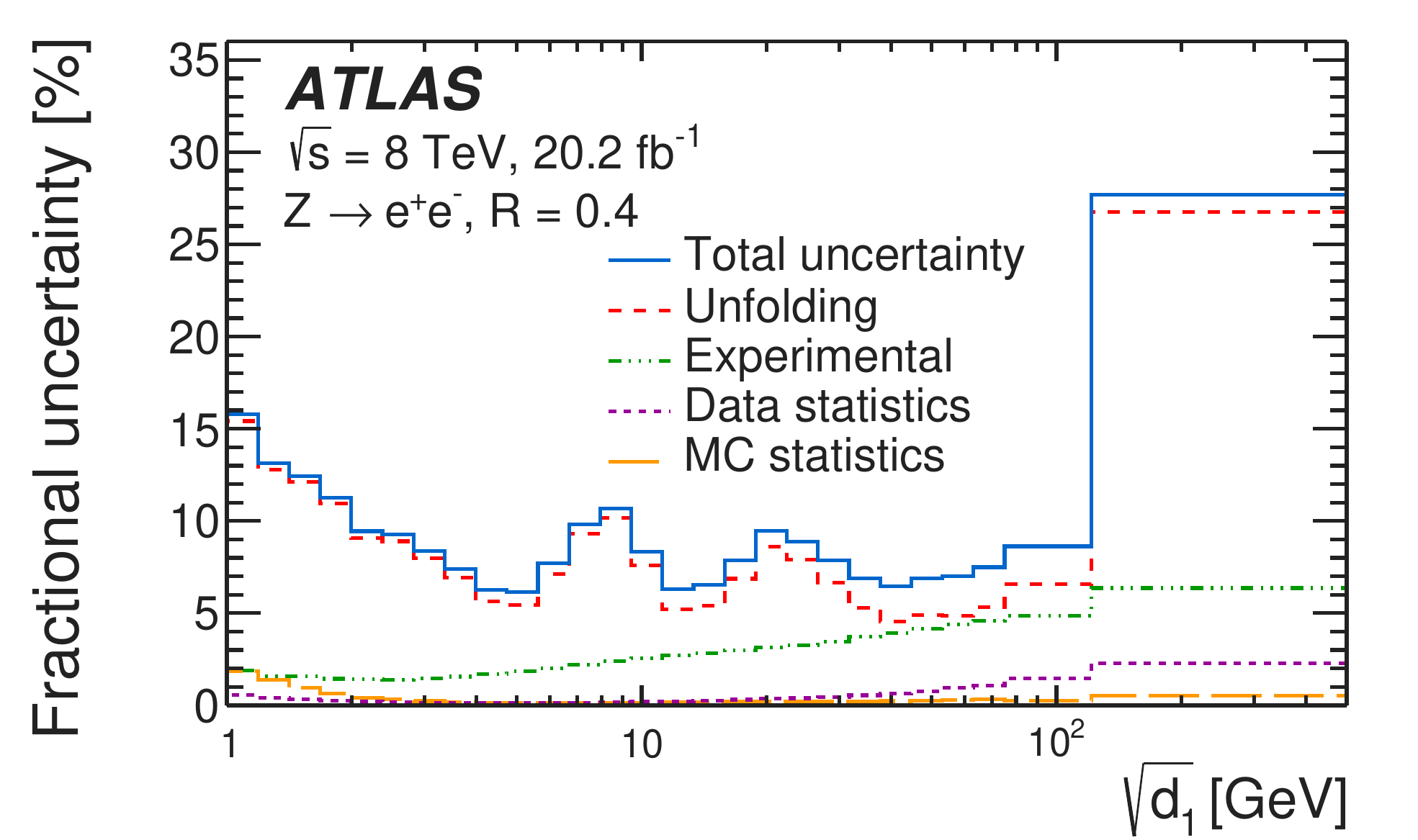}
    }
  }
  \subfloat[electron, $R=1.0$, charged+neutral]{
    \shortstack{
      \includegraphics[width=0.45\textwidth]{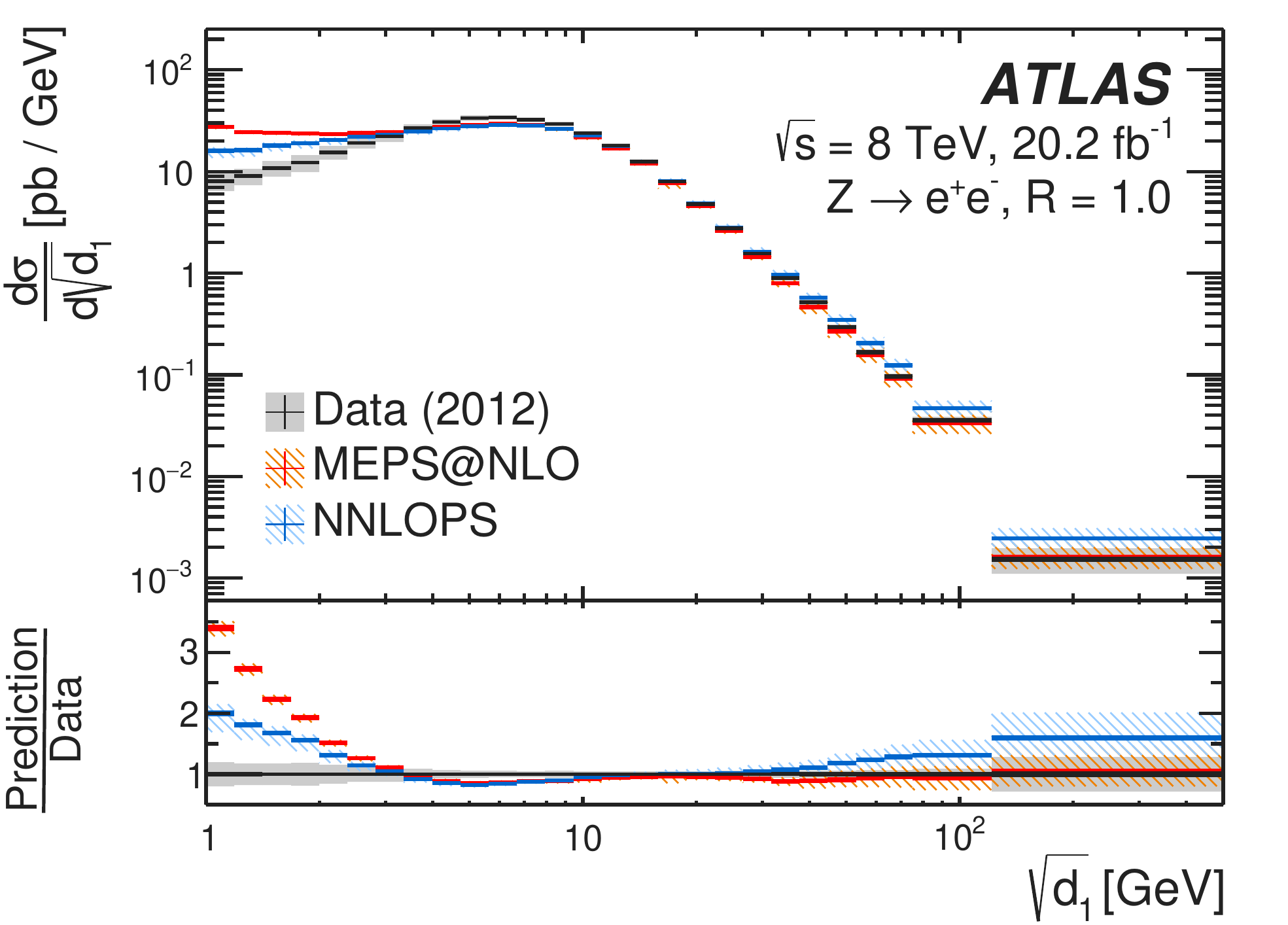}\\
      \includegraphics[width=0.45\textwidth]{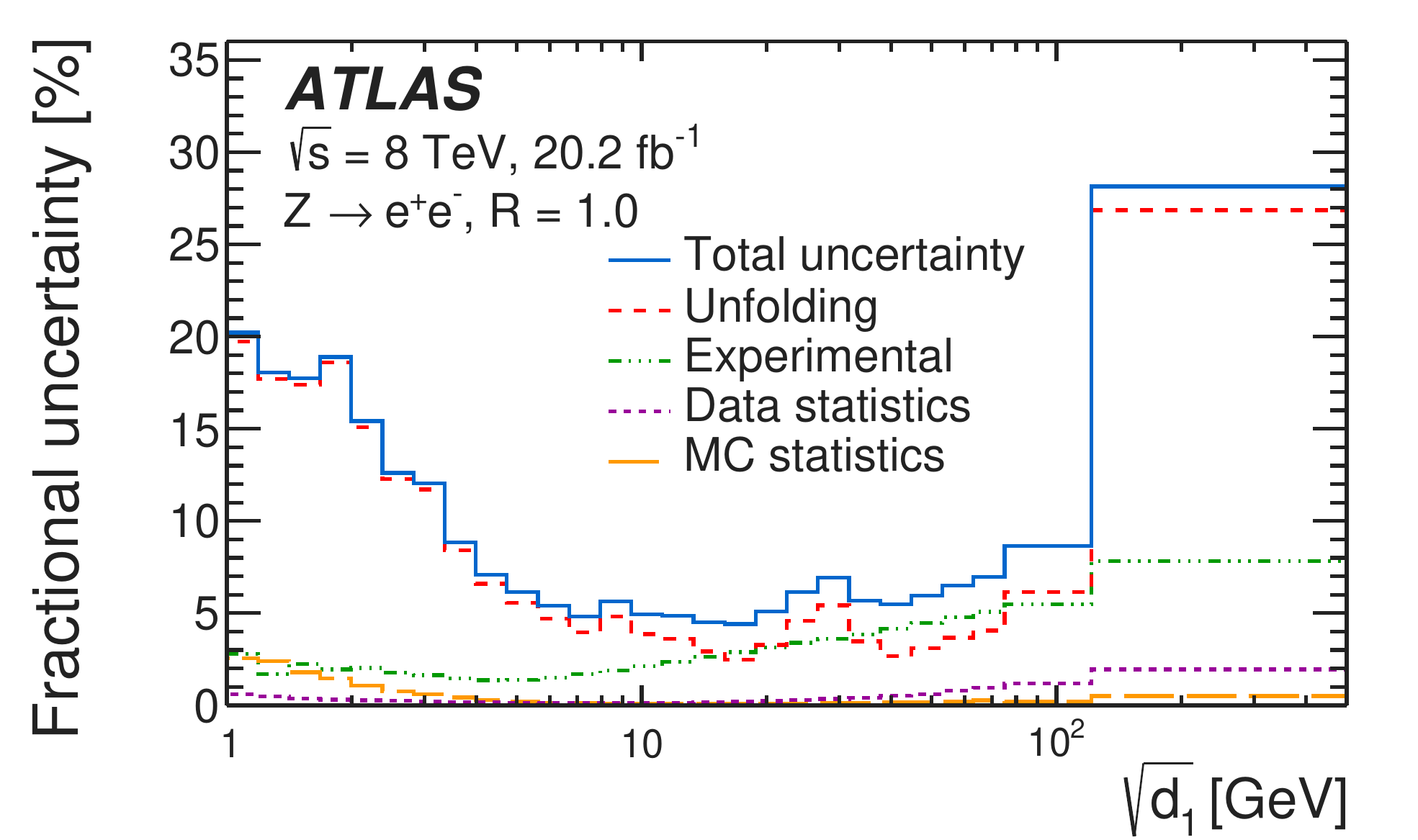}
    }
  }

  \caption{Unfolded $\sqrt{d_1}$ splitting scale for four different variations
    of the analysis as described in the sub-captions.
    The upper panels show the distributions in the same style as
    Figure~\ref{fig:el04nominal}.
    The lower panels show breakdowns of the systematic uncertainties in data in
    the same style as Figure~\ref{fig:el04nominal_systs}.
  }
  \label{fig:variations}
\end{figure}

\FloatBarrier

\section{Conclusions}
\label{sec:conclusion}

Differential cross sections are measured as a function of the splitting scales
in the \kt\ algorithm applied to the hadronic activity in $Z\to \ell\ell$
events.
The measurements use an 8\,\TeV\ proton--proton collison data set recorded by the ATLAS detector at the LHC with
an integrated luminosity of 20.2 fb$^{-1}$.

Splitting scales up to the seventh order are measured for both
the $Z\to e^+e^-$ and $Z\to \mu^+\mu^-$ channels and for jet-radius parameters
of $R=0.4$ and $R=1.0$.
Charged-particle tracks are used for the analysis.

The measurements are corrected for detector effects.
In addition to the charged-particle final state, a Monte Carlo simulation-based
extrapolation to the charged- and neutral-particle final state is provided.
The final results are compared to two state-of-the-art
theoretical predictions, one including NNLO accuracy matrix elements, and a
second prediction with multi-leg NLO merging. Significant deviations from data 
are found for both predictions, in the perturbative as well as the non-perturbative
regimes.

The measurements of splitting scales are sensitive to the hard perturbative modelling
at high scale values as well as soft hadronic activity at lower values. They thus
provide a valuable input complementary to standard jet measurements, in particular in
the transition region.
With their specific identification of QCD jet
evolution they provide means to constrain and potentially tune Monte Carlo event
generators.

\section*{Acknowledgements}


We thank CERN for the very successful operation of the LHC, as well as the
support staff from our institutions without whom ATLAS could not be
operated efficiently.

We acknowledge the support of ANPCyT, Argentina; YerPhI, Armenia; ARC, Australia; BMWFW and FWF, Austria; ANAS, Azerbaijan; SSTC, Belarus; CNPq and FAPESP, Brazil; NSERC, NRC and CFI, Canada; CERN; CONICYT, Chile; CAS, MOST and NSFC, China; COLCIENCIAS, Colombia; MSMT CR, MPO CR and VSC CR, Czech Republic; DNRF and DNSRC, Denmark; IN2P3-CNRS, CEA-DSM/IRFU, France; SRNSF, Georgia; BMBF, HGF, and MPG, Germany; GSRT, Greece; RGC, Hong Kong SAR, China; ISF, I-CORE and Benoziyo Center, Israel; INFN, Italy; MEXT and JSPS, Japan; CNRST, Morocco; NWO, Netherlands; RCN, Norway; MNiSW and NCN, Poland; FCT, Portugal; MNE/IFA, Romania; MES of Russia and NRC KI, Russian Federation; JINR; MESTD, Serbia; MSSR, Slovakia; ARRS and MIZ\v{S}, Slovenia; DST/NRF, South Africa; MINECO, Spain; SRC and Wallenberg Foundation, Sweden; SERI, SNSF and Cantons of Bern and Geneva, Switzerland; MOST, Taiwan; TAEK, Turkey; STFC, United Kingdom; DOE and NSF, United States of America. In addition, individual groups and members have received support from BCKDF, the Canada Council, CANARIE, CRC, Compute Canada, FQRNT, and the Ontario Innovation Trust, Canada; EPLANET, ERC, ERDF, FP7, Horizon 2020 and Marie Sk{\l}odowska-Curie Actions, European Union; Investissements d'Avenir Labex and Idex, ANR, R{\'e}gion Auvergne and Fondation Partager le Savoir, France; DFG and AvH Foundation, Germany; Herakleitos, Thales and Aristeia programmes co-financed by EU-ESF and the Greek NSRF; BSF, GIF and Minerva, Israel; BRF, Norway; CERCA Programme Generalitat de Catalunya, Generalitat Valenciana, Spain; the Royal Society and Leverhulme Trust, United Kingdom.

The crucial computing support from all WLCG partners is acknowledged gratefully, in particular from CERN, the ATLAS Tier-1 facilities at TRIUMF (Canada), NDGF (Denmark, Norway, Sweden), CC-IN2P3 (France), KIT/GridKA (Germany), INFN-CNAF (Italy), NL-T1 (Netherlands), PIC (Spain), ASGC (Taiwan), RAL (UK) and BNL (USA), the Tier-2 facilities worldwide and large non-WLCG resource providers. Major contributors of computing resources are listed in Ref.~\cite{ATL-GEN-PUB-2016-002}.

\clearpage
\appendix
\part*{Appendix}
\addcontentsline{toc}{part}{Appendix}
\section{Results for the electron channel with $R=1.0$}
\label{sec:app_el10nominal}

This section contains results obtained with a jet radius parameter $R=1.0$ in the electron channel.

\begin{figure}[htbp]
  \centering
  \includegraphics[width=0.45\textwidth]{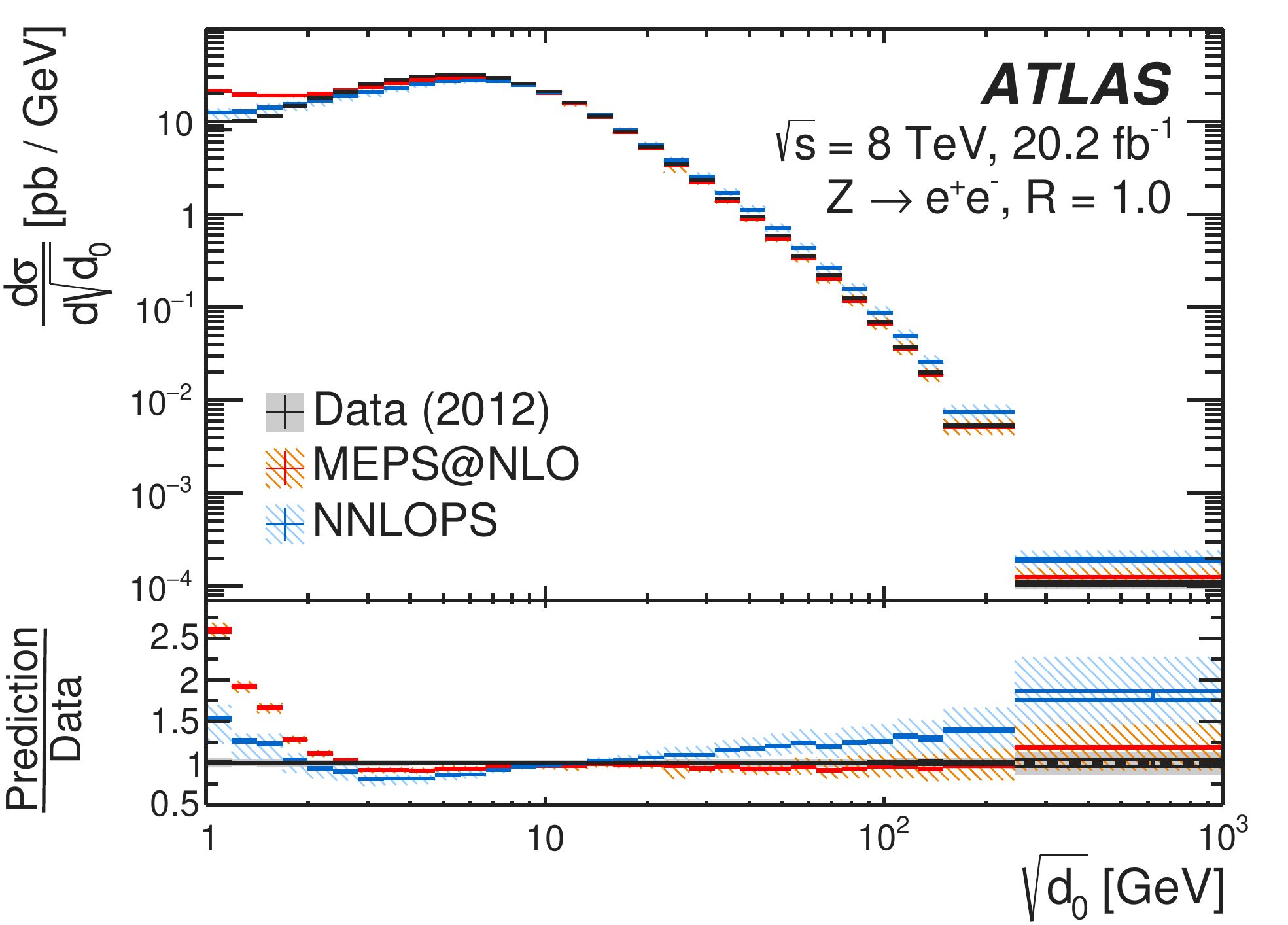}%
  \includegraphics[width=0.45\textwidth]{unfolded_r10/el_D1_kT4_inclusive}\\
  \includegraphics[width=0.45\textwidth]{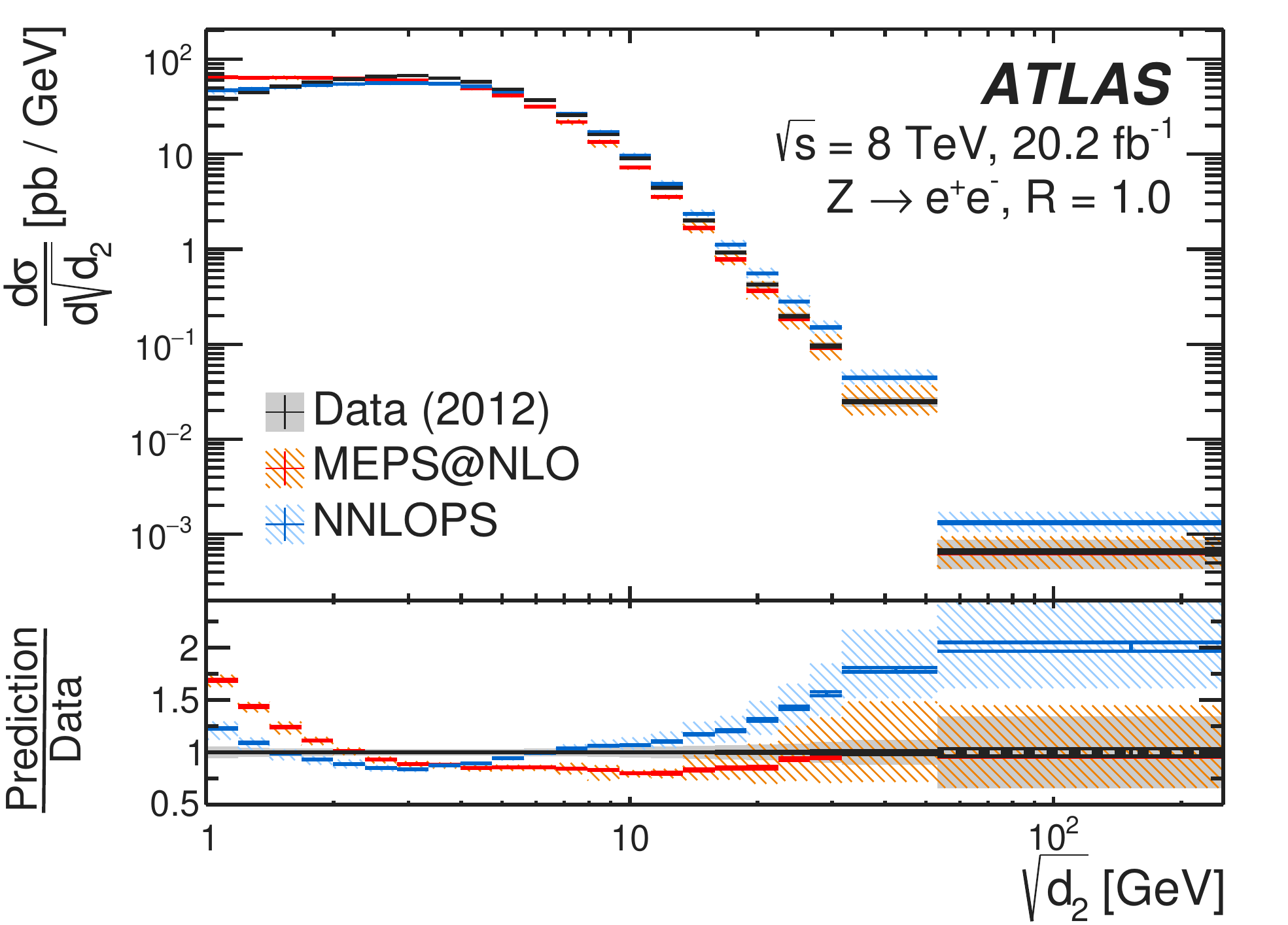}%
  \includegraphics[width=0.45\textwidth]{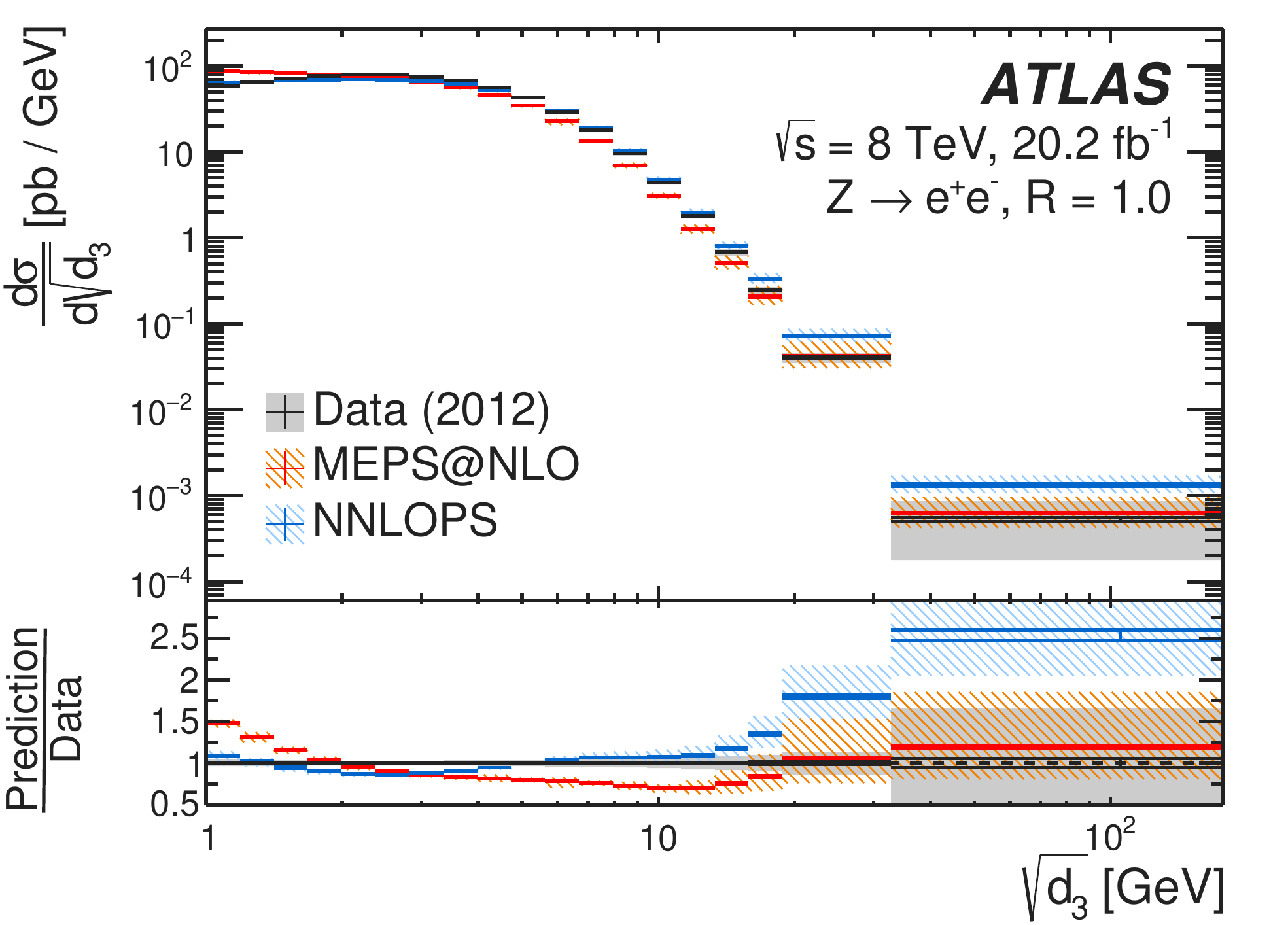}\\
  \includegraphics[width=0.45\textwidth]{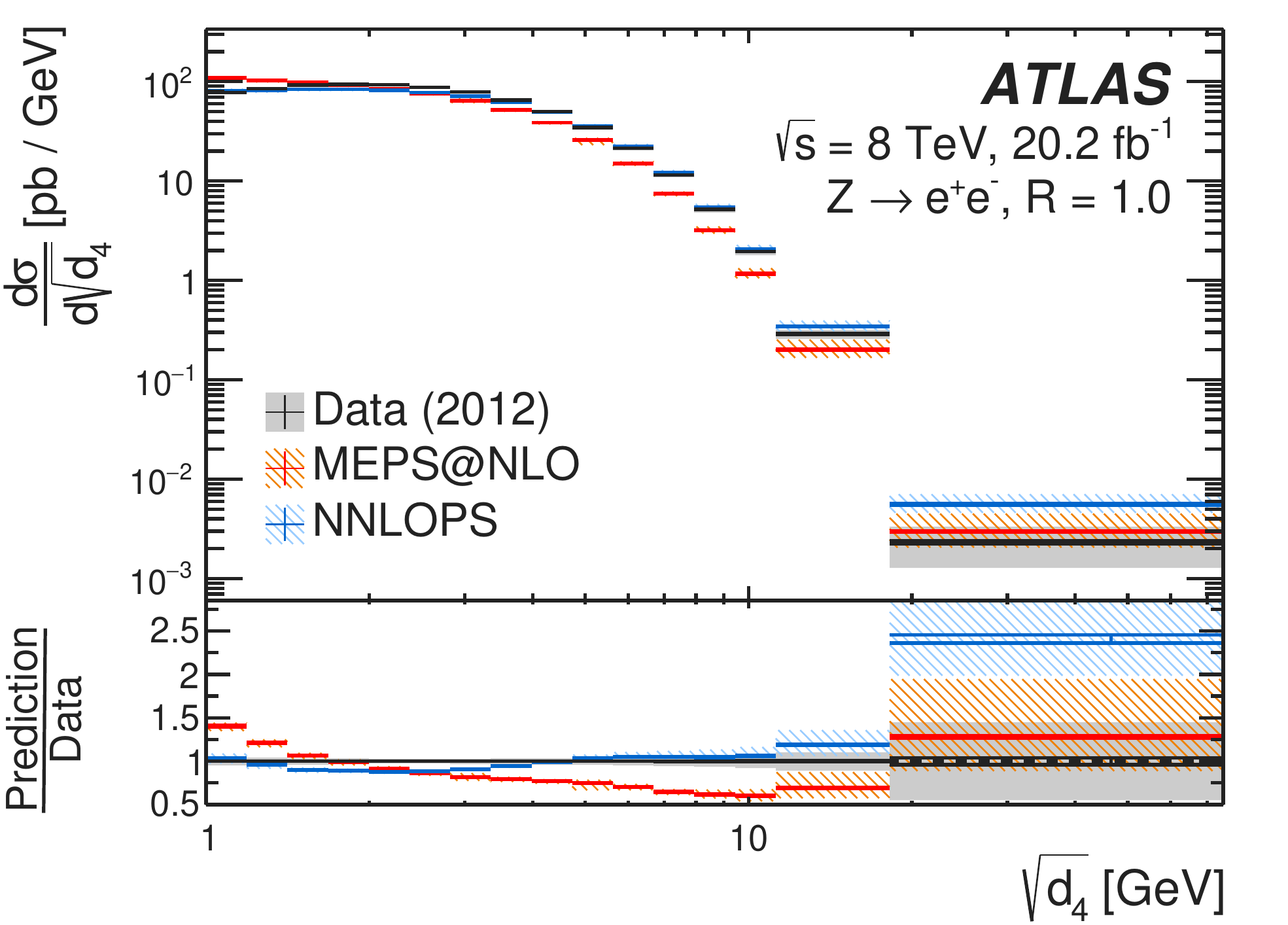}%
  \includegraphics[width=0.45\textwidth]{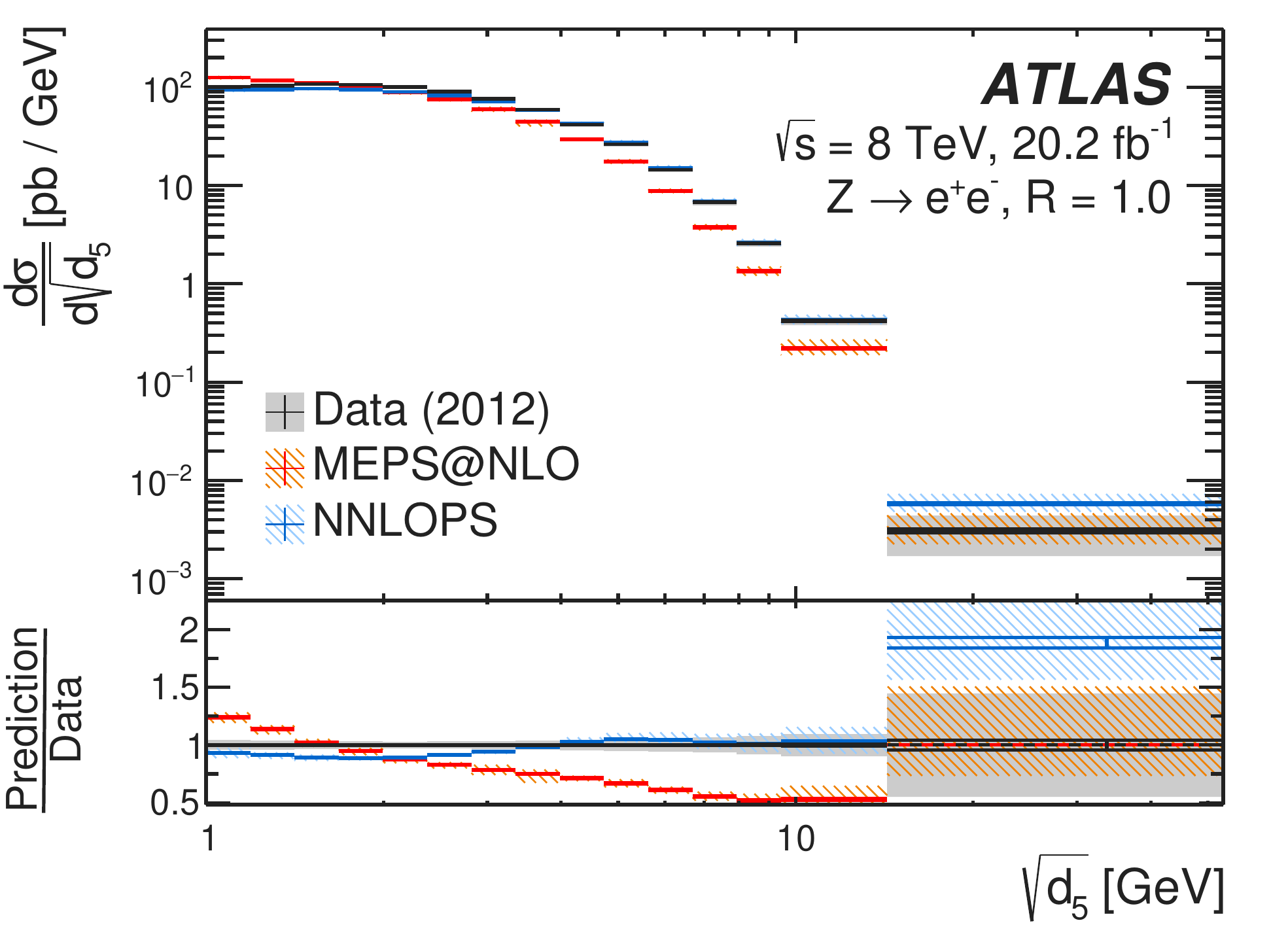}\\
  \includegraphics[width=0.45\textwidth]{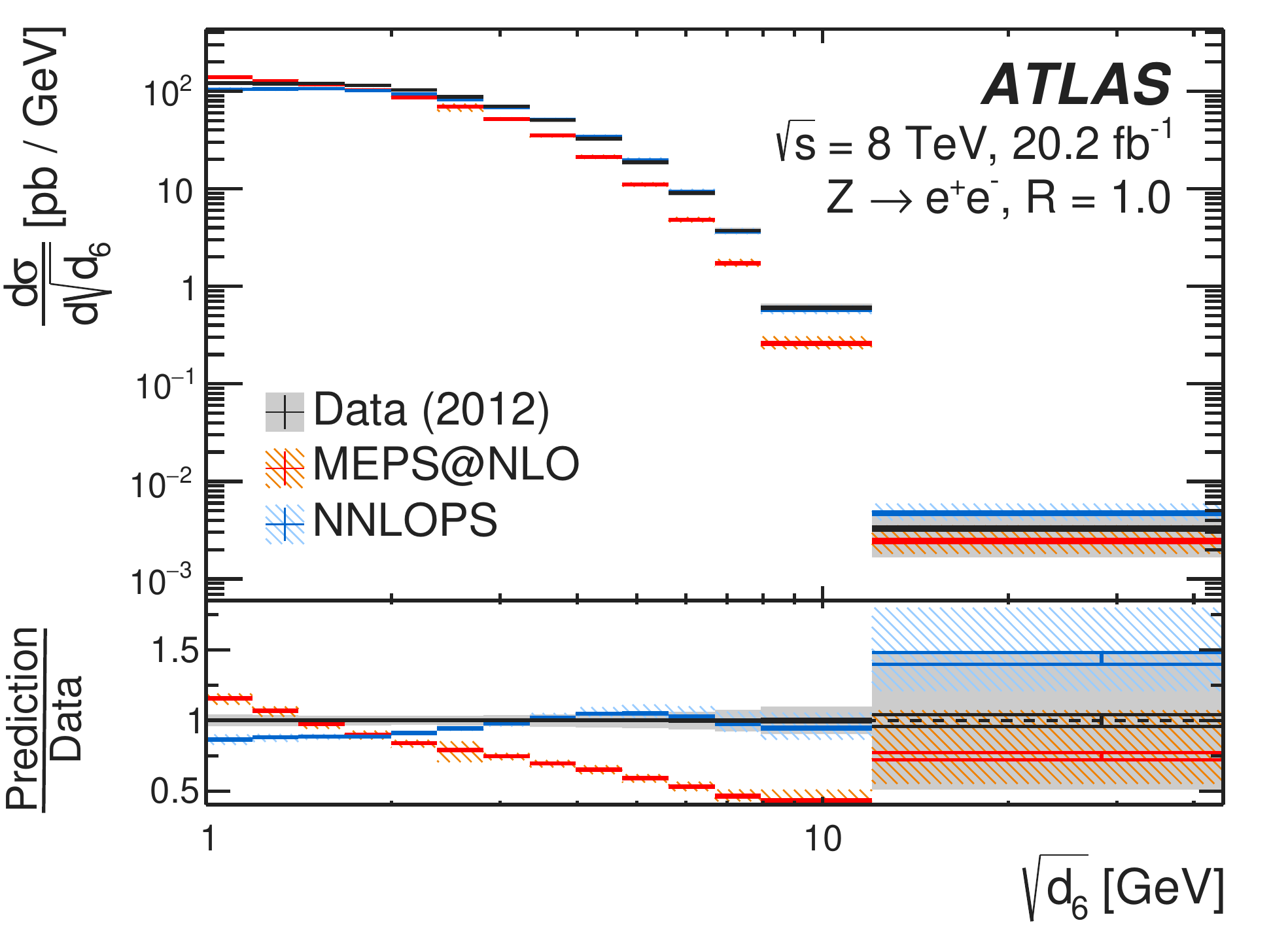}%
  \includegraphics[width=0.45\textwidth]{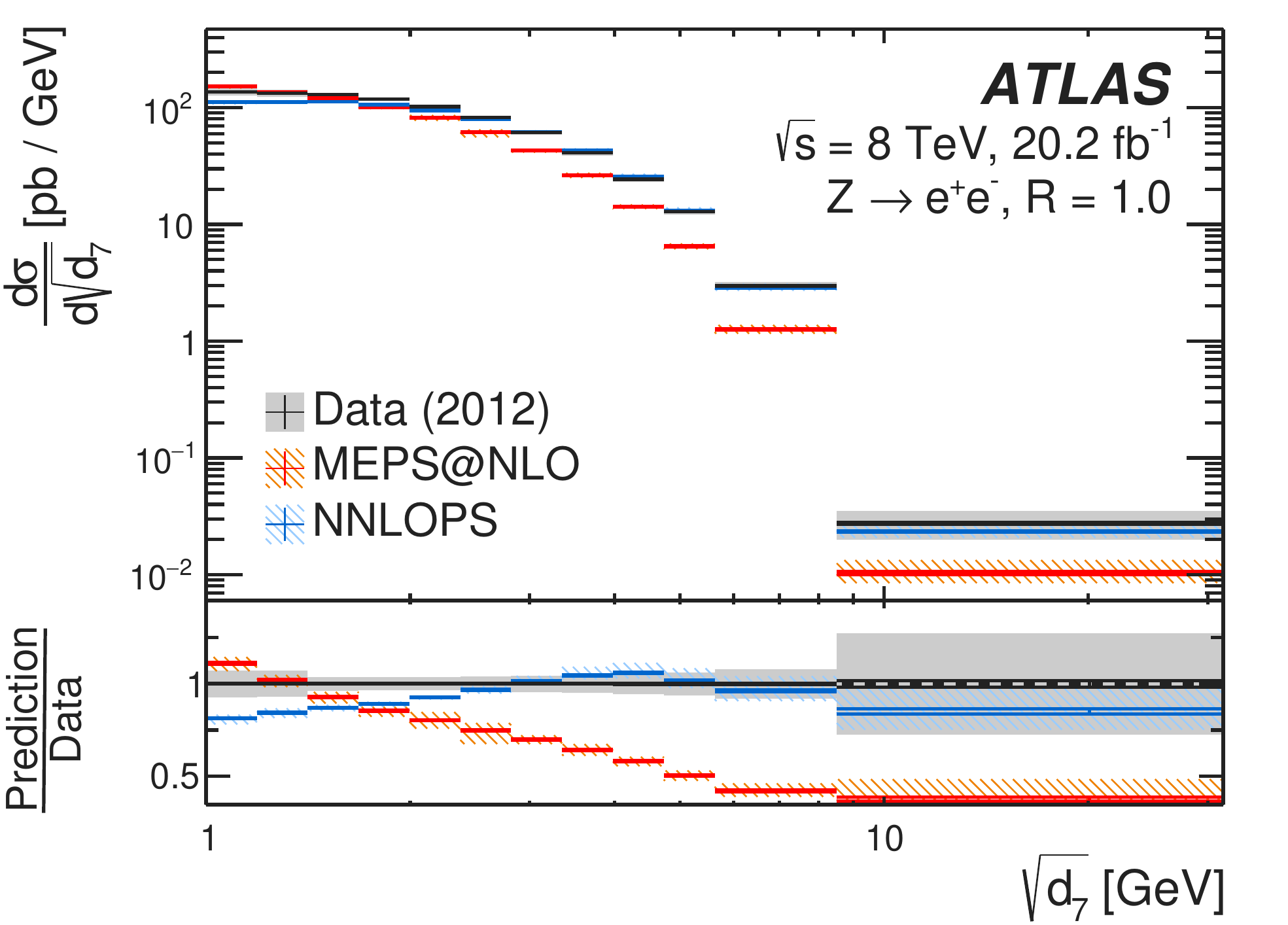}
  \caption{Charged-only distributions for the eight leading splitting scales
    in the electron channel using the jet-radius parameter $R=1.0$.
    The size of the error bars reflects the statistical uncertainty, while the combined statistical
    and systematic uncertainty is indicated by the grey band.
    Theoretical predictions from \sherpa\ with NLO multijet merging (``MEPS@NLO'') and from \powheg$+$\pythia\ with NNLO matching (``NNLOPS'') are displayed.
  }
\end{figure}

\clearpage
\section{Results for the muon channel with $R=0.4$ and $R=1.0$}
\label{sec:muonresults}

This section contains results obtained with a jet-radius parameter $R=0.4$ and $R=1.0$ in the muon channel.

\begin{figure}[htbp]
  \centering
  \includegraphics[width=0.45\textwidth]{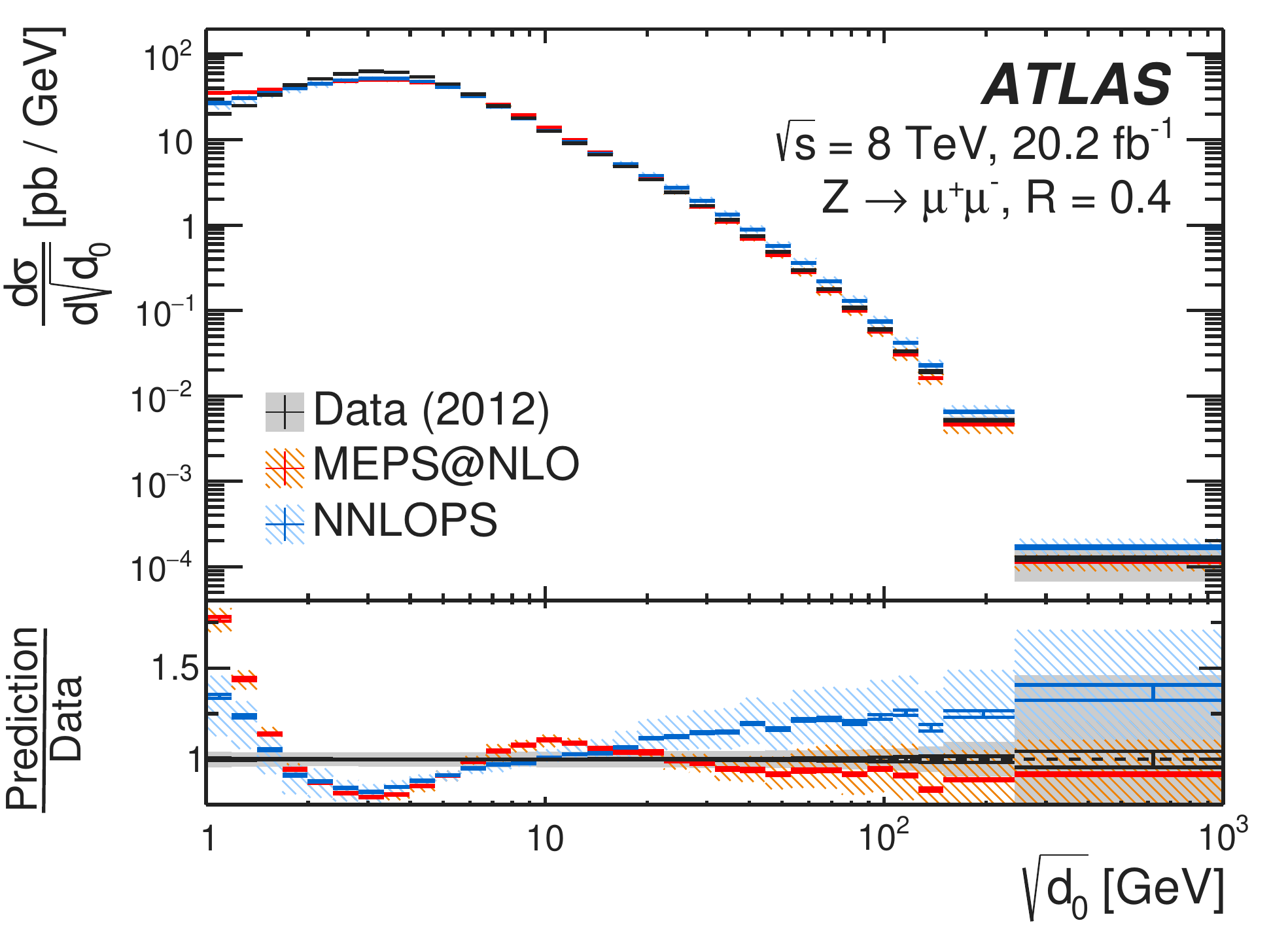}%
  \includegraphics[width=0.45\textwidth]{unfolded/mu_D1_kT4_inclusive}\\
  \includegraphics[width=0.45\textwidth]{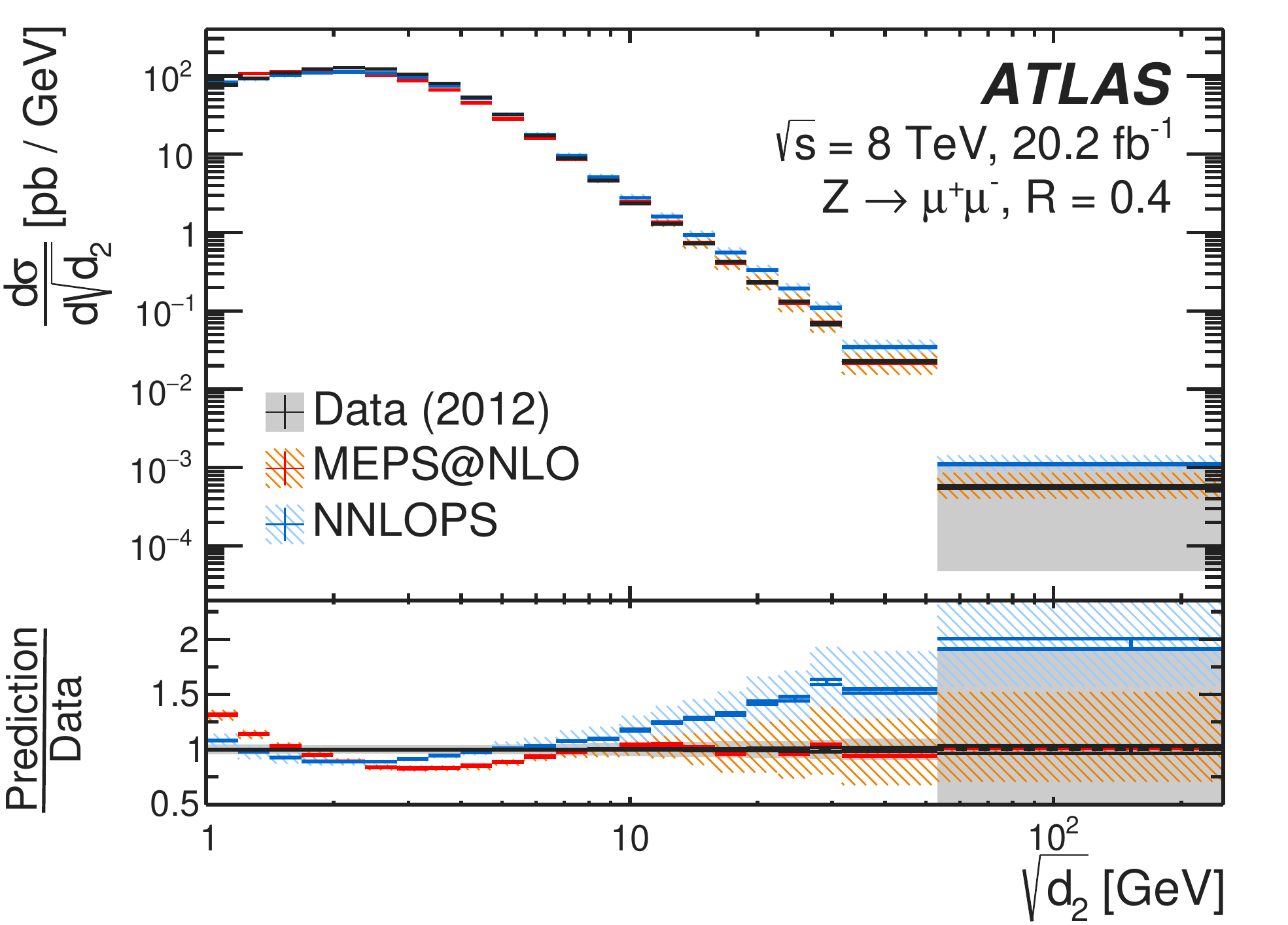}%
  \includegraphics[width=0.45\textwidth]{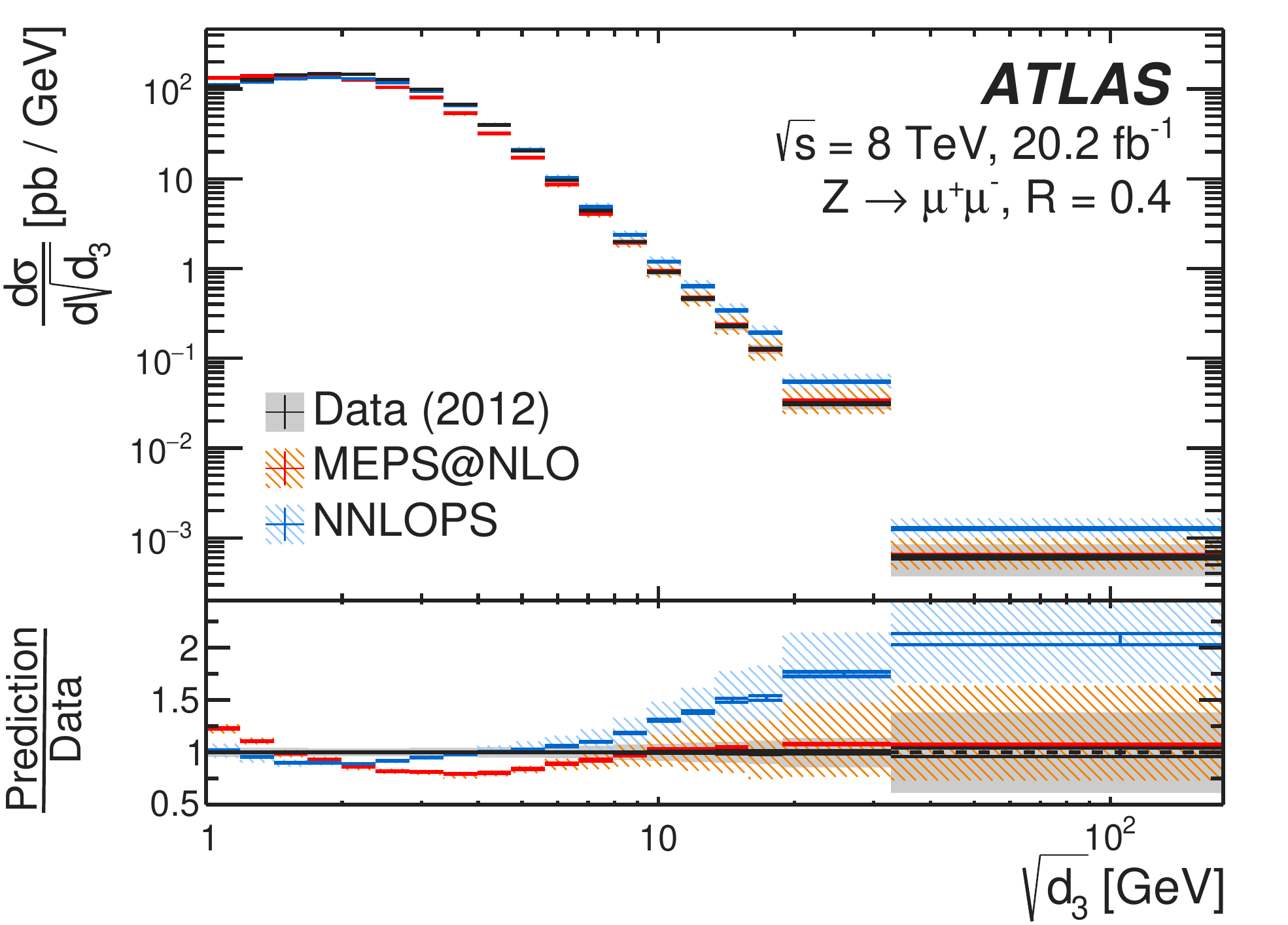}\\
  \includegraphics[width=0.45\textwidth]{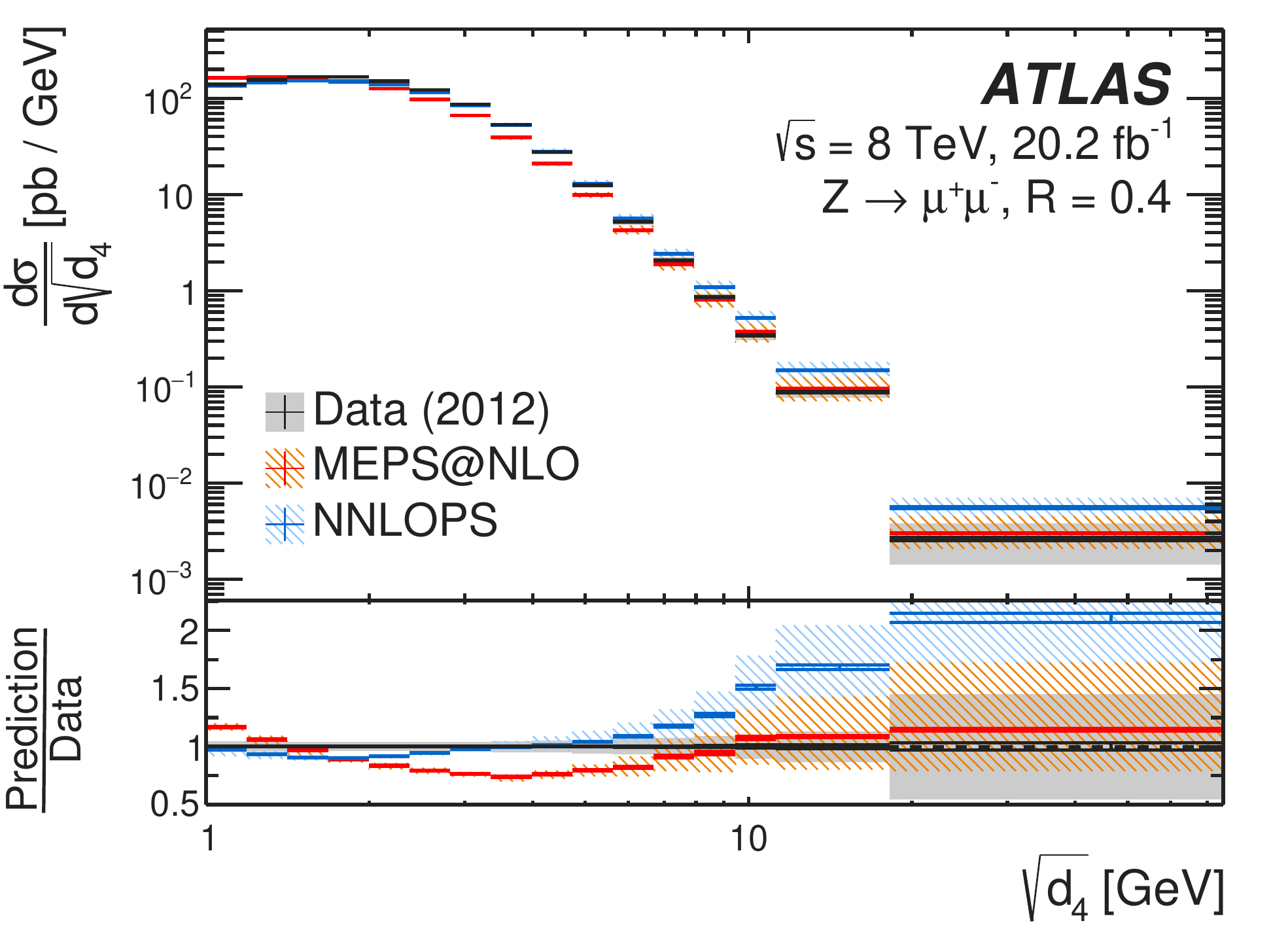}%
  \includegraphics[width=0.45\textwidth]{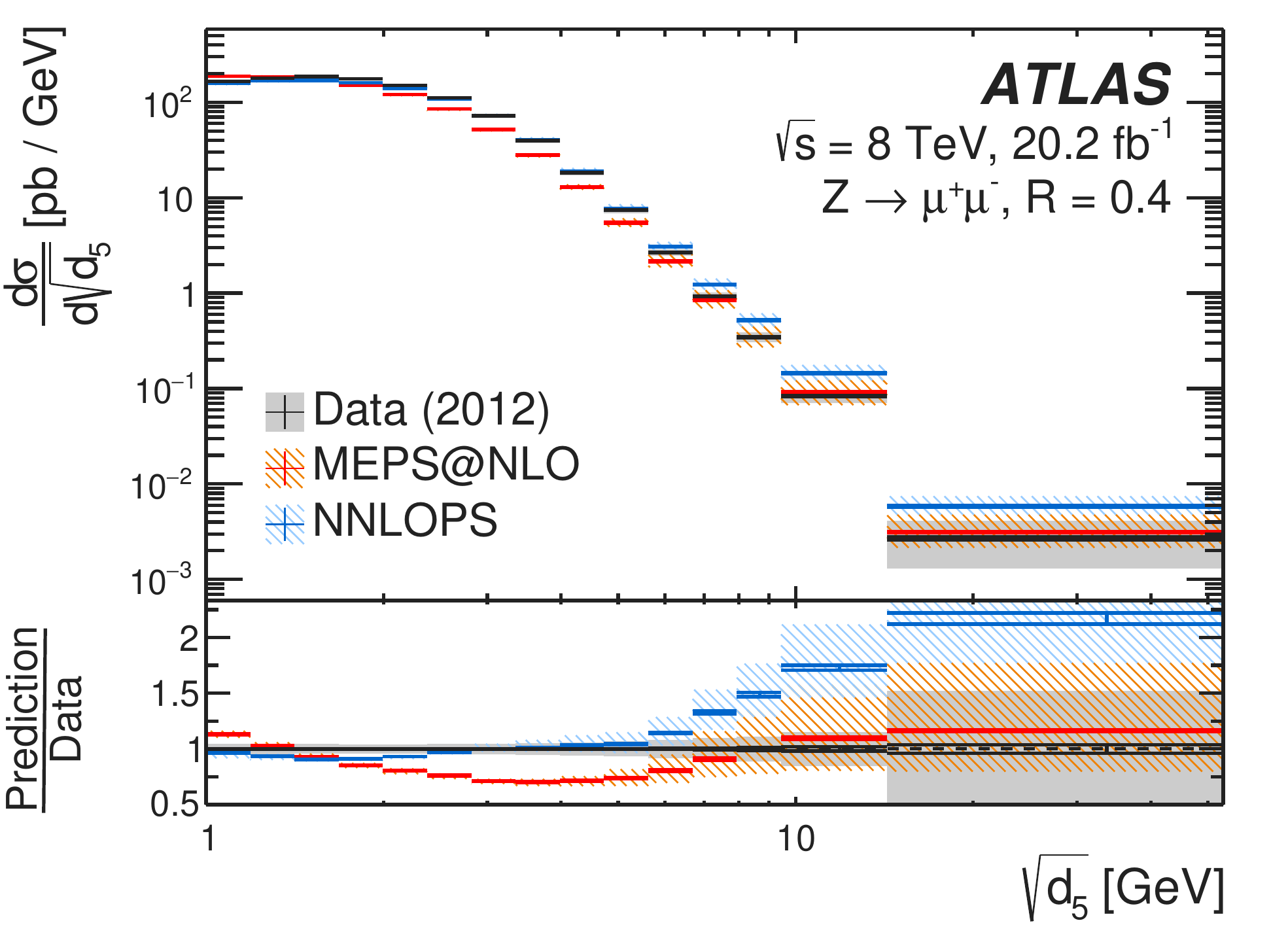}\\
  \includegraphics[width=0.45\textwidth]{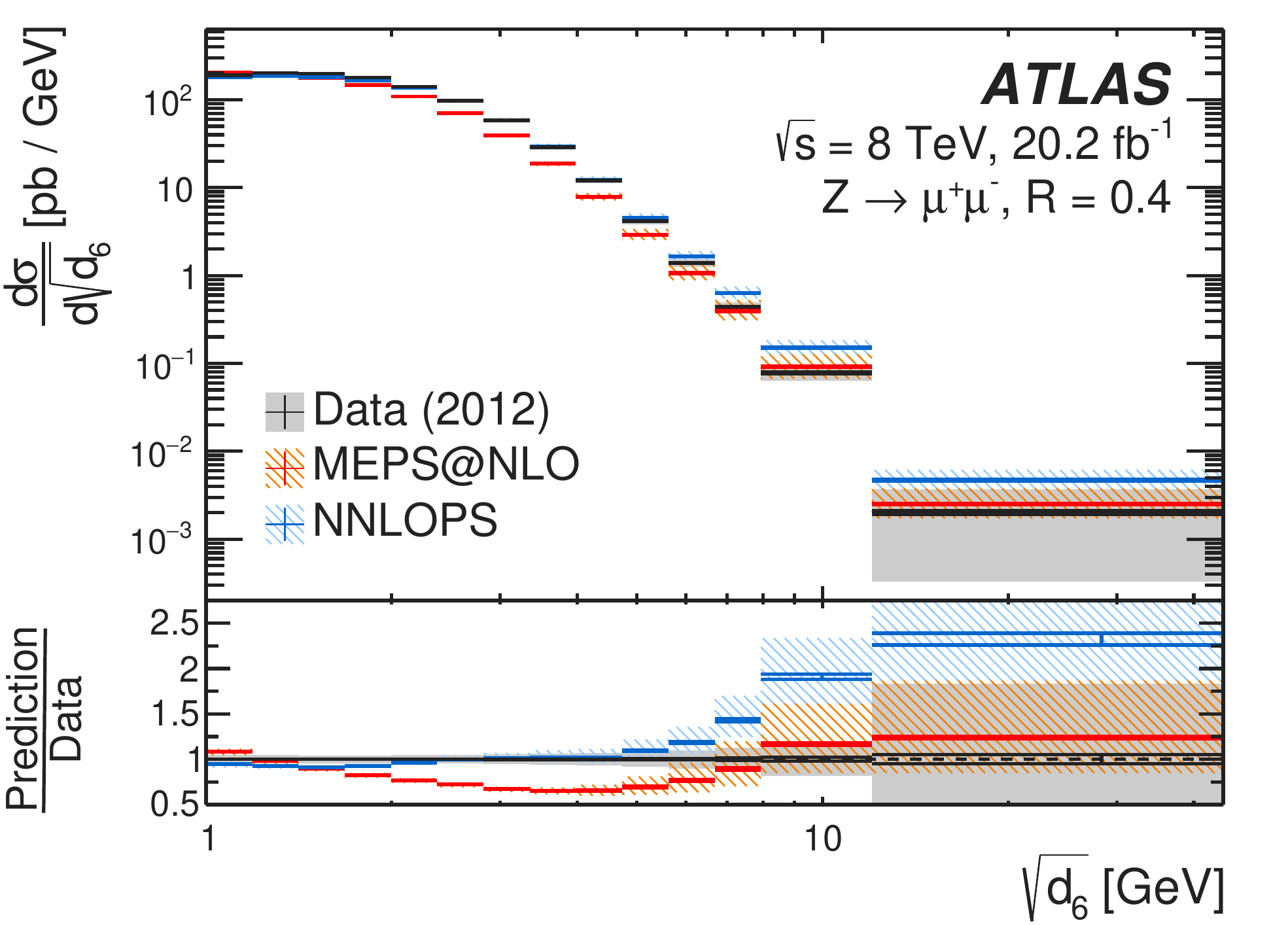}%
  \includegraphics[width=0.45\textwidth]{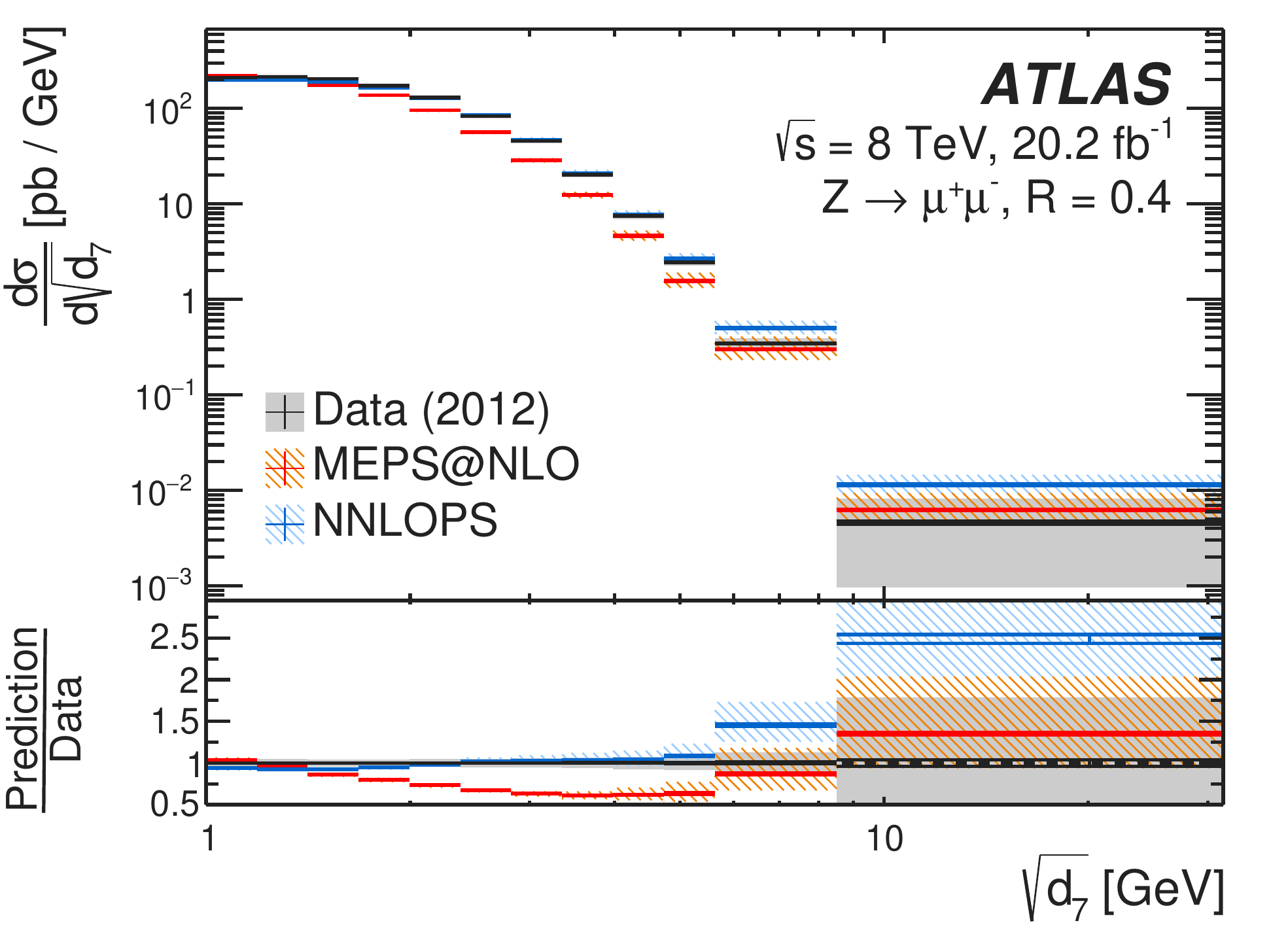}
  \caption{Charged-only distributions for the eight leading splitting scales
    in the muon channel using the jet-radius parameter $R=0.4$.
    The size of the error bars reflects the statistical uncertainty, while the combined statistical
    and systematic uncertainty is indicated by the grey band.
    Theoretical predictions from \sherpa\ with NLO multijet merging (``MEPS@NLO'') and from \powheg$+$\pythia\ with NNLO matching (``NNLOPS'') are displayed.
  }
\end{figure}

\begin{figure}[htbp]
  \centering
  \includegraphics[width=0.45\textwidth]{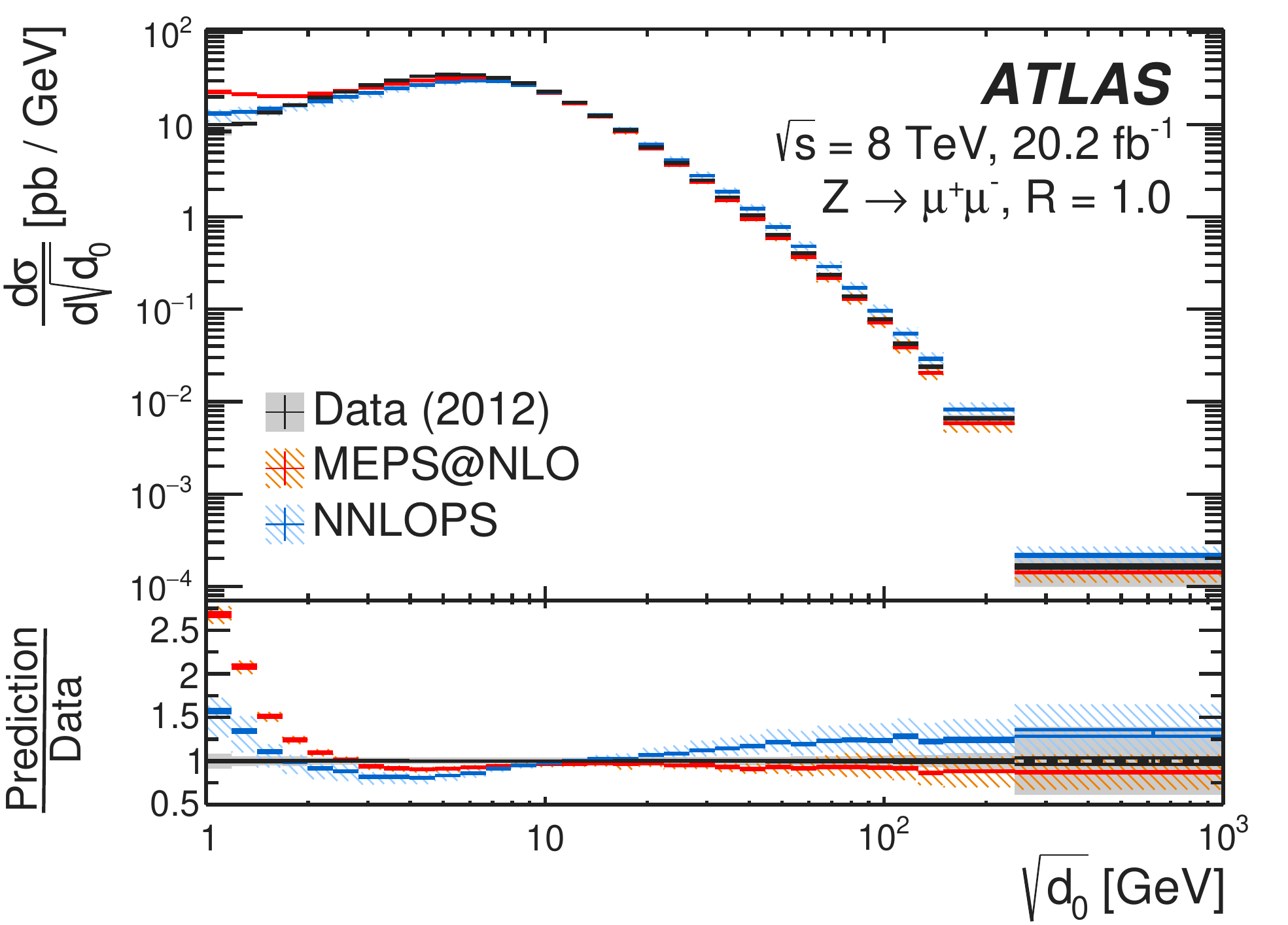}%
  \includegraphics[width=0.45\textwidth]{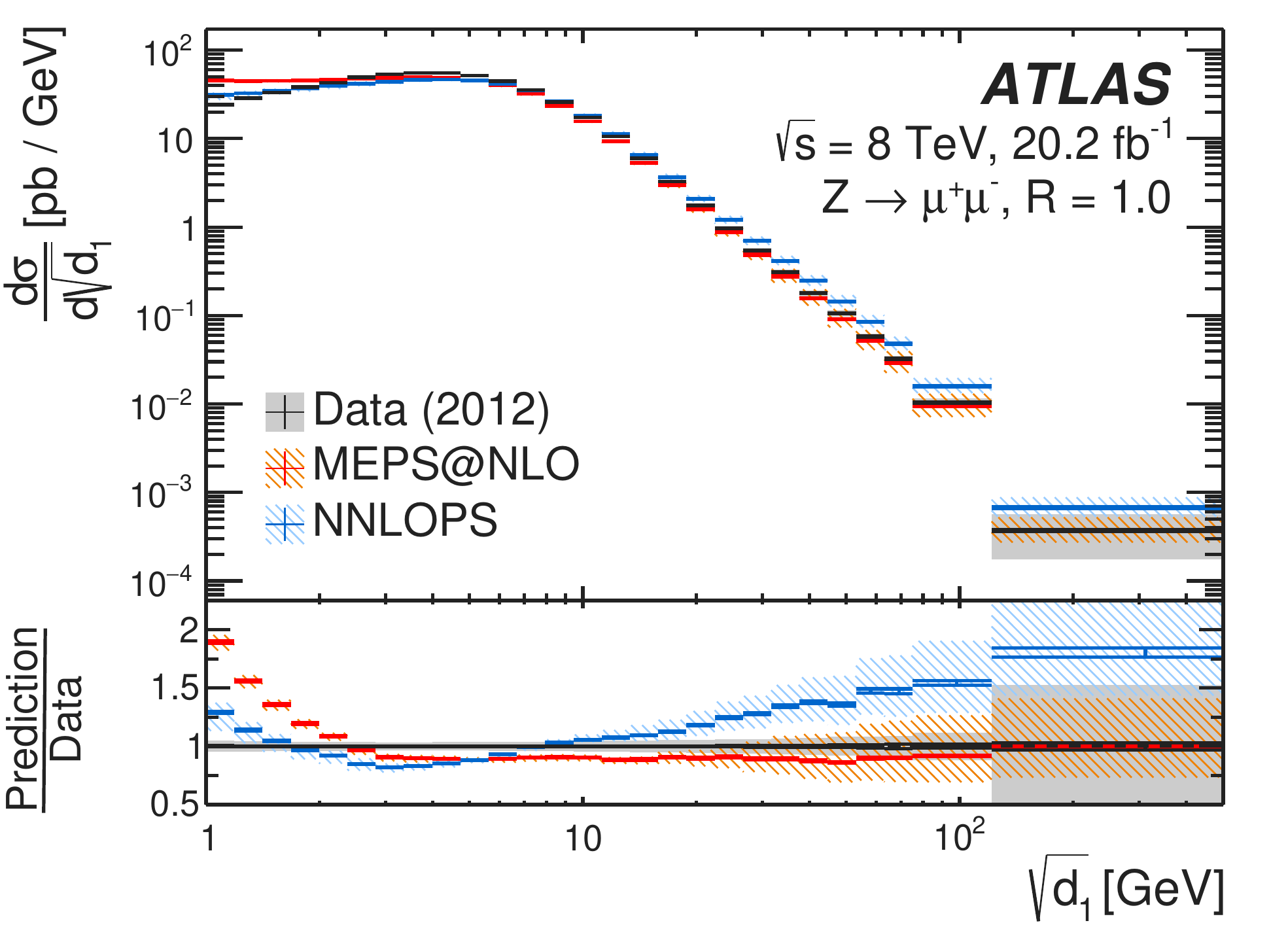}\\
  \includegraphics[width=0.45\textwidth]{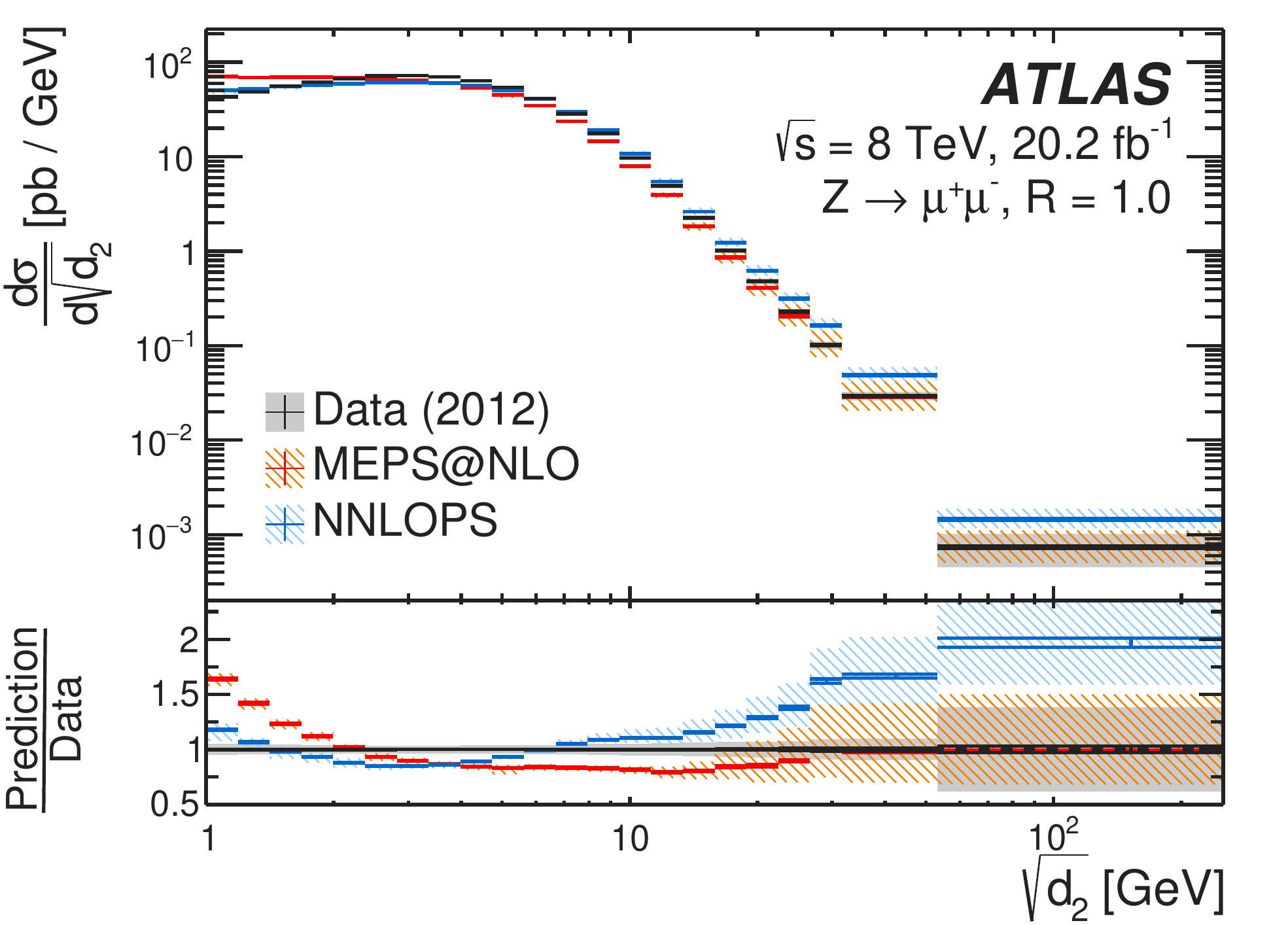}%
  \includegraphics[width=0.45\textwidth]{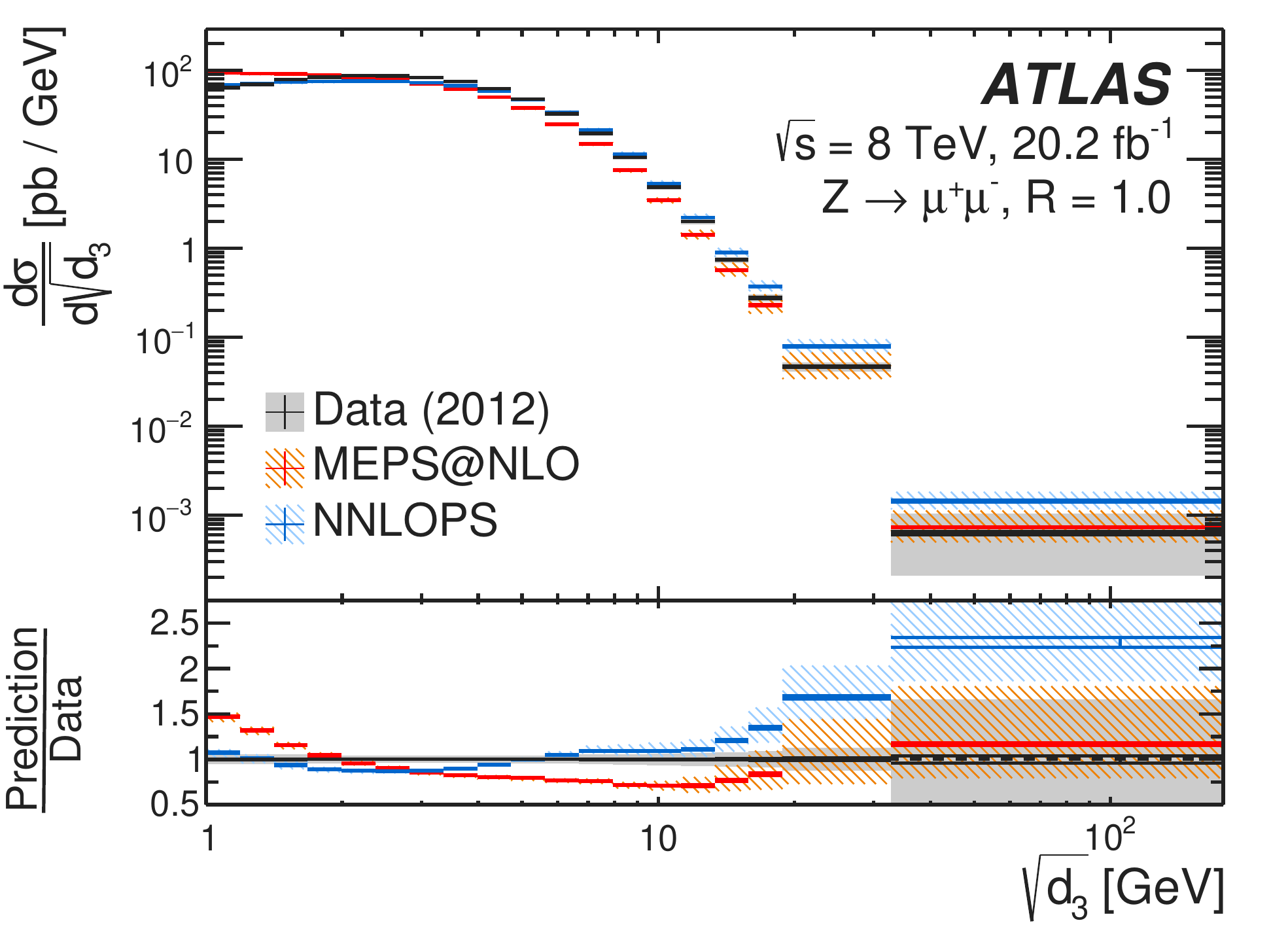}\\
  \includegraphics[width=0.45\textwidth]{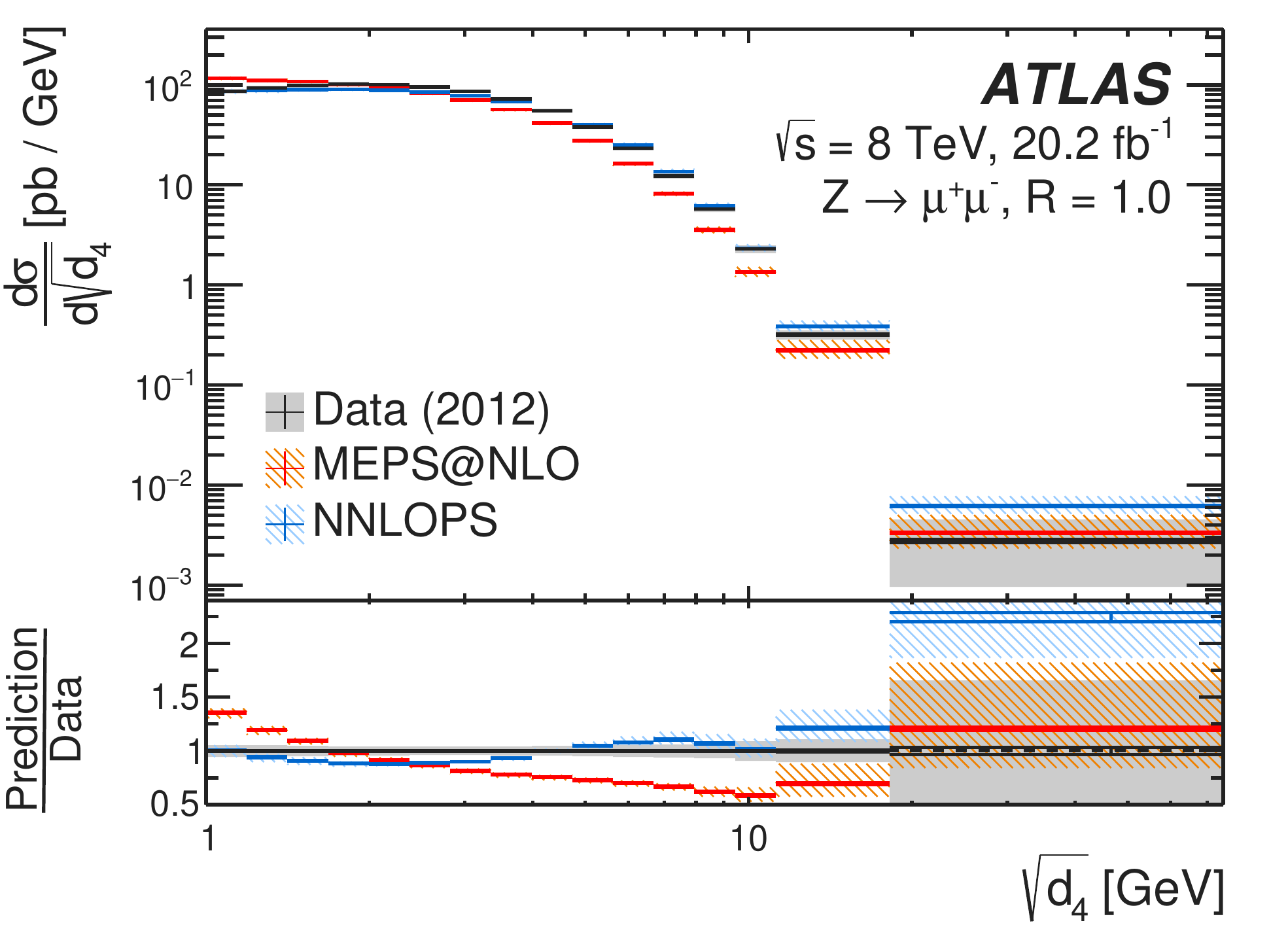}%
  \includegraphics[width=0.45\textwidth]{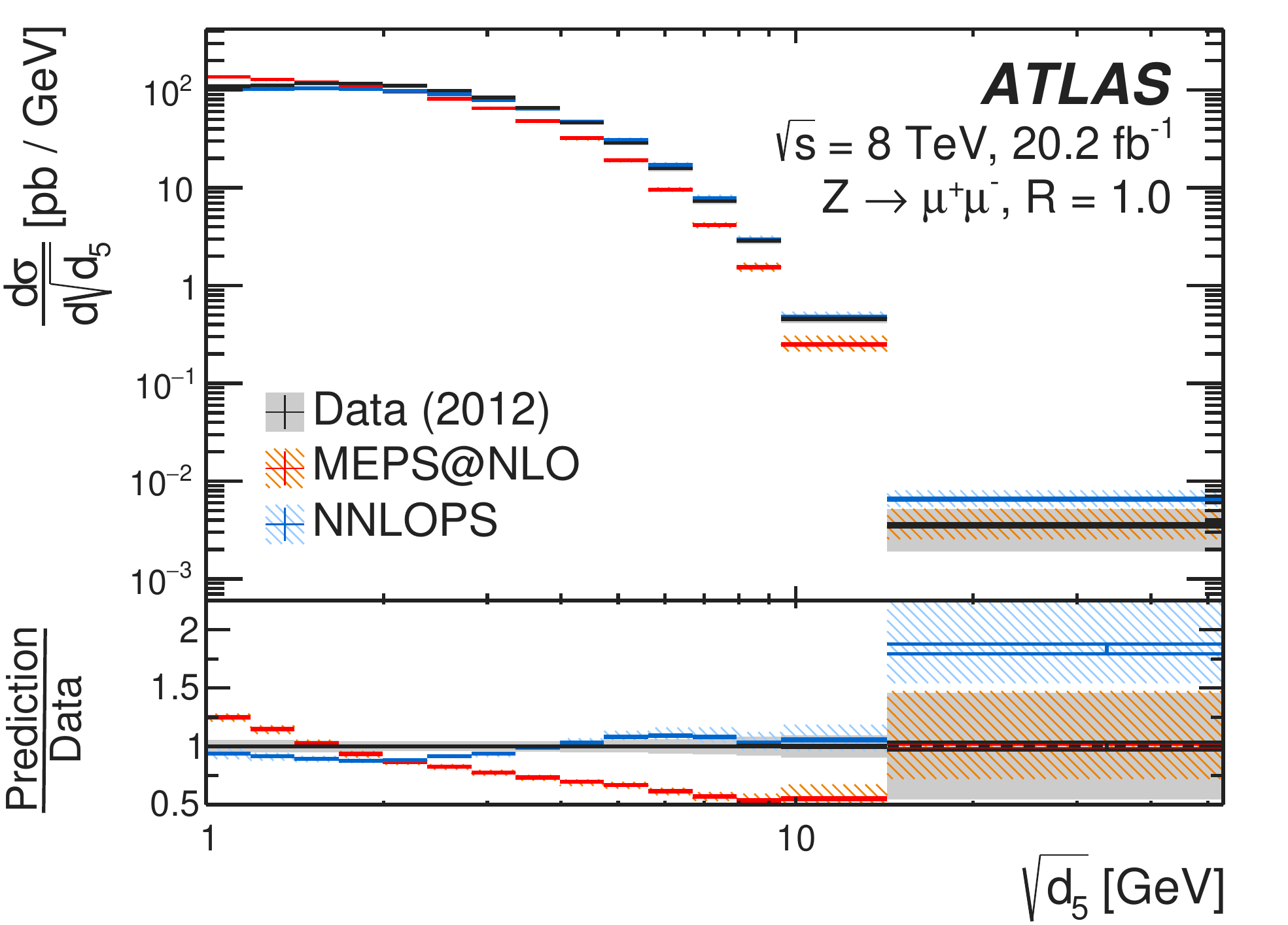}\\
  \includegraphics[width=0.45\textwidth]{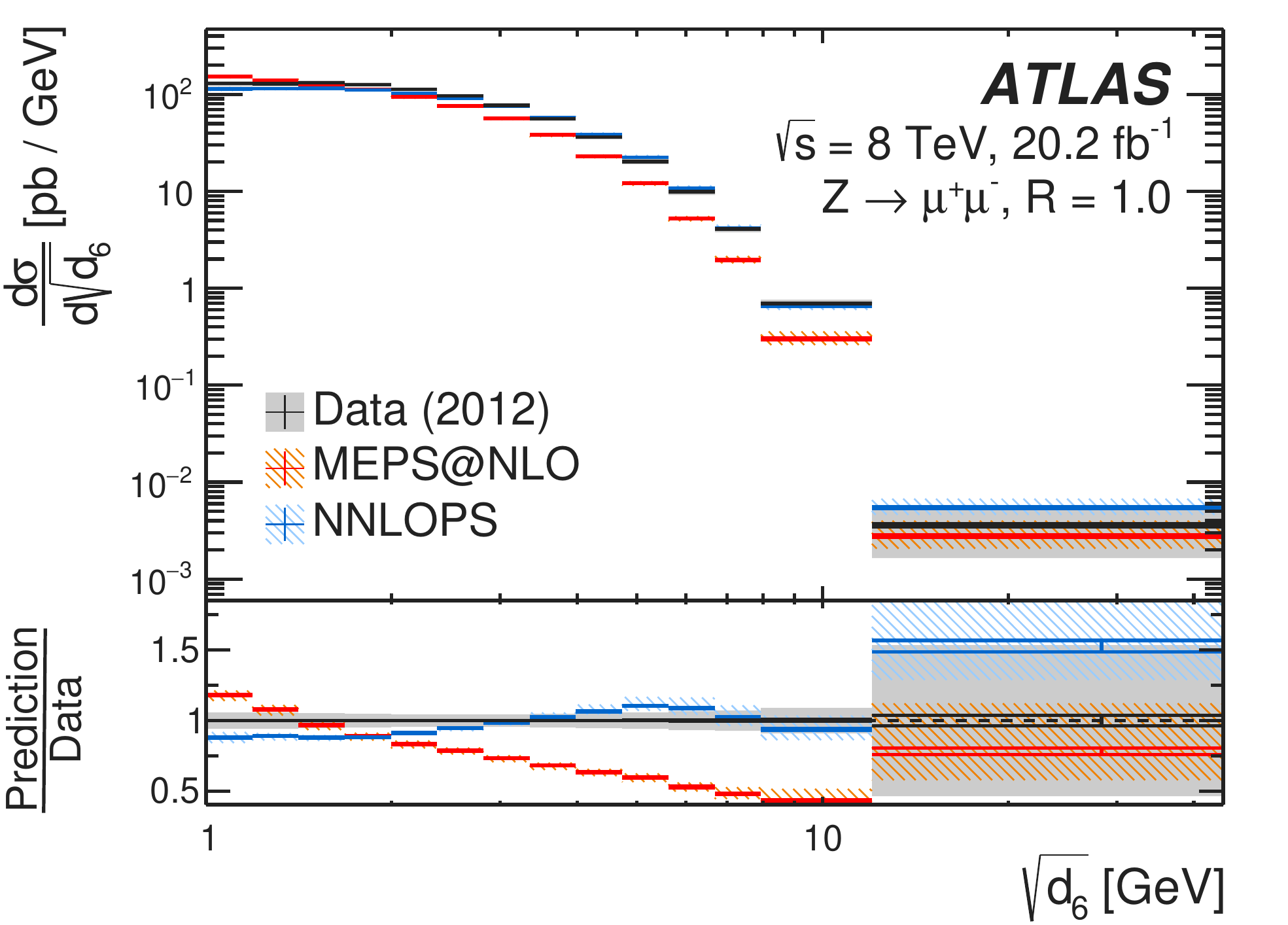}%
  \includegraphics[width=0.45\textwidth]{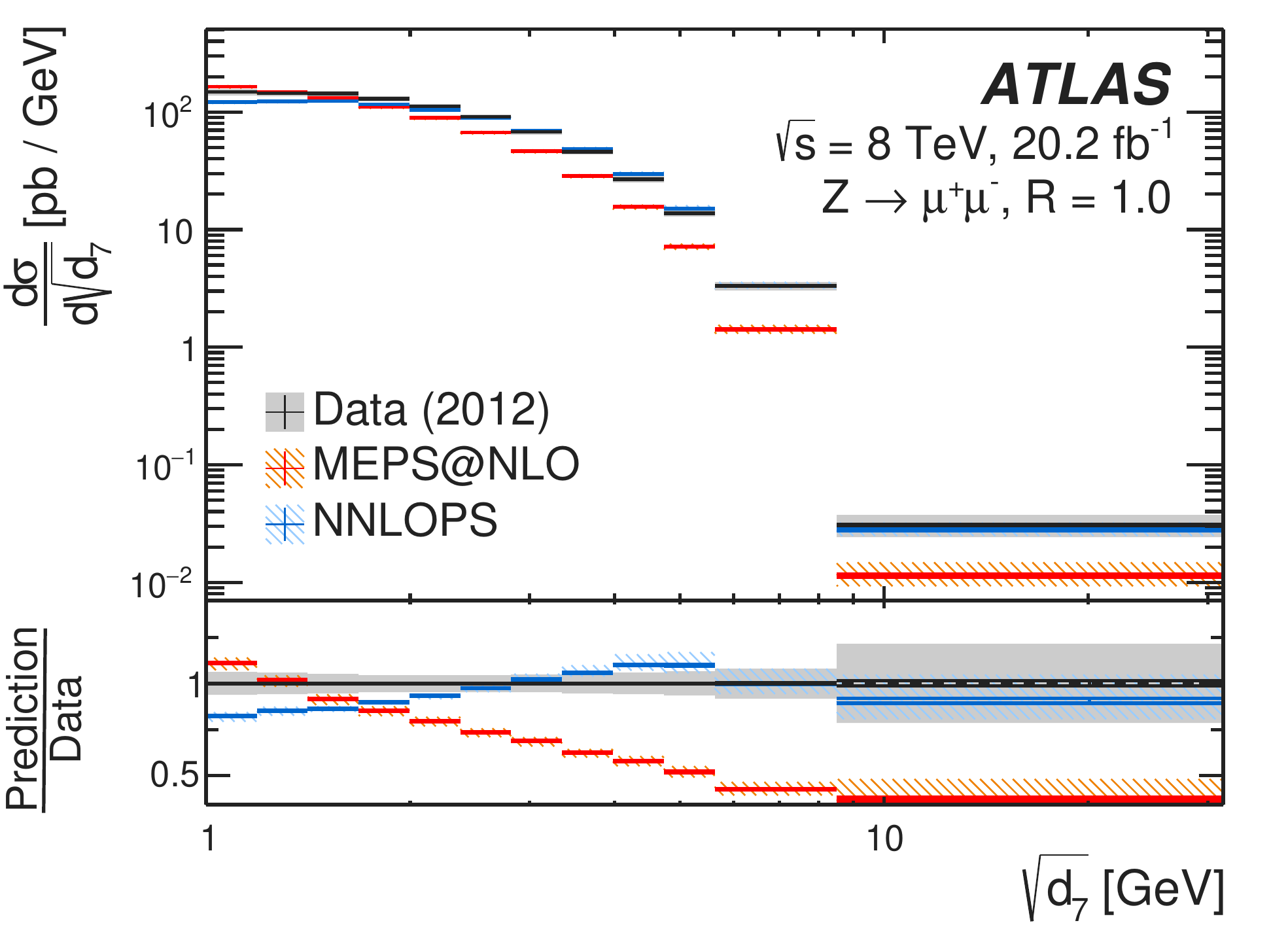}
  \caption{Charged-only distributions for the eight leading splitting scales
    in the muon channel using the jet-radius parameter $R=1.0$.
    The size of the error bars reflects the statistical uncertainty, while the combined statistical
    and systematic uncertainty is indicated by the grey band.
    Theoretical predictions from \sherpa\ with NLO multijet merging (``MEPS@NLO'') and from \powheg$+$\pythia\ with NNLO matching (``NNLOPS'') are displayed.
}
\end{figure}

\clearpage

\printbibliography

\newpage 
\begin{flushleft}
{\Large The ATLAS Collaboration}

\bigskip

M.~Aaboud$^\textrm{\scriptsize 137d}$,
G.~Aad$^\textrm{\scriptsize 88}$,
B.~Abbott$^\textrm{\scriptsize 115}$,
J.~Abdallah$^\textrm{\scriptsize 8}$,
O.~Abdinov$^\textrm{\scriptsize 12}$$^{,*}$,
B.~Abeloos$^\textrm{\scriptsize 119}$,
S.H.~Abidi$^\textrm{\scriptsize 161}$,
O.S.~AbouZeid$^\textrm{\scriptsize 139}$,
N.L.~Abraham$^\textrm{\scriptsize 151}$,
H.~Abramowicz$^\textrm{\scriptsize 155}$
% ...
\bigskip
\\
% ...
$^{8}$ Department of Physics, The University of Texas at Arlington, Arlington TX, United States of America\\
% ...
$^{12}$ Institute of Physics, Azerbaijan Academy of Sciences, Baku, Azerbaijan\\
% ...
$^{88}$ CPPM, Aix-Marseille Universit{\'e} and CNRS/IN2P3, Marseille, France\\
% ...
$^{115}$ Homer L. Dodge Department of Physics and Astronomy, University of Oklahoma, Norman OK, United States of America\\
% ...
$^{119}$ LAL, Univ. Paris-Sud, CNRS/IN2P3, Universit{\'e} Paris-Saclay, Orsay, France\\
% ...
$^{137}$ $^{(a)}$ Facult{\'e} des Sciences Ain Chock, R{\'e}seau Universitaire de Physique des Hautes Energies - Universit{\'e} Hassan II, Casablanca; $^{(b)}$ Centre National de l'Energie des Sciences Techniques Nucleaires, Rabat; $^{(c)}$ Facult{\'e} des Sciences Semlalia, Universit{\'e} Cadi Ayyad, LPHEA-Marrakech; $^{(d)}$ Facult{\'e} des Sciences, Universit{\'e} Mohamed Premier and LPTPM, Oujda; $^{(e)}$ Facult{\'e} des sciences, Universit{\'e} Mohammed V, Rabat, Morocco\\
% ...
$^{139}$ Santa Cruz Institute for Particle Physics, University of California Santa Cruz, Santa Cruz CA, United States of America\\
% ...
$^{151}$ Department of Physics and Astronomy, University of Sussex, Brighton, United Kingdom\\
% ...
$^{155}$ Raymond and Beverly Sackler School of Physics and Astronomy, Tel Aviv University, Tel Aviv, Israel\\
% ...
$^{161}$ Department of Physics, University of Toronto, Toronto ON, Canada\\
% ...
$^{a}$ Also at Department of Physics, King's College London, London, United Kingdom\\
$^{b}$ Also at Institute of Physics, Azerbaijan Academy of Sciences, Baku, Azerbaijan\\
$^{c}$ Also at Novosibirsk State University, Novosibirsk, Russia\\
$^{d}$ Also at TRIUMF, Vancouver BC, Canada\\
$^{e}$ Also at Department of Physics {\&} Astronomy, University of Louisville, Louisville, KY, United States of America\\
% ...
$^{*}$ Deceased
\end{flushleft}


\end{document}